\documentclass[longauth]{aa}

\usepackage{txfonts} 
\usepackage{amsmath,amssymb}
\usepackage{natbib}
\usepackage[squaren,Gray]{SIunits}
\usepackage{color, colortbl}
\usepackage{array}
\usepackage{longtable,lscape}
\usepackage{multirow}
\usepackage[dvipsnames]{xcolor}
\usepackage{scrextend}
\usepackage{graphicx}
\usepackage[flushleft,online]{threeparttable}
\usepackage[normalem]{ulem}
\usepackage{comment}
\usepackage{booktabs} 
\usepackage{multirow} 
\usepackage{soul} 

\newcommand{\Mjup}{M$_{\text{Jup}}$}

\newcommand{\bPic}{$\beta$ Pictoris}
\newcommand{\hr}{HR~4796}
\newcommand{\tyc}{TYC~9340-437-1}
\newcommand{\qEri}{$q^1$~Eri}

\newcommand {\micron}{\unit{}{\micro\meter}}

\usepackage{natbib,twoopt}
\usepackage[colorlinks=true,breaklinks=true]{hyperref}
\hypersetup{citecolor=blue}
\bibpunct{(}{)}{;}{a}{}{,} 
\makeatletter
\newcommandtwoopt{\citeads}[3][][]{\href{http://adsabs.harvard.edu/abs/#3}%
{\def\hyper@linkstart##1##2{}%
\let\hyper@linkend\@empty\citealp[#1][#2]{#3}}}
\newcommandtwoopt{\citepads}[3][][]{\href{http://adsabs.harvard.edu/abs/#3}%
{\def\hyper@linkstart##1##2{}%
\let\hyper@linkend\@empty\citep[#1][#2]{#3}}}
\newcommandtwoopt{\citetads}[3][][]{\href{http://adsabs.harvard.edu/abs/#3}%
{\def\hyper@linkstart##1##2{}%
\let\hyper@linkend\@empty\citet[#1][#2]{#3}}}
\newcommandtwoopt{\citeyearads}[3][][]%
{\href{http://adsabs.harvard.edu/abs/#3}
{\def\hyper@linkstart##1##2{}%
\let\hyper@linkend\@empty\citeyear[#1][#2]{#3}}}
\makeatother

\begin{document} 

   \title{The ALMA survey to Resolve exoKuiper belt Substructures (ARKS)}
  \subtitle{V: Comparison between scattered light and thermal emission}

\author{ 
J. Milli\inst{1} \and 
J. Olofsson\inst{2} \and 
M. Bonduelle\inst{1} \and 
R. Bendahan-West\inst{3} \and 
J. P. Marshall\inst{4} \and 
E. Choquet\inst{5} \and 
A. A. Sefilian\inst{6} \and 
Y. Han\inst{7} \and 
B. Zawadzki\inst{8} \and 
S. Mac Manamon\inst{9} \and 
E. Mansell\inst{8} \and 
C. del Burgo\inst{10,11} \and 
J. M. Carpenter\inst{12} \and 
A. M. Hughes\inst{8} \and 
M. Booth\inst{13} \and 
E. Chiang\inst{14} \and 
S. Ertel\inst{6,15} \and 
Th. M. Esposito\inst{16,17} \and 
Th. Henning\inst{18} \and 
J. Hom\inst{6} \and 
M. R. Jankovic\inst{19} \and 
A. V. Krivov\inst{20} \and 
J. B. Lovell\inst{21} \and 
P. Luppe\inst{9} \and 
M. A. MacGregor\inst{22} \and 
S. Marino\inst{3} \and 
B. C. Matthews\inst{23,24} \and 
L. Matr\`a\inst{9} \and 
A. Mo\'or\inst{25} \and 
N. Pawellek\inst{25,26} \and 
T. D. Pearce\inst{27} \and 
S. P\'erez\inst{28,29,30} \and 
V. Squicciarini\inst{3} \and 
P. Weber\inst{28,29,30} \and 
D. J. Wilner\inst{21} \and 
M. C. Wyatt\inst{31}
}

\institute{
Univ. Grenoble Alpes, CNRS, IPAG, F-38000 Grenoble, France
    email{julien.milli@univ-grenoble-alpes.fr} \and
European Southern Observatory, Karl-Schwarzschild-Strasse 2, 85748 Garching bei M\"unchen, Germany \and
Department of Physics and Astronomy, University of Exeter, Stocker Road, Exeter EX4 4QL, UK \and
Academia Sinica Institute of Astronomy and Astrophysics, 11F of AS/NTU Astronomy-Mathematics Building, No.1, Sect. 4, Roosevelt Rd, Taipei 106319, Taiwan \and
Aix Marseille Universit\'e, CNRS, CNES,  LAM, Marseille, France \and
Department of Astronomy and Steward Observatory, The University of Arizona, 933 North Cherry Ave, Tucson, AZ, 85721, USA \and
Division of Geological and Planetary Sciences, California Institute of Technology, 1200 E. California Blvd., Pasadena, CA 91125, USA \and
Department of Astronomy, Van Vleck Observatory, Wesleyan University, 96 Foss Hill Dr., Middletown, CT, 06459, USA \and
School of Physics, Trinity College Dublin, the University of Dublin, College Green, Dublin 2, Ireland \and
Instituto de Astrof\'isica de Canarias, Vía L\'actea S/N, La Laguna, E-38200, Tenerife, Spain \and
Departamento de Astrof\'isica, Universidad de La Laguna, La Laguna, E-38200, Tenerife, Spain \and
Joint ALMA Observatory, Avenida Alonso de C\'ordova 3107, Vitacura 7630355, Santiago, Chile \and
UK Astronomy Technology Centre, Royal Observatory Edinburgh, Blackford Hill, Edinburgh EH9 3HJ, UK \and
Department of Astronomy, University of California, Berkeley, Berkeley, CA 94720-3411, USA \and
Large Binocular Telescope Observatory, The University of Arizona, 933 North Cherry Ave, Tucson, AZ, 85721, USA \and
Astronomy Department, University of California, Berkeley, CA 94720, USA \and
SETI Institute, Carl Sagan Center, 339 Bernardo Ave., Suite 200, Mountain View, CA 94043, USA \and
Max-Planck-Institut f\"ur Astronomie, K\"onigstuhl 17, 69117 Heidelberg, Germany \and
Institute of Physics Belgrade, University of Belgrade, Pregrevica 118, 11080 Belgrade, Serbia \and
Astrophysikalisches Institut und Universit\"atssternwarte, Friedrich-Schiller-Universit\"at Jena, Schillerg\"asschen 2-3, 07745 Jena, Germany \and
Center for Astrophysics | Harvard \& Smithsonian, 60 Garden St, Cambridge, MA 02138, USA \and
Department of Physics and Astronomy, Johns Hopkins University, 3400 N Charles Street, Baltimore, MD 21218, USA \and
Herzberg Astronomy \& Astrophysics, National Research Council of Canada, 5071 West Saanich Road, Victoria, BC, V9E 2E9, Canada \and
Department of Physics \& Astronomy, University of Victoria, 3800 Finnerty Rd, Victoria, BC V8P 5C2, Canada \and
Konkoly Observatory, HUN-REN Research Centre for Astronomy and Earth Sciences, MTA Centre of Excellence, Konkoly-Thege Mikl\'os \'ut 15-17, 1121 Budapest, Hungary \and
Institut f\"ur Astrophysik (IfA), University of Vienna, T\"urkenschanzstra\ss{}e 17, A-1180 Vienna, Austria \and
Department of Physics, University of Warwick, Gibbet Hill Road, Coventry CV4 7AL, UK \and
Departamento de Física, Universidad de Santiago de Chile, Av. V\'ictor Jara 3493, Santiago, Chile \and
Millennium Nucleus on Young Exoplanets and their Moons (YEMS), Chile \and
Center for Interdisciplinary Research in Astrophysics Space Exploration (CIRAS), Universidad de Santiago, Chile \and
Institute of Astronomy, University of Cambridge, Madingley Road, Cambridge CB3 0HA, UK
}

\date{Received: 21 July 2025 ; accepted: 3 December 2025 }

 
  \abstract
   {
   Debris discs are analogues to our own Kuiper belt around main-sequence stars and are therefore referred to as exoKuiper belts. They have been resolved at high angular resolution at wavelengths spanning the optical/near-infrared to the submillimetre-millimetre regime. Short wavelengths can probe the light scattered by such discs, which is dominated by micron-sized dust particles,  while millimetre wavelengths can probe the thermal emission of millimetre-sized particles. Determining differences in the dust distribution between millimetre- and micron-sized dust is fundamental to revealing the dynamical processes affecting the dust in debris discs. 
   }
   {
  We aim to compare the scattered light from the discs of the `ALMA survey to Resolve exoKuiper belt Substructures' (ARKS) with the thermal emission probed by ALMA. We focus on the radial distribution of the dust, and we also put constraints on the presence of giant planets in those systems. 
  }
    {
     We used high-contrast scattered light observations obtained with VLT/SPHERE, GPI, and the {\em HST} to uniformly study the dust distribution in those systems and compare it to the dust distribution extracted from the ALMA observations carried out in the course of the ARKS project. We also set constraints on the presence of planets by using these high-contrast images combined with exoplanet evolutionary models.
   }
   {
   Fifteen of the 24 discs comprising the ARKS sample are detected in scattered light, with \tyc{} being imaged for the first time at near-infrared wavelengths. For six of those 15 discs, the dust surface density seen in scattered light peaks farther out compared to that observed with ALMA. These six discs except one are known to also host cold CO gas. Conversely, the systems without significant offsets are not known to host gas, except one. Moreover, with our scattered light near-infrared images, we achieve typical sensitivities to planets from 1 to 10 $M_\text{Jup}$ beyond 10 to 20 au, depending on the system age and distance. 
   }
   {
    This observational study suggests that the presence of gas in debris discs may affect the small and large grains differently, pushing the small dust to greater distances where the gas is less abundant.  
   }



   \maketitle
%
 
\section{Introduction}
\label{sec_intro}

Debris discs, analogues to our own Kuiper belt, have been detected around more than a thousand main-sequence stars \citep{Cao2023}, mainly through the infrared excess emitted by cold circumstellar dust (typically $T\leq100$~K). This dust is thought to originate from kilometre-sized or larger planetesimals orbiting in a birth ring and colliding to produce smaller and smaller debris \citep{Wyatt2008}. This continuous collisional cascade explains the presence of short-lived dust particles after millions or billions of years around a star. \\
The first debris disc was detected by the {\em InfraRed Astronomical Satellite} \citep[IRAS,][]{Aumann1984}. Before the 2000s, the limited angular resolution and contrast capabilities of ground- and space-based instruments led to the resolved optical or near-infrared imaging of only two such systems: \bPic{} and \hr{} \citep{Smith1984,Schneider1999}. Recent advances in high-contrast imaging have enabled observations of many debris discs in scattered light, either in total intensity or polarimetry \citep[e.g.][]{Schneider2014,Esposito2020,Ren2023} using the \textit{Hubble Space Telescope} ({\em HST}), Spectro-Polarimetric High-contrast Exoplanet REsearch \citep[SPHERE;][]{Beuzit2019} on the \emph{Very Large Telescope} (\emph{VLT}), and the Gemini Planet Imager \citep[GPI;][]{Macintosh2014}. Such optical or near-infrared observations are sensitive to the starlight scattered by particles with sizes similar to or smaller than the wavelength, which dominate the scattering cross-section \citep{Thebault2019} and still have a sufficient scattering efficiency. 

In parallel, over the past decade, millimetre and submillimetre interferometry has yielded images of such systems, including the recent REsolved ALMA and SMA Observations of Nearby Stars \cite[REASONS;][]{Matra2025} survey, opening the path for a population study based on high spatial resolution images of 74 of such systems. The subsequent `ALMA survey to Resolve exo\-Kuiper belt Substructures' \cite[ARKS;][]{overview_arks} acquired the highest angular resolution on a subsample of 24 discs, reaching, for some systems, a similar resolution as optical or near-infrared imaging ($\sim40$~mas). As observations of dust thermal emission are most sensitive to particles with sizes similar to the observational wavelength \citep{Hughes2018}, the high resolution offered by the ARKS program presents the possibility to study, in detail, where the millimetre-sized dust grains are located. Comparing ALMA observations (millimetre-sized grains) to scattered light images (micron-sized particles or below, at the bottom of the collisional cascade) therefore allows mapping of the spatial distribution of grains with different sizes in the disc. 

In optically thin discs, the orbits of the dust particles depend on various mechanisms on top of the stellar gravitational field. These include radiation and wind forces from the central star \citep{Burns1979}, collisions, gravitational perturbation due to planets, stellar companions \citep{Mustill2009,Nesvold2016,Farhat2023} or even the debris disc itself \citep{Sefilian2024}, in addition to gas drag \citep{Takeuchi2001}. Small grains are typically affected more strongly by these non-gravitational forces than their larger counterparts \citep{Pawellek2019}. 

The outward force induced by the star's radiation pressure is usually defined by its ratio, called $\beta$, to the opposing gravitational pull from the star. The value of $\beta$ determines the nature of the orbits. In particular, the smallest dust grains created by collisions have a higher $\beta$ value than larger grains, resulting in more elliptical orbits \cite[with an eccentricity $e\sim\beta/(1-\beta)$,][]{Strubbe2006}.
This effect naturally creates a dust size segregation that extends beyond the birth ring, with smaller dust particles seen at larger distances \citep{Krivov2010}. This can result in an extended halo of small dust particles, which are detectable in scattered light with sensitive wide-band optical imagers \cite[e.g.][for HR~4796]{Schneider2018} or with polarimetry \cite[e.g.][for HD~129590]{Olofsson2024}.

Conversely, Poynting-Robertson drag, which causes dust grains to lose angular momentum due to stellar radiation, can induce inward migration \citep{Kennedy2015,Pawellek2019,Su2024,Sommer2025}. In late-type stars, strong stellar winds may also induce an inward migration, possibly even exceeding the Poynting-Robertson drag effect \citep{Plavchan2005,Pawellek2019}.

In the birth ring, where the dust density is the highest, the high collision rate will also impact the dust population through the grinding of bigger grains into smaller particles. Other dynamical effects in the debris disc may be due to the gravitational perturbation of planetary companions or gas drag for discs hosting a sufficient amount of gas. We now know of more than 20 debris discs that contain gas, both hot \citep{Rebollido2018} and cold \citep{Iglesias2018,Cataldi2023}. 

Theoretical studies have shown that small and large grains may undergo different migration in the presence of cold gas at a few tens of au \citep{Takeuchi2001,Krivov2009}. To test those theories against observational evidence, we propose using the ARKS sample to compare in detail the surface density of the millimetre-sized dust detected with ALMA in thermal emission with that of the micron-sized dust seen in scattered light with optical or near-infrared imagers.

The remainder of the paper proceeds as follows. After presenting the scattered light observations of the ARKS sample and the methodology in Sect. \ref{sec_obs}, we present the detections in Sect. \ref{sec_detections}. We compare the dust surface density as estimated from the scattered light images and thermal emission in Sect. \ref{sec_comparison}. We  place constraints on the presence of planets in Sect. \ref{sec_sensitivity_to_planets} and discuss the results in Sect. \ref{sec_discussion} before concluding in Sect. \ref{sec_conclusions}.

\section{The ARKS scattered light sample and methods}
\label{sec_obs}

\subsection{Description of the ARKS sample and scattered light observations}

The 24 discs comprising the ARKS sample are listed in Table~\ref{tab_targets} of App. \ref{app_ARKS_sample} along with important stellar properties. We refer the reader to \citet{overview_arks} for a comprehensive description of the sample and its selection. All targets were observed with various high-contrast imagers. In particular, all of them were scrutinised by the SPHERE instrument and its near-infrared imager IRDIS \citep{Beuzit2019}. 
The shaded rows in Table~\ref{tab_targets} show the 15 discs detected in scattered light and analysed in this paper. All but one of the scattered light detections are archival observations that have already been presented in the literature. The last column of Table \ref{tab_targets} lists those references. The only exception is \tyc, for which we present the first scattered-light detection obtained with the {\em HST} in Appendix \ref{app_TYC}. 
For each target in this study, we used the instrument and observation that achieve the highest angular resolution while maintaining a sufficient signal-to-noise ratio (S/N) to allow for a meaningful analysis. In most cases, we used observations taken with SPHERE - IRDIS, either in linear polarisation or in total intensity.  In one case, we used \emph{GPI} in its polarimetric mode. For the faintest discs, or those with a low inclination, they could only be detected with the \emph{HST}, with one of the following three instruments: the \emph{Advanced Camera for Surveys} (ACS), the \emph{Near Infrared Camera and Multi-Object Spectrometer} (NICMOS) or the \emph{Space Telescope Imaging Spectrograph} (STIS). 

\subsection{Surface density extraction}

The surface brightness of a disc scales differently with the distance to the star $r$ in scattered light and thermal emission. Assuming a constant scattering efficiency across the disc, the scattered light scales as $1/r^2$ while the thermal emission scales as $1/\sqrt{r}$ in the Rayleigh-Jeans blackbody approximation of a dust temperature profile $T = 278.3 L_{\star}^{1/4} r^{-1/2}$ \citep{Wyatt2008}, where $T$ is in Kelvin, $L_{\star}$ is the stellar luminosity in solar units and $r$ is in au. To be able to consistently compare the radial profiles from scattered light and thermal emission, we decided to compare the surface density profiles rather than the surface brightness profiles.

\subsubsection{ALMA}
\label{subsec_ALMA}

For ALMA, we used the surface density as estimated with the non-parametric technique called \emph{frank}\footnote{For one target, \bPic, \emph{frank} is not reliable because it does not handle multiple pointings, so we used the alternative non-parametric technique \emph{rave} operating in the image space.} \citep{Jennings2020,Terrill2023}. Compared to the alternative non-parametric technique \emph{rave} \citep{Han2022,Han2025}, which uses images obtained with the \emph{clean} image reconstruction technique when fitting to ALMA data, which can be imperfect, \emph{frank} works directly in visibility space and can be more sensitive to very sharp features. As a double-check, we also compared the surface density with that extracted from a double power-law parametric modelling, because we employed the same functional form for the scattered light modelling. This form is often employed to measure the steepness of the inner and outer profiles and infer the dynamical excitation of the disc. For single belts, the double power-law parametric modelling is consistent with the non-parametric approach, but this form does not fit well all the data, especially structured belts or multiple-ring which are better fit by more complex parametric forms (triple power law, double gaussian, possibly including gaps). In all cases (parametric or non-parametric approaches), the 1D radial brightness profiles of discs are converted to surface density profiles assuming a blackbody equilibrium temperature profile, stellar luminosities and the mass opacity as described in \citet{overview_arks}, i.e. 1.9\,cm$^2$\,g$^{-1}$ for ALMA Band 7, 1.3\,cm$^2$\,g$^{-1}$ for ALMA Band 6.
A full description of the surface density extraction from the ALMA data is provided in \citet{rad_arks} and \citet{ver_arks}. 

\subsubsection{Scattered light}

For the scattered light observations, we fitted a scattered light model to the disc images  where the dust surface density is parameterised with a double power-law, following the model introduced in \cite{Augereau2001} and detailed in Appendix \ref{app_parametrisation}. This is a simple ray-tracing model without any dust physics. It assumes a constant scattering efficiency across the disc, and a non-isotropic scattering phase function (either a parametric form or an empirical scattering phase function). Because retrieving the total intensity image of a disc is different from retrieving the polarised intensity, the fitting procedure is also specific to each case. 

In polarised intensity (written \emph{pI} in Table \ref{tab_targets}, which applies to HD~121617, AU Mic, \bPic, HR~4796, HD~145560 and HD~32297), we used the image corresponding to the azimuthal Stokes parameter $Q_\phi$ to directly fit a disc model convolved with the instrumental Point-Spread Function (PSF), following the methodology described in \citet{Olofsson2020}.

For observations carried out in total intensity (written \emph{I} in Table \ref{tab_targets}), the processing is different whether the image comes from the \emph{VLT - SPHERE}, from \emph{HST - ACS} or from \emph{HST - NICMOS}. For SPHERE images (HD~131835, HD~131488, AU Mic, HD~15115, 49~Ceti, HD~61005, HD~32297, HR~4796), the post-processing technique that allows the disc detection is angular differential imaging (ADI). In this case, finding the best model requires forward modelling, a technique where the disc model is subtracted from the data, which are then re-processed to analyse the residual image. These steps are repeated by adapting the disc model until the residuals are minimised. 
For the {\em HST - ACS} image of HD~10647 ($q^1$~Eri), we directly fitted a disc model to the image presented in \citet{Lovell2021}. We did not perform any forward modelling, as the PSF subtraction technique applied is roll subtraction with a roll amplitude of $\sim25^\circ$. This amplitude is sufficiently large with respect to where the disc is detected to not be biased by self-subtraction artefacts \cite[see][for details on the data reduction]{Lovell2021}. For the {\em HST - NICMOS} image of \tyc, the post-processing technique to reveal the disc is reference differential imaging (see Appendix \ref{app_TYC}), and the same forward modelling procedure as for the \emph{VLT - SPHERE} images is applied. 

There are three discs for which the dust surface densities were already extracted and presented in previous publications, namely HD~92945, HD~107146, and 49~Ceti \citep{Golimowski2011,Ertel2011,Choquet2016}. Therefore, we did not carry out dedicated forward modelling for these systems. For 49~Ceti, the exact same double-power law modelling was already presented in \citet{Choquet2016} based on VLT - SPHERE and {\em HST} - NICMOS data. We therefore re-used the published results, and show in this paper the best model detailed in their Table 1 for the VLT - SPHERE dataset, as both images had a similar S/N but VLT - SPHERE provides the highest angular resolution. For HD~92945, we used the published non-parametric surface density extracted in \citet{Golimowski2011}, as this disc contains radial substructures that are not well represented by a double-power law profile. For HD~107146, the surface density was extracted in \citet{Ertel2011} with a parametric modelling akin to a double-power law profile.

We determined the best models with a Nested Sampling Monte Carlo framework \cite[implemented in the Python package \emph{PyMultiNest},][]{Buchner2014} using uniform priors. In particular, we did not use any prior information based on the ALMA best-fit models \citep{overview_arks}. This was done on purpose to not  bias the results. The only exception is \tyc, for which the S/N is poor, and therefore we had to fix the inclination and position angle to the best value found with ALMA. The best-fitting parameters are shown in Appendix \ref{app_modelling_results} along with their associated uncertainty. For most discs, the inclination and position angle extracted from the ALMA and scattered light images are compatible at a $3\sigma$ level. However, there are six exceptions, highlighted in grey in Table \ref{tab_modelling_results}, corresponding to discs inclined by $>74^\circ$: \bPic,  HD~32297, HD~15115, AU~Mic, HD~131835 and $q^1$ Eri. A discussion of these differences is given in Appendix \ref{app_modelling_results}.

\section{The ARKS discs detected in scattered light}
\label{sec_detections}

\begin{figure*}
    \centering
   \includegraphics[height=\hsize]{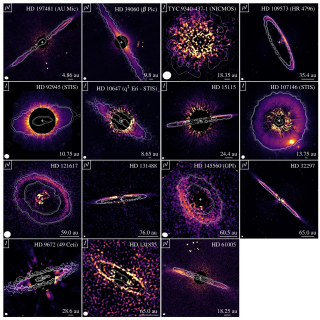}
   \caption{ARKS discs detected in scattered light. The observations were made using the VLT - SPHERE instrument, unless otherwise specified. The discs denoted by `$I$' were observed in total intensity, whereas the ones with `$pI$' were observed in polarised light. The white line at the bottom right corner represents an angular scale of 0.5\farcs, with the corresponding projected distance in au labelled in each panel. The white contours show the ALMA continuum observations, with three contour levels at a S/N of 3, 5 and 7 $\sigma$ \cite[see the corresponding ALMA image in Paper I Fig. 3 ][for each disc]{overview_arks}. The white ellipse at the bottom left represents the ALMA image resolution and the white cross is the stellar position from GAIA DR3. For HD~131835, we show the image post-processed with the PACO algorithm \citep{Flasseur2018}, best highlighting the two rings. In all panels, North is up, East to the left.} 
    \label{fig_mosaici}
    \end{figure*}

\begin{figure*}
    \centering
   \includegraphics[width=\textwidth]{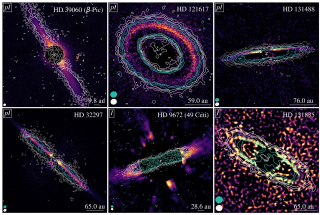}
   \caption{Discs from the ARKS sample that have gas detections. The scattered light observations are the same as in Fig.~\ref{fig_mosaici}. Overlaid on the scattered light images are the moment 0 maps for the gas observed with ALMA, using the robust values shown in \citet{gas_arks}, i.e. 0.5 for \bPic, HD 121617, 49 Ceti, HD 131835 and 2.0 for HD 131488 and HD 32297. The $^{12}$CO (J=3-2 line) is plotted in white and the $^{13}$CO (J=3-2 line) in green, with the exception of \bPic ~where only the $^{12}$CO (J=2-1 line) is plotted, as this disc is not detected in $^{13}$CO. For every disc, there are 3 contour levels, corresponding to 3, 5, and 7 $\sigma$. The ellipses at the bottom left (white for $^{12}$CO and green for $^{13}$CO) represent the beam size for the ALMA observations. In all panels, North is up, East to the left.} 
    \label{fig_mosaicgas}
    \end{figure*}

\subsection{Detection versus non-detection}

Among the 24 discs of the ARKS sample, we detected 15 of them in scattered light and have nine non-detections (shaded vs non-shaded rows in Table~\ref{tab_targets}, respectively). Fig. \ref{fig_mosaici} shows a mosaic of the scattered light images, overlaid with contours from the ALMA continuum. 
From the detections vs non-detections, some trends already emerge. Among the ten highly inclined discs (last ten rows in Table \ref{tab_targets}), all except one (HD~14055, which has the lowest fractional luminosity $L_\mathrm{disc}/L_* <10^{-4}$) are detected in scattered light. However, only six out of the 14 moderately inclined discs are recovered in scattered light, and these six discs are among the seven brightest in terms of fractional luminosity. More quantitatively, the faintest highly inclined disc detected in scattered light is $q^1$~Eri (HD~10647) with a fractional luminosity of $2.6\times10^{-4}$ and the faintest moderately inclined disc detected in scattered light is HD~92945 with a fractional luminosity of $6.6\times10^{-4}$ \citep{Matra2025}. Previous scattered light surveys have indeed already shown that detections are strongly correlated with high infrared excess and highly inclined discs \citep{Esposito2020, Engler2025}. This is a combination of two effects: more dust intercepts the line of sight for inclined systems, and the scattering phase function generally favours forward scattering \citep{Hapke2012} 

\subsection{Systems with CO gas detections}

Six of the 15 discs detected in scattered light also contain CO gas detected with ALMA and they are highlighted in bold face in Table \ref{tab_targets}. We therefore also compared the dust surface density with the CO intensity extracted from the ALMA data, as detailed in the companion ARKS paper by \citet{gas_arks}. The scattered light images are overlaid with the contours of the ALMA $^{12}$CO and $^{13}$CO (J=3-2) moment 0 maps (velocity integrated intensity) in Fig. \ref{fig_mosaicgas} (for \bPic~this is the J=2-1 transition). Five of them are considered gas-rich, whereas HD 39060 (\bPic) is considered gas-poor ($M_\mathrm{CO} \lesssim 10^{-4} M_\oplus$). 

\section{Comparison of the radial profiles}
\label{sec_comparison}

The radial profiles of the dust surface density are shown in Fig. \ref{fig_radial_profile_comparison} and \ref{fig_radial_profile_comparison_cont}. The uncertainty on the scattered light profile is the $1$-$\sigma$ error estimated from the Monte Carlo fitting process\footnote{We sampled 100 profiles from the Monte Carlo posterior distribution parametrising the surface density, and defined the lower and upper uncertainty shown in Fig. \ref{fig_radial_profile_comparison} and \ref{fig_radial_profile_comparison_cont} as the $16^\text{th}$ and $84^\text{th}$ percentile of the surface density for each distance in au.}. The uncertainty on the ALMA surface density is the $1$-$\sigma$ error  estimated from the \emph{frank} algorithm\footnote{For $\beta$ Pic where \emph{rave} was used instead of \emph{frank} to extract the ALMA dust surface density, the uncertainty is the $1$-$\sigma$ error  estimated from \emph{rave}.} \citep{rad_arks}.

Table \ref{tab_offsets} displays the peak locations of the dust surface density. For the surface density extracted from the scattered light image, the peak location can be computed analytically based on the dust density parametrisation (see Appendix \ref{app_parametrisation_scatt_light}and Eq. \ref{eq_max_surface_density}). The same can be done with the surface density extracted from the ALMA image (see Appendix \ref{app_parametrisation_alma} and Eq. \ref{eq_max_surface_density_alma}). For the ALMA surface density extracted with the non-parametric approaches \emph{frank} (or \emph{rave} for \bPic), we defined the uncertainty on the peak location as the range of separations where the $1\sigma$ upper bound profile of the surface density is above the lower bound of the maximum surface density. Based on these values and uncertainties for the maximum surface density, we computed the relative offset between scattered light and ALMA. We describe the comparison to the ALMA parametric profiles only when there is disagreement with the ALMA non-parametric profiles.

We can distinguish three categories of discs, depending on this offset. When a zero offset is compatible with the $1\sigma$ uncertainty, we classify the offset as not significant (group described in Sect. \ref{sec_groupA}). In the other cases, the offset can be either positive (meaning micron-sized dust is seen further out than millimetre-sized dust, described in Sect. \ref{sec_groupB}) or negative (described in Sect. \ref{sec_groupC}, for a single disc HD~61005). 

 \begin{figure*}
    \centering
   \includegraphics[width=0.46\hsize]{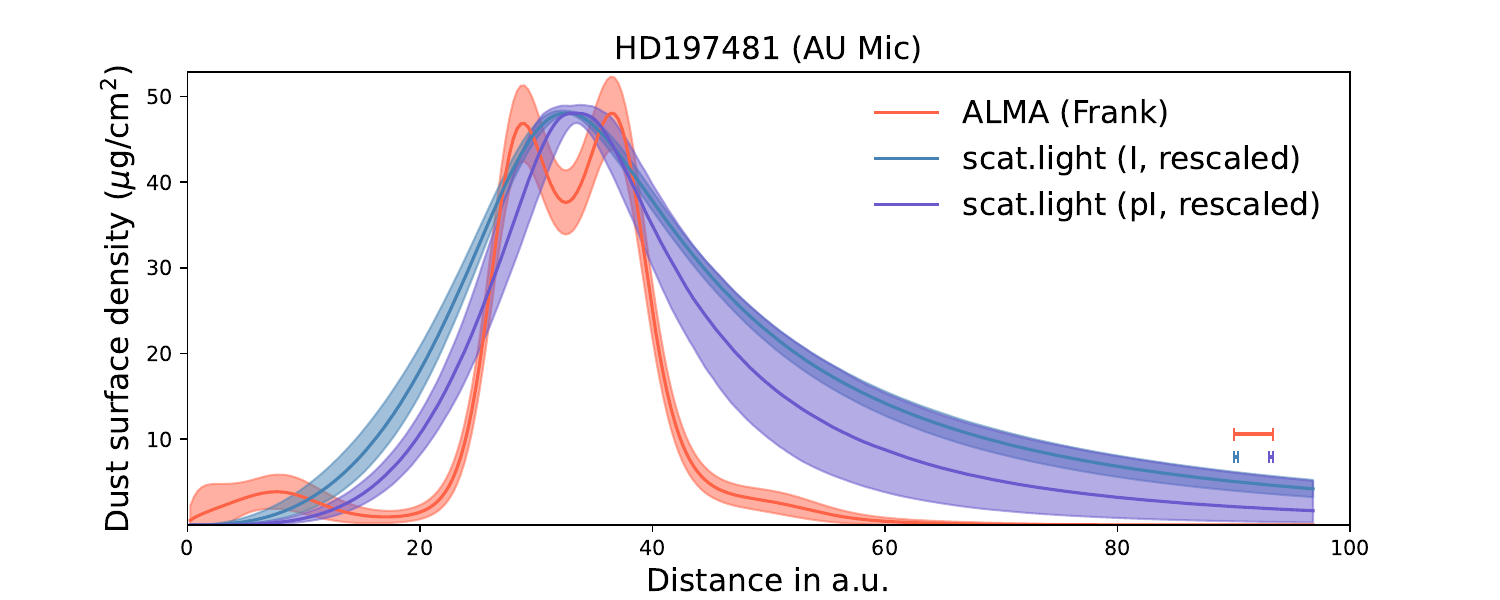}
   \includegraphics[width=0.46\hsize]{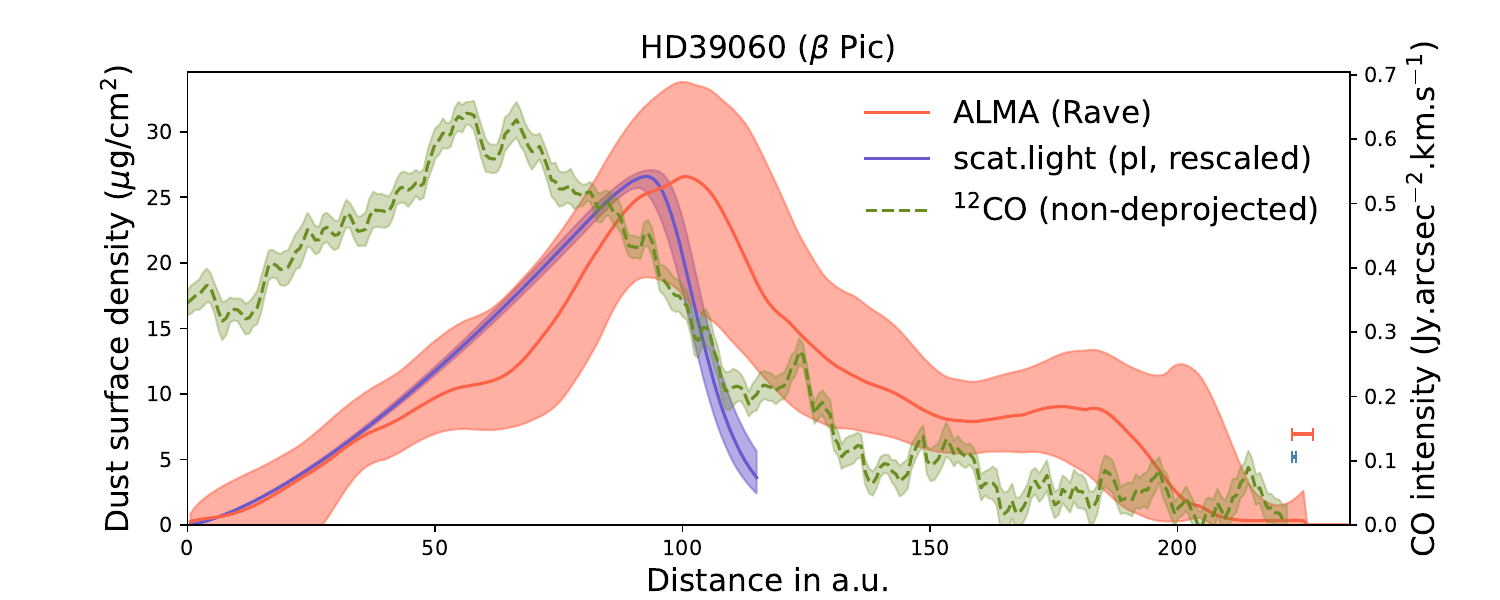}
   \includegraphics[width=0.46\hsize]{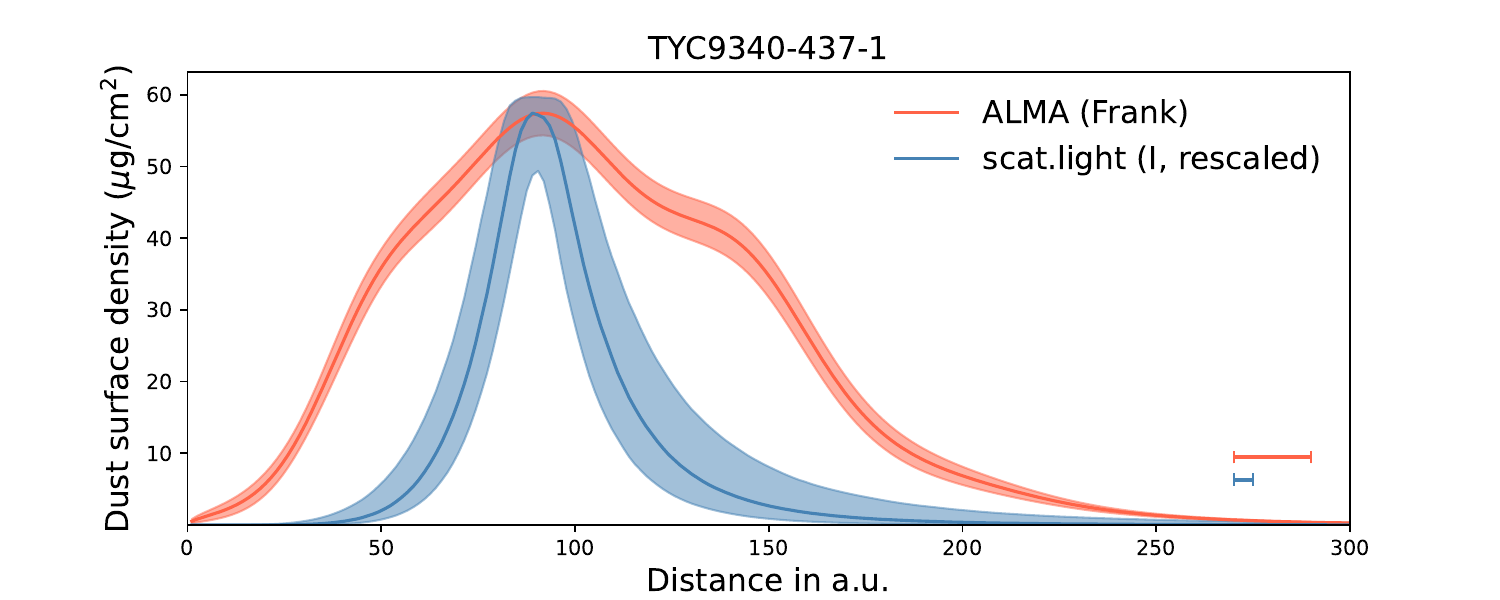}
   \includegraphics[width=0.46\hsize]{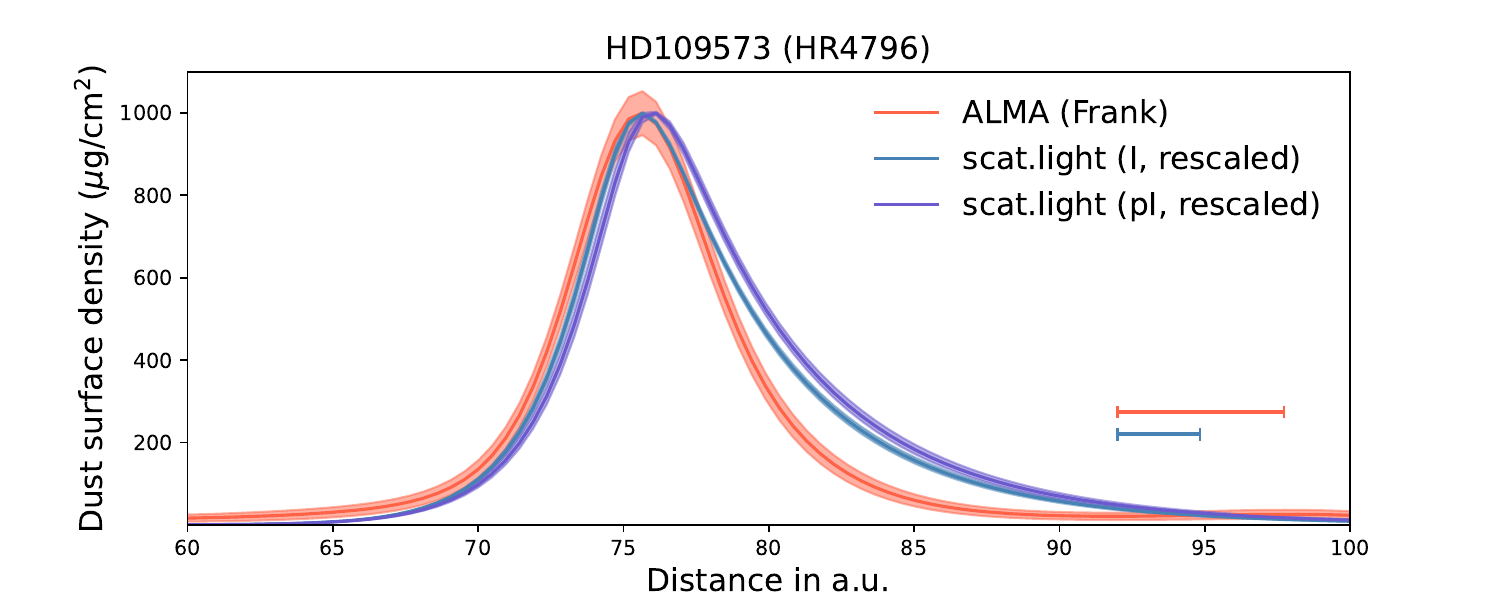}
   \includegraphics[width=0.46\hsize]{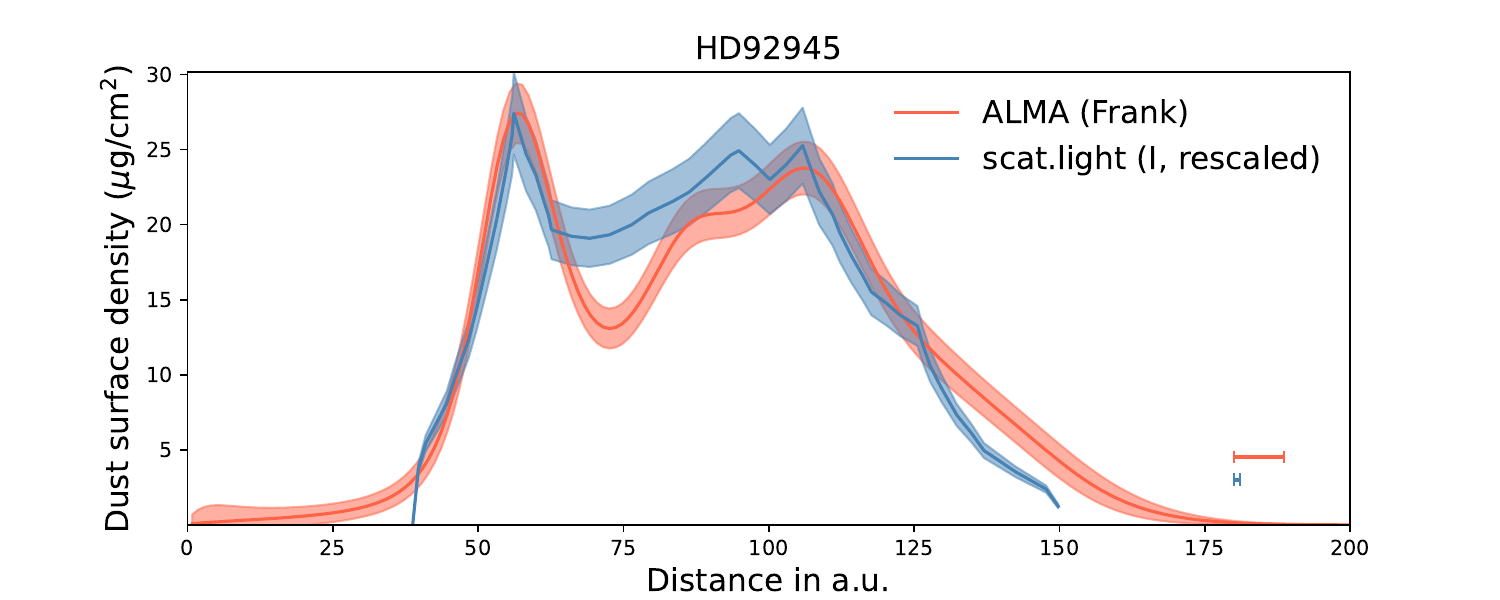}
   \includegraphics[width=0.46\hsize]{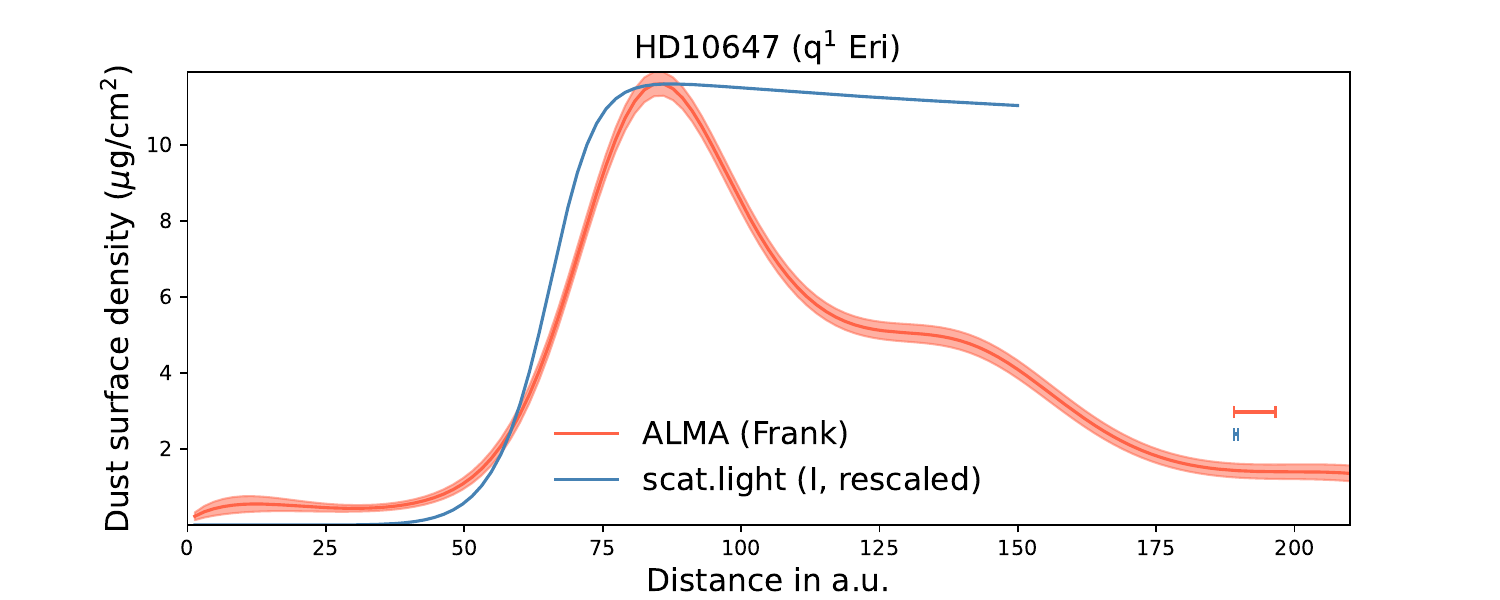}
   \includegraphics[width=0.46\hsize]{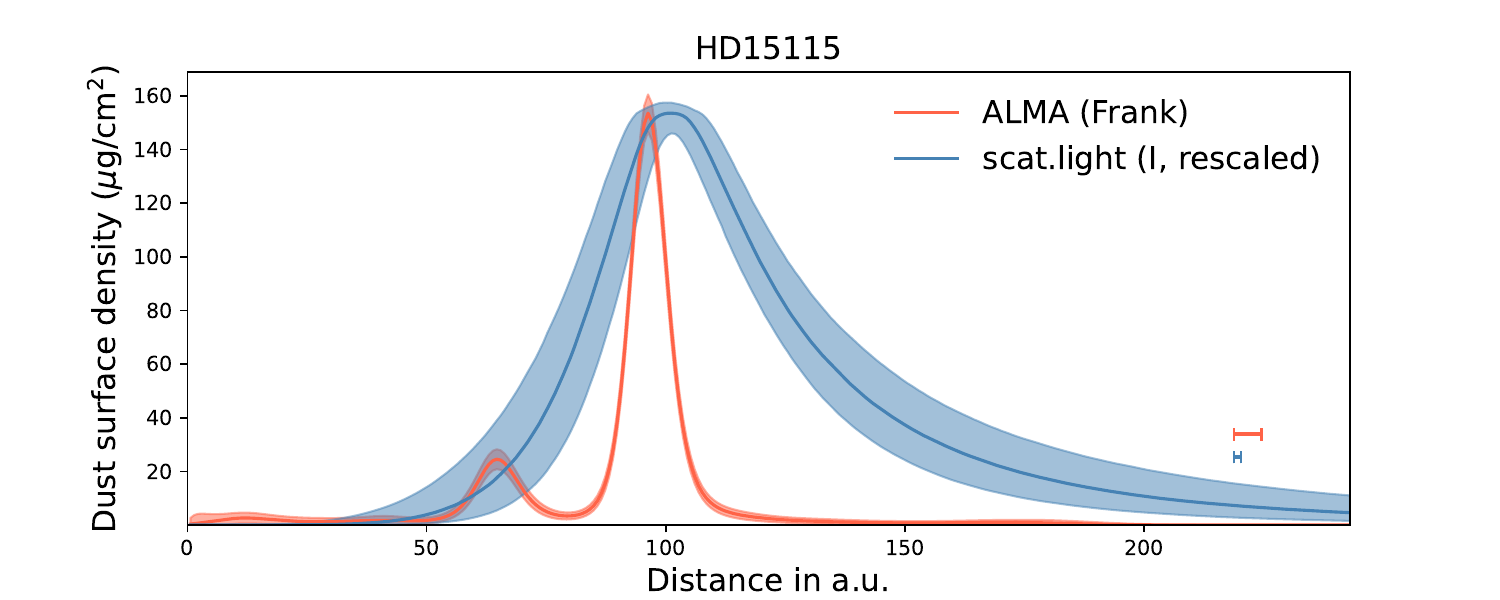}
   \includegraphics[width=0.46\hsize]{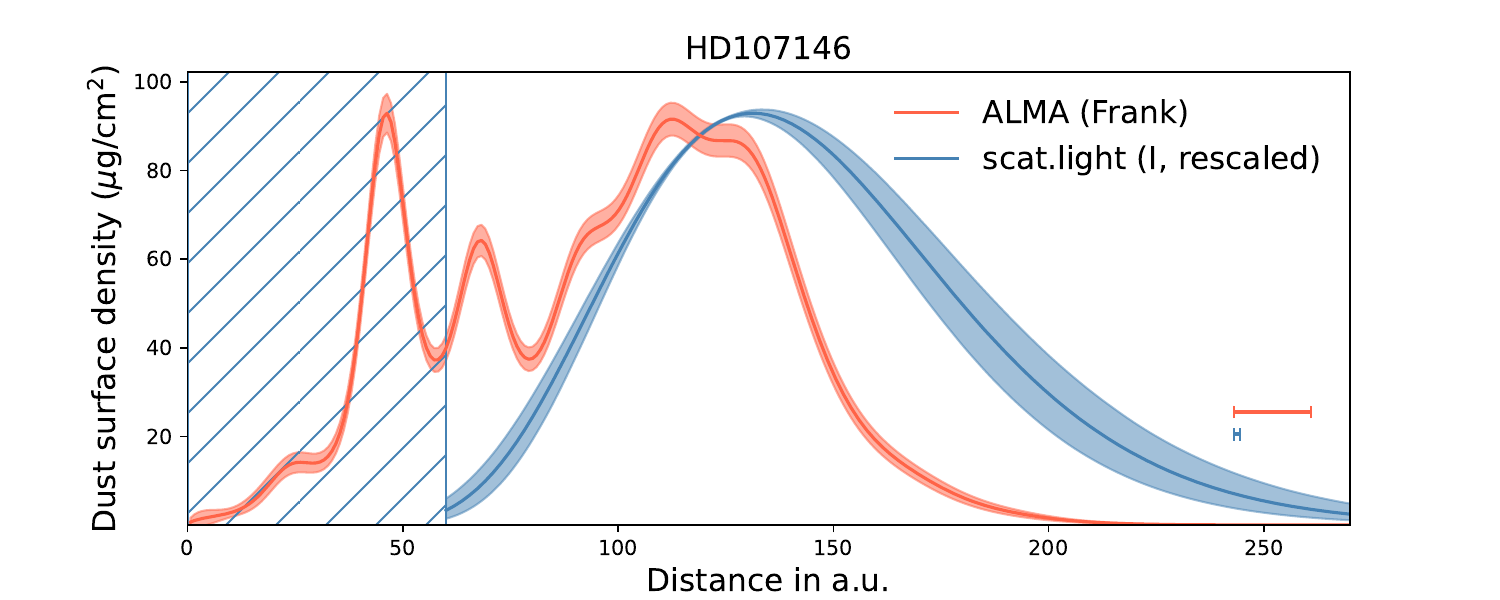}
   \includegraphics[width=0.46\hsize]{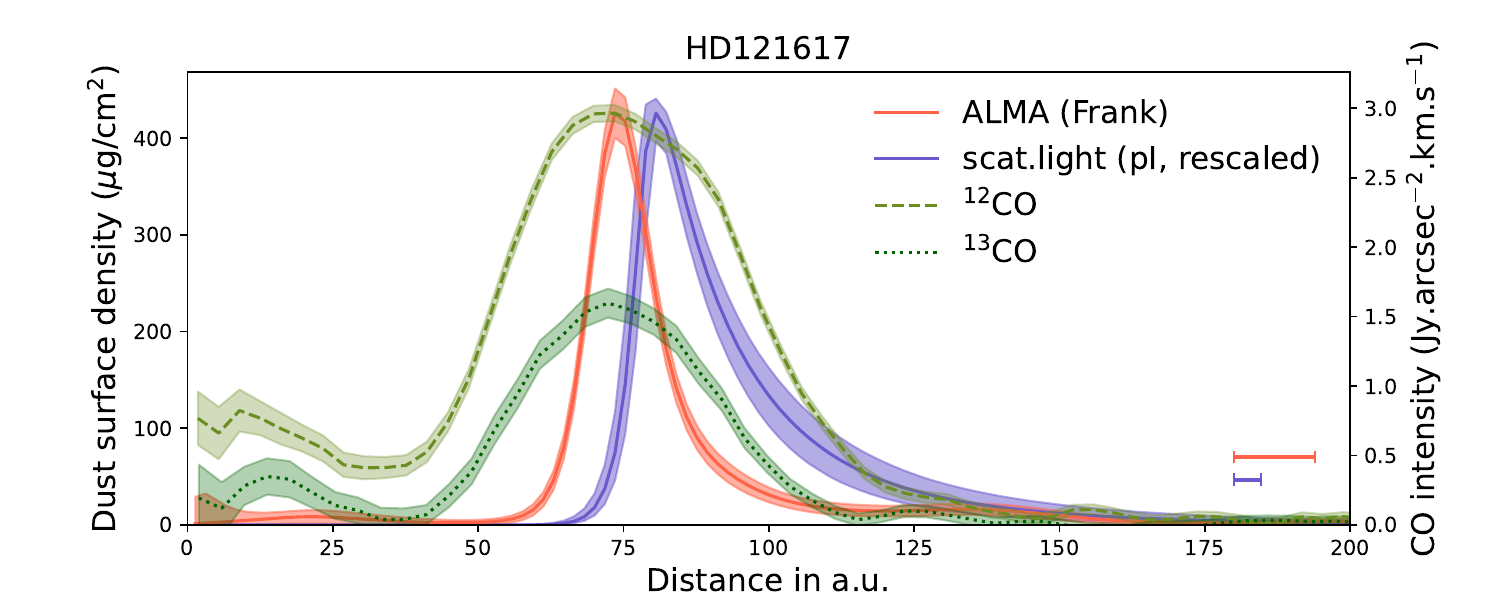}
   \includegraphics[width=0.46\hsize]{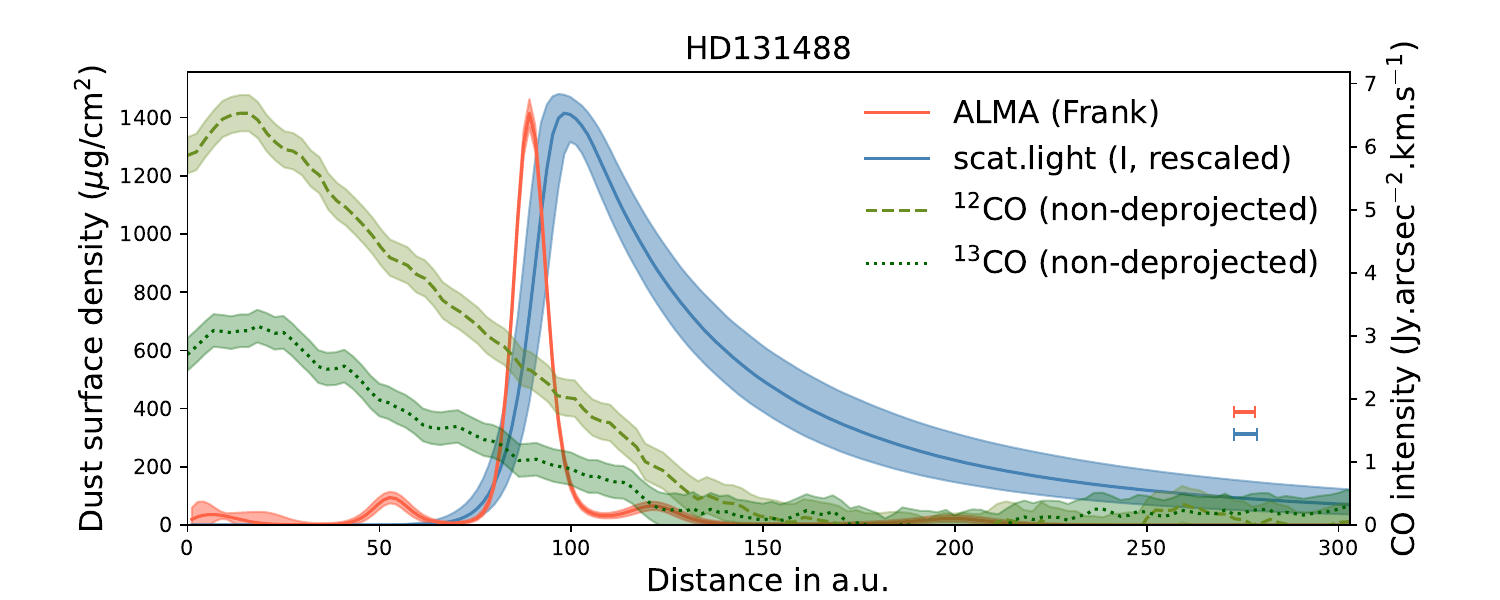}
   \includegraphics[width=0.46\hsize]{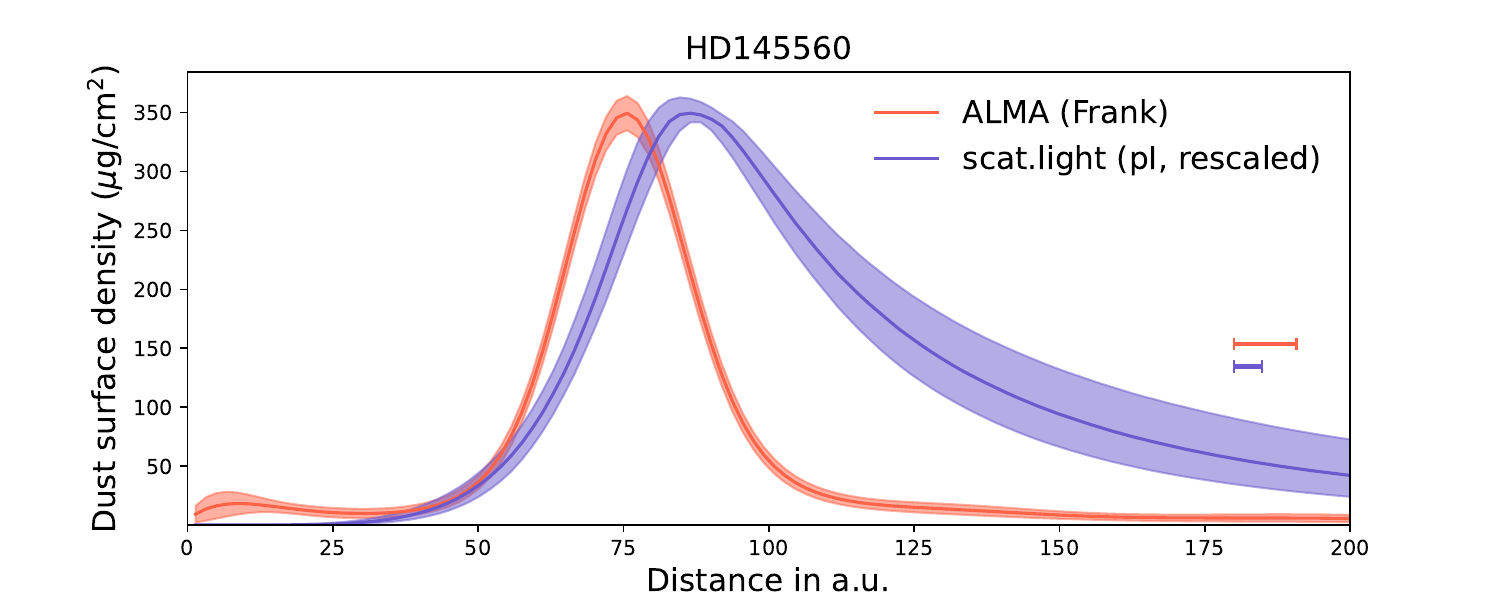}
   \includegraphics[width=0.46\hsize]{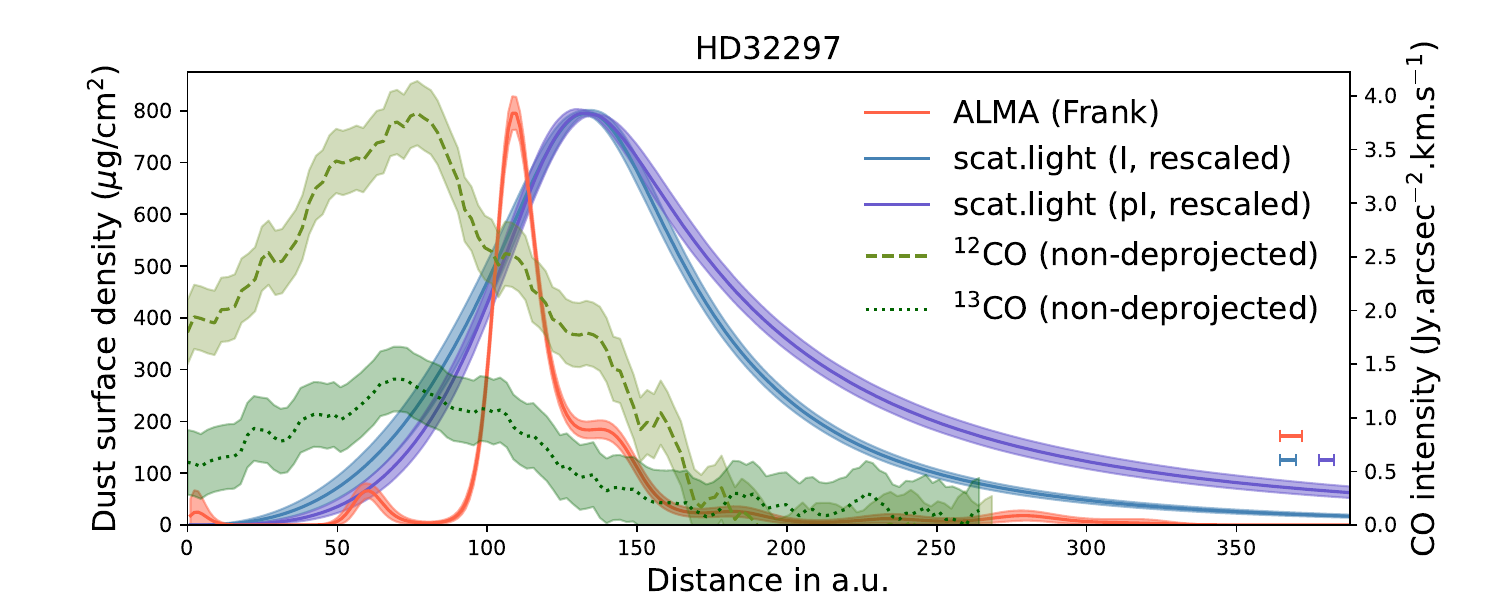}
   \includegraphics[width=0.46\hsize]{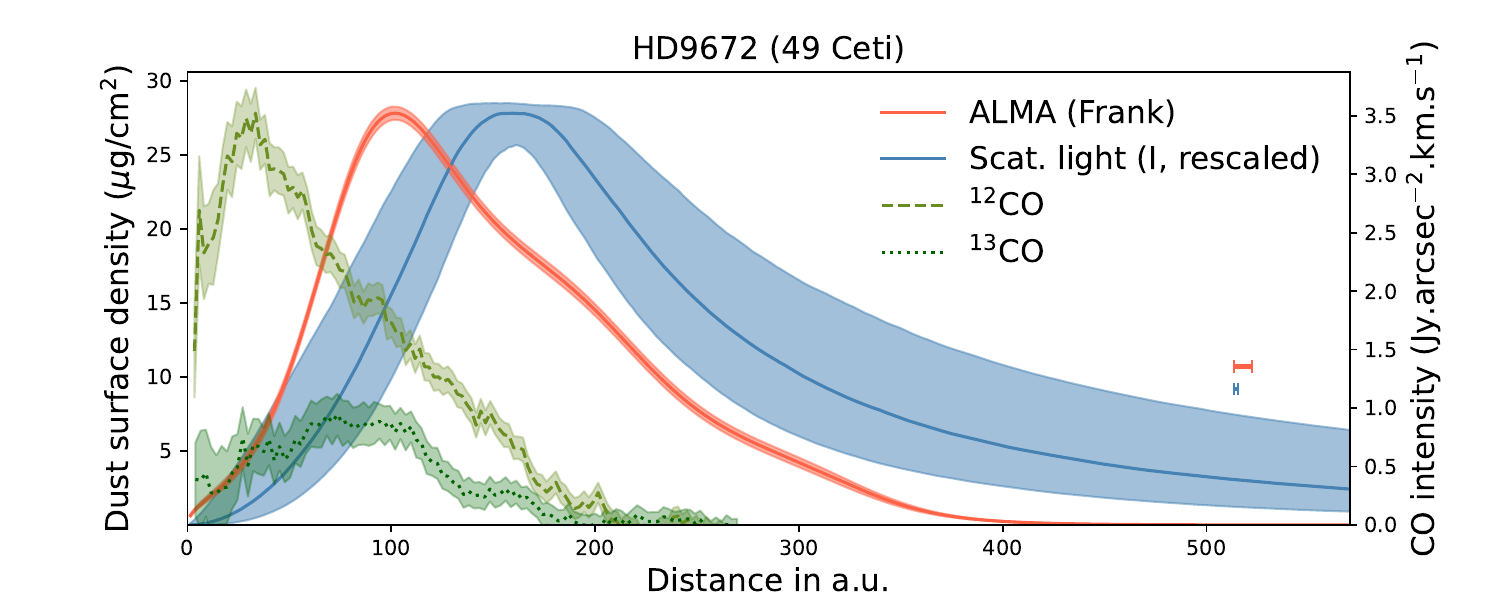}
   \includegraphics[width=0.46\hsize]{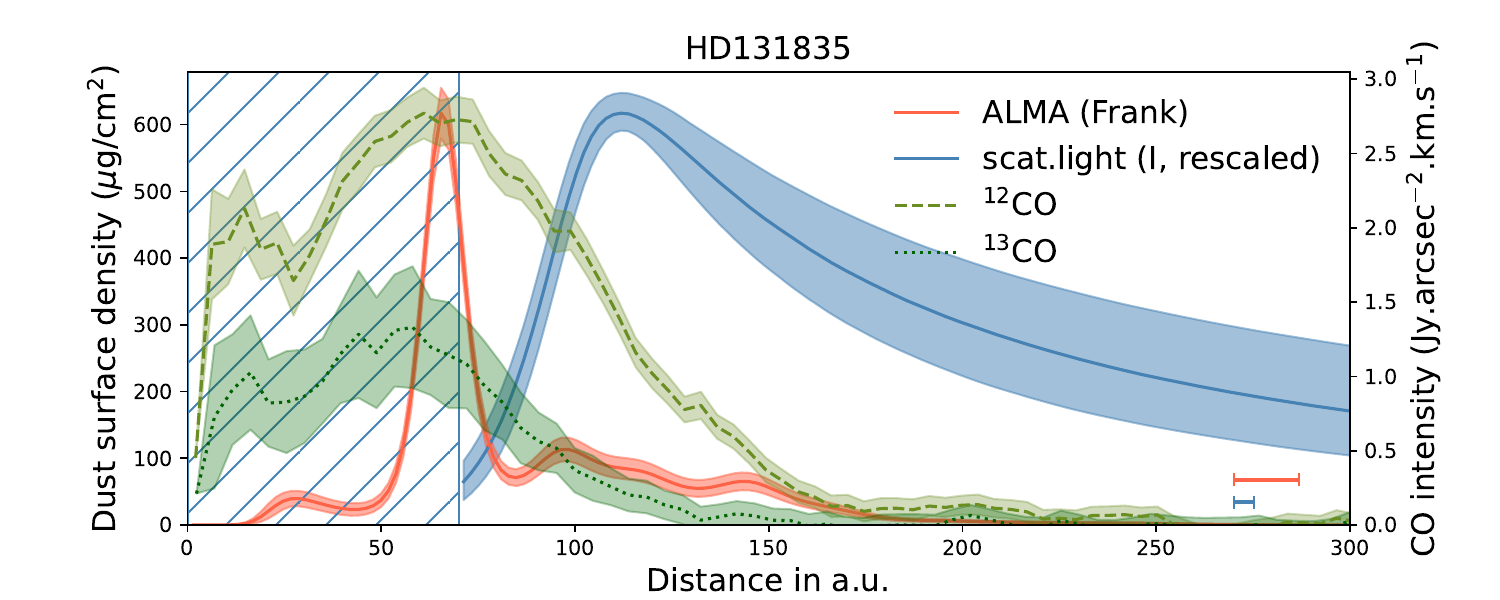}
   \caption{Comparison of the dust surface densities extracted from the ALMA continuum image and from the scattered light image in total intensity (I) or polarised intensity (pI). The scattered light profiles are rescaled to the ALMA profiles for display. For systems with CO detected, the $^{12}$CO or $^{13}$CO intensity profile is overplotted. Blue hatched regions correspond to inner regions where no scattered light modelling could be performed. The scale in the bottom right-hand corner in each panel indicates the ALMA and scattered light image resolution.}
    \label{fig_radial_profile_comparison}
    \end{figure*}

 \begin{figure}
    \centering
   \includegraphics[width=\hsize]{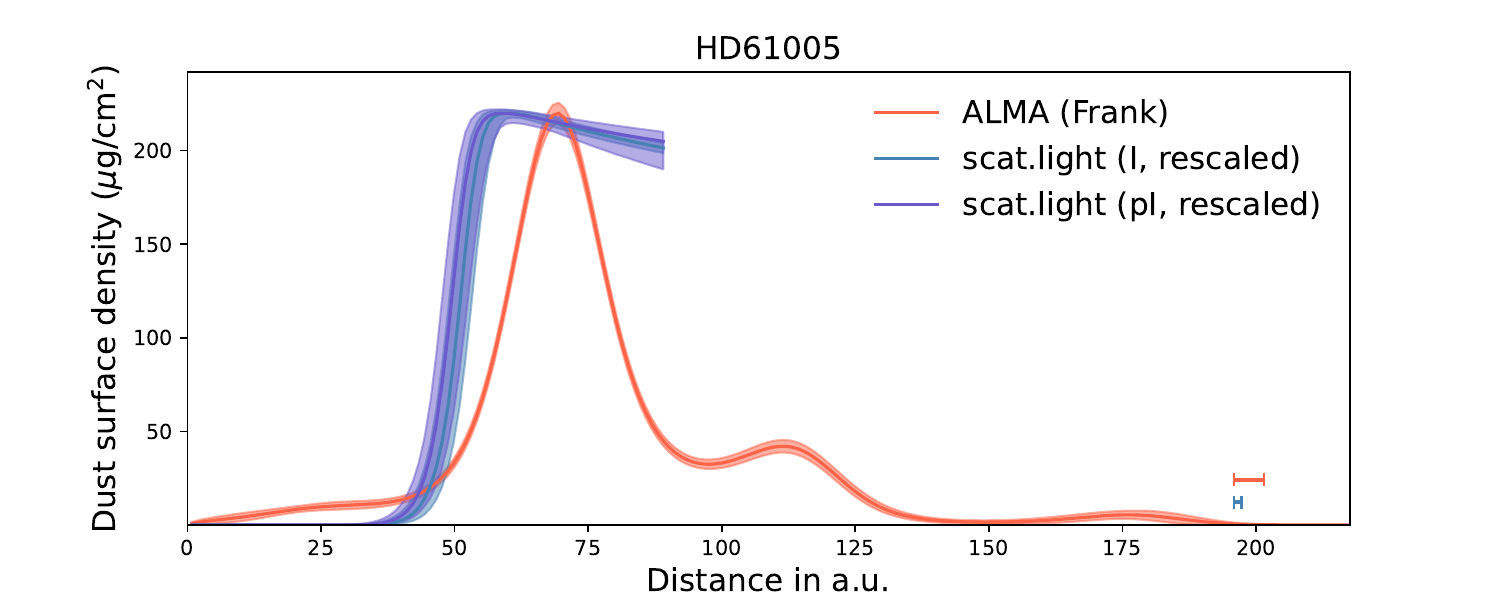}
   \caption{Continued from Fig. \ref{fig_radial_profile_comparison}.} 
    \label{fig_radial_profile_comparison_cont}
    \end{figure}

\begin{table*}
    \caption{Radii of the maximum dust surface density measured in scattered light (second column) and in the ALMA continuum (third and fifth columns) together with the corresponding relative offset (fourth and sixth columns).}
    \centering
    \begin{tabular}{l c c c c c}
    \hline
    \hline
      Name  & Scatt. light & ALMA (\emph{frank}$^a$) & rel. offset & ALMA (double power law) & rel. offset  \\
            &  au  & au & \% & au & \% \\
    \hline
    \multicolumn{6}{l}{Disc with no detectable offset in scattered light}\\
HD~197481 & $32.6\pm0.6$ & $36.4^{+2.0}_{-2.5}$ $^{(b)}$ & $-10.5^{+6.4}_{-5.1}$ & $31.8^{+1.1}_{-0.6}$ & $2.6\pm3.2$ \\
\rowcolor{gray!20}HD~39060 & $91.8^{+1.7}_{-1.9}$ & $100.9^{+34.7}_{-63.7}$ &  $-9^{+34}_{-31}$ & $114.6\pm5.0$  &  $-20.0\pm4.0$\\
\tyc & $90.0\pm4.9$ & $ 91.9^{+15.6}_{-15.9}$ & $-2\pm17$ & $ 97.8^{+6.0}_{-7.0}$ & $-7.9\pm7.8$ \\
HD~109573 & $75.6\pm0.1$ & $75.7\pm1.1$ & $0.0\pm1.4$ & $76.7^{+0.3}_{-0.4}$ & $-1.4\pm0.5$ \\
HD~92945 & $56.1^{+4.4}_{-1.9}$ & $56.5^{+4.6}_{-3.5}$ & $-0.7^{+10.0}_{-8.8}$ & $53.3^{+1.3}_{-1.2}$ & $6.8^{+8.7}_{-4.2}$ \\
HD~10647 & $87.3\pm0.1$ & $86.0^{+4.6}_{-1.6}$ & $1.5^{+1.9}_{-5.5}$ & $75.6^{+1}_{-0.5}$  & $16.9\pm1.2$ \\
HD~15115 & $100.9^{+3.8}_{-4.0}$ & $96.3\pm1.6$ & $4.7^{+4.3}_{-4.6}$ & $99.5\pm0.4$ & $1.3^{+3.8}_{-4.1}$ \\ 
HD~107146 & $131.3^{+2.7}_{-1.8}$ & $112.1^{+19.9}_{-5.6}$ & $17.1^{+6.4}_{-20.8}$ & NA & NA  \\
    \hline
    \multicolumn{6}{l}{Discs with an outward offset $>1\sigma$ in scattered light}\\
\rowcolor{gray!20}HD~121617 & $80.6^{+1.4}_{-1.1}$ & $73.6^{+3.6}_{-2.0}$ & $9.6^{+3.5}_{-5.6}$ & $69.6^{+0.7}_{-0.5}$ & $15.8\pm2.1$\\
\rowcolor{gray!20}HD~131488 & $98.7\pm3.6$ & $89.1^{+1.6}_{-1.4}$ & $10.8^{+4.6}_{-4.3}$ & $89.7^{+0.4_{-0.3}}$ & $10\pm4$ \\
HD~145560 & $86.6\pm2.5$ & $75.6^{+4.2}_{-4.4}$ & $14.5^{+7.5}_{-7.2}$ & $73.8\pm1.2$ & $16.2\pm3.9$ \\
\rowcolor{gray!20}HD~32297 & $133.3\pm2.1$  & $108.4^{+1.9}_{-2.2}$  & $23.0^{+3.2}_{-2.9}$ & $106.6^{+0.7}_{-0.5}$  & $25.0\pm2.1$\\
\rowcolor{gray!20}HD~9672 & $159.4\pm19$ & $102.3\pm11$ & $55.8^{+25.8}_{-24.4}$ & $106.1\pm4$ & $49.3\pm19$ \\
\rowcolor{gray!20}HD~131835 & $112.2^{+3.6}_{-3.2}$ & $65.4^{+4.0}_{-2.1}$ & $71.5^{+7.8}_{-11.6}$ & $61.9\pm1$ & $81\pm7$ \\
     \hline
    \multicolumn{6}{l}{Disc with an inward offset $>1\sigma$ in scattered light}\\
HD~61005 & $59.7^{+2.7}_{-2.4}$ & $69.5^{+2.5}_{-2.8}$ & $-14.1^{+5.2}_{-4.6}$  & $66.2\pm0.6$ & $-10.3\pm3.4$\\
\hline
\end{tabular}
\label{tab_offsets}    

\tablefoot{
Relative offsets are with respect to ALMA, and a positive offset indicates the scattered light surface density peaks at a larger distance. The discs are grouped in three categories, within which they are sorted with increasing relative offsets. The six discs with a CO detection are indicated with grey shading. \\
\tablefoottext{a}{For $\beta$ Pic, the \emph{rave} profile was used instead of \emph{frank} (see text for details).}
\tablefoottext{b}{For HD~197481, ALMA (\emph{frank}) reveals two peaks. The maximum surface density is in the second peak, which is the value provided here, and it explains the negative offset.}
\tablefoottext{c}{For HD~107146, ALMA (\emph{frank}) also displays three peaks, the offset is computed relative to the third peak, which corresponds to the ring seen in scattered light.}
}

\end{table*}

    \subsection{Discs with no detectable offset}
\label{sec_groupA}

For eight out of the 15 considered discs, we found no detectable offset within our error bars. We outline these discs below.

\noindent \textbf{HD~197481 (AU Mic)} 

The main belt is resolved with ALMA into two separate peaks, the second one having a slightly higher surface density based on the \emph{frank} extraction as reported in Table \ref{tab_offsets}. The parametric modelling employed in scattered light does not allow this level of detail to be captured, and the surface density extracted in scattered light (total and polarised intensity) peaks at $\sim33$ au, which is exactly in the middle of the two peaks seen with ALMA. When the ALMA observation is parametrised with the double power-law surface density profile as in scattered light, we find a relative offset of $2.6\pm3.2\%$, and therefore we cannot offer a conclusion on the significance of an offset for AU Mic.

\noindent \textbf{HD~39060 (\bPic)}

The ALMA dataset of \bPic{} used multiple pointings, and the \emph{frank} algorithm does not appropriately deal with this observational setting. Therefore, we used instead the surface density extracted with the \emph{rave} algorithm (see Fig. \ref{fig_radial_profile_comparison}). 
In scattered light, we used the high S/N image of the disc obtained from deep polarimetric imaging in the H band originally intended to measure the polarisation fraction of the planet \bPic{} b \citep{vanHolstein2021}. Additionally, polarised images are not subject to self-subtraction \citep{Milli2012,Milli2014}, which gives more confidence in the extracted surface density.
However, due to the large uncertainties on the ALMA dust surface density extracted from \emph{rave}, the measured offset of $-8.6^{+22.1}_{-26.2}\%$ is not significant. While Fig. \ref{fig_radial_profile_comparison} may seem to show that the dust surface density seen in scattered light drops to zero beyond $\sim115$ au, one should bear in mind that the SPHERE - IRDIS field of view does not go beyond that separation, but dust is detected up to several thousands of au in the \bPic{} system \citep{Janson2021}. The $1/r^2$ dependence of the scattered light intensity also makes it very challenging to probe the outermost regions. 

\noindent \textbf{\tyc} 

The S/N is poor, both in the {\em HST} - NICMOS scattered light image and in the ALMA continuum image (see Fig. \ref{fig_mosaici}). Therefore, the relative offset of $-2\pm17\%$ has large uncertainty and is not significant. We also note that the surface density profiles in Fig. \ref{fig_radial_profile_comparison} may seem narrower in scattered light compared to the ALMA continuum. However, the poor S/N does not allow us to confidently determine whether there is a significant difference in width. Thus, a higher sensitivity scattered light observation is required to confidently compare both profiles.

\noindent \textbf{HD~109573 (HR~4796)}

The ring around HR~4796 is very narrow and slightly eccentric \citep{Milli2017} and the location of the peak surface density is identical between the scattered light measurements (both total and polarised intensity) and the ALMA continuum extracted with \emph{frank}. The steep inner edge is similar between ALMA and SPHERE: $\alpha_\text{in}=34^{+5}_{-3}$ for the ALMA surface density and $\alpha_\text{in}=35.7\pm0.3$ for the SPHERE surface density. The smoother outer profile in scattered light ($\alpha_\text{out}=-17.6\pm0.5$) compared to ALMA ($\alpha_\text{out}=-21.7\pm1.8$) is real and likely due to the smallest dust particles being very sensitive to the stellar radiation pressure forming an extended halo already known from {\em HST} - STIS optical observations \citep{Schneider2018}.

\noindent \textbf{HD~92945}

The dust surface density profile shown in Fig. \ref{fig_radial_profile_comparison} is the best non-parametric model proposed by \citet{Golimowski2011} to reproduce the disc surface brightness of the {\em HST} - ACS image, assuming a 10\% uncertainty on this profile. The shape of this profile very closely matches the dust surface density extracted from the ALMA continuum, although with a slightly shallower gap. The peak positions reported in Table \ref{tab_offsets} correspond to the innermost ring, but both peak locations are compatible within error bars, showing no offset between the micron-sized and millimetre-sized dust.

\noindent \textbf{HD~10647 (\qEri)} 

The best scattered light model shows a very smooth outer profile with a power-law exponent $\alpha_\text{out}=-1.1$ up to $\sim150$~au, where the disc signal becomes too faint. Conversely, the ALMA continuum falls off steeply beyond 90~au but shows an extended halo centered at $\sim160$~au with a FWHM of $\sim100$~au \citep{rad_arks}. Unlike most other systems, the disc is only detected in scattered light in the optical with {\em HST} - ACS in the F606W filter (central wavelength 589~nm), likely more sensitive to small eccentric submicron dust particles, which may tentatively explain the smooth surface density. 
Despite those differences, the offset in the peak surface density between scattered light and ALMA, as extracted from \emph{frank}, is not significant. There is a significant offset with the ALMA peak profile when parametrised with a double power-law ($16.9\pm1.2\%$, see Table \ref{tab_offsets}). However, this parametrisation is not the best functional form fitting the ALMA data \citep{rad_arks} compared to the double Gaussian, which peaks at $88\pm4$~au and for which no significant offset is detectable.

\noindent \textbf{HD~15115 (the needle\footnote{The term needle was coined by \citet{Kalas2007} because the disc is edge-on and very thin.})}

HD~15115 is a double-ringed disc \citep{MacGregor2019} seen close to edge-on. The polarised intensity of this disc is faint \citep{Engler2019}; therefore we used the SPHERE - IRDIS J-band total intensity image presented in \citet{Engler2019} for the surface density modelling.  
In the ALMA (\emph{frank)} profile, the second peak at $96\pm2$\,au is six times higher than the first peak at $65\pm4$\,au, so we computed the relative offset between SPHERE and ALMA with respect to this second (and highest) peak. While the tentative offset of $4.7^{+4.3}_{-4.6}\%$ is just at $1\sigma$, the offset computed with the ALMA double power law modelling of $1.3^{+3.8}_{-4.1}\%$ is not significant. Though this parametric modelling did not provide a good fit to the data, because of the second ring seen with ALMA, the double power law parametrisation with a gap is the preferred fit \citep{rad_arks}, and it does not show an offset either ($3.9^{+4.3}_{-4.5}$\%). We therefore cannot confirm the presence of an offset in this system. However, upon the assumption that the dust is being produced in the first and fainter ALMA ring, then the offset would be $56\pm10$\%, and it would be statistically significant.

\noindent \textbf{HD~107146}

HD~107146 is a wide belt extending from 40 to 140~au with a gap at 56~au and a second one at 78~au. The ALMA \emph{frank} profile suggests that an intermediate ring may be present within the gap \cite[see companion paper][]{rad_arks}. The disc is only detected in scattered light with the {\em HST} instruments \cite[ACS, NICMOS and STIS,][]{Ertel2011,Schneider2014}. Using the ACS filter F606W (deemed more reliable than the F814W due to instrumental artefacts at short separations), \citet{Ertel2011} modelled the surface density of the disc with a parametric form (a product of a power-law and an exponential function analogous to Planck’s law).  The fitted profile is shown in Fig. \ref{fig_radial_profile_comparison}. Due to the large coronagraph and the starlight residuals, the region within 60~au (orange-shaded region in Fig. \ref{fig_radial_profile_comparison}) is not reliable in the ACS image used to extract the surface density profile \citep{Ertel2011}. More recent STIS images shown in \citet{Schneider2014} and reproduced in Fig. \ref{fig_mosaici}\footnote{The clump seen in the South of the image is a background galaxy.} show some level of emission in the inner $2\arcsec$, suggesting that the region within 60~au may not be devoid of material as shown with ALMA, but we cannot confidently confirm this statement. We focus instead on the outer ring. An offset of  $17.1^{+6.4}_{-20.8}\%$ is detected in the peak location between the scattered light and ALMA continuum. This offset is not statistically significant, and therefore we cannot rule out the fact that millimetre- and micron-sized dust grains are spatially collocated based on these observations. We note that the scattered light profile extends to larger separations, similar to $q^1$ Eri, and this again may be due to the fact that ACS images at optical wavelengths are sensitive to the halo of small grains extending farther from the star.

\subsection{Discs with an outward offset in scattered light}
\label{sec_groupB}

For six out of the 15 considered discs, we find significant offsets, as outlined below.

\noindent \textbf{HD~121617}

HD~121617 is a narrow ring which shows a clear offset of $9.6^{+3.5}_{-5.6}\%$ between the polarised scattered light of the disc and the ALMA continuum surface density extracted with \emph{frank}. This shift is also confirmed with the double power law parametrisation, and even more significant ($15.8\pm2.1\%$). This mismatch is visible with the naked eye in the image itself due to the low inclination of the disc (Fig. \ref{fig_mosaici}). CO gas is detected in this system and both $^{12}$CO and $^{13}$CO peak at a radius smaller than the location of the maximum scattered light peak \cite[$74.4$~au vs $80.6^{+1.4}_{-1.1}$~au, see][]{gas_arks,line_arks}. 
This indicates that the millimetre-sized grains surface density peaks at a location consistent with the $^{13}$CO, which is also the gas pressure maximum \citep{hd121617_arks}. On the other hand, the micron-sized grains traced by scattered light are shifted outward, nearly 1sigma outward from the pressure maximum (based on the $^{13}$CO best-fit radially Gaussian model).This is a location where the gradient is such that micron-sized dust could get stalled there if gas drag combined with radiation pressure play a dominant role in the grain dynamics, assuming sufficiently high total gas densities are present \citep{vortex_arks}.

The scattered light image shows no azimuthal asymmetry (beyond the forward/backward side asymmetry, which is an effect of the anisotropy of scattering and not an azimuthal asymmetry in the surface density). This in contrast to the ALMA image, which is discussed in the companion ARKS papers \citet{asym_arks,hd121617_arks, vortex_arks}.

\noindent \textbf{HD~131488}

HD~131488 is another narrow ring \citep{Pawellek2024} with a significant offset of $10.8^{+4.6}_{-4.3}\%$ detected between the total intensity scattered light and the ALMA continuum. The offset is consistent whether the ALMA surface density is extracted with \emph{frank} or with the double power law parametrisation. Because of the close to edge-on geometry, the CO intensity profile cannot be reliably deprojected \citep{gas_arks}, but the non-deprojected profile still allows us to conclude that most of the CO resides inside the dust ring, with the micron-sized dust peaking close to the outer edge of the CO disc.

\noindent \textbf{HD~145560}

Despite the faintness of the disc in scattered light \citep{Hom2020}, with a detection only in polarised intensity and not in total intensity, an offset of $14.5^{+7.5}_{-7.2}$\% at a $2\sigma$ confidence level is measured for HD~145560. This result should be taken with caution because of the poor S/N of the disc. This is the only disc with an offset that is not known to contain CO gas. Gas may still be present but undetected in the system: the $3\sigma$ upper limit on the integrated line flux of $^{12}$CO (3-2) is $67$ mJy~km/s, for a comparison, HD~131835, at a similar distance and age but with an earlier spectral type A2IV compared to F5V has a $^{12}$CO (3-2) integrated line flux of $1.9\pm0.2$ Jy~km/s \citep{gas_arks}.

\noindent \textbf{HD~32297}

HD~32297 is an edge-on disc \citep{Bhowmik2019,Olofsson2022_HD32297} with a significant offset of $23.0^{+3.2}_{-2.9}\%$ detected. This offset is consistent with both the total and polarised intensity images, and with the ALMA \emph{frank} and the double power law parametrisation. We note that the location of the maximum surface density for micron-sized dust corresponds also to a small plateau in the millimetre-sized dust density seen with ALMA. HD~32297 is a CO-rich system, with the CO peak located interior to the dust ring. Due to the edge-on geometry of the disc, the CO line intensity cannot be deprojected, but the non-deprojected CO profile visible in Fig. \ref{fig_mosaici} shows that the peak of the micron-sized dust particles occurs at a radius where the CO has dropped by more than 50\%, similar to the case of HD~131488 discussed above.

\noindent \textbf{HD~9672 (49~Ceti)}

A significant offset of $55.8^{+25.8}_{-24.4}\%$ is detected for 49~Ceti. Despite a large error bar due to the faintness of the disc in scattered light, the offset of $49.3\pm19\%$ using the double power-law rather than \emph{frank} confirms this large offset. Both the SPHERE - IRDIS model presented here and the {\em HST} - NICMOS model presented in \citet{Choquet2016} confirm this offset. 49 Ceti is another gas-rich disc, with CO detected within the dust ring (see Fig. \ref{fig_radial_profile_comparison}). The deprojected $^{12}$CO intensity shows that the peak micron-sized dust surface density occurs where the $^{12}$CO line intensity has already dropped by more than a factor of seven compared to its peak value.

\noindent \textbf{HD~131835}

This disc shows the largest offset among the ARKS targets, with a value of $71.5^{7.8}_{-11.6}\%$. It is the subject of a dedicated companion paper \citep{hd131835_arks} because of its complex structure. It hosts at least two rings detected in scattered light at 66 and 96~au \citep{Feldt2017}, the outer one being the brightest in scattered light (see Fig. \ref{fig_mosaici} and Appendix \ref{app_HD131835} for more details taking into account possible post-processing biases), while ALMA only sees one prominent ring at  $\sim65$~au. In this paper, we focus on and model the outer ring only, because the S/N is not sufficient to perform a full exploration of the inner and outer ring parameters. The micron-sized dust surface density shown in Fig. \ref{fig_radial_profile_comparison} only represents this outer ring, while regions within 70~au have been hatched. The peak in surface density from the scattered light modelling is located at $112.2^{+3.6}_{-3.2}$ au\footnote{This location of the peak surface density is larger than the value obtained with the best model presented in \citet{Feldt2017} and Eq. \ref{eq_max_surface_density}, which is $105\pm4$~au after taking into account the updated distance to the star of 129.7~pc compared to 123~pc in \citet{Feldt2017}. Both values are still compatible within $1\sigma$.}. 

The ARKS paper \citet{hd131835_arks} explores two scenarios to explain this large offset: two planetesimal belts that appear very different in ALMA and SPHERE due to very different collisional properties, and a single planetesimal belt at 65 au with micron-sized dust pushed to $>100$ au by gas drag, creating the outer ring.

Intriguingly, the ALMA \emph{frank} surface density profile shows a secondary peak with a lower amplitude slightly inward from the peak surface density of the outer ring, at 98~au. If there is indeed an outer parent belt distinct from the inner one at 98~au, the offset seen in scattered light would become $14.5^{+5.9}_{-5.7}\%$, a value similar to the typical offset also found for the three other CO-rich discs HD~121617, HD~131488, and HD~32297. 
Comparing the peak surface density location with the deprojected CO intensity, we find, as for the other gas-rich discs, that the CO is mostly located within the dust rings, and that the scattered light peak radius occurs when the CO intensity has significantly dropped by more than half its maximum value. However we note that \emph{frank} modeling is not very robust for low S/N emission and oscillations can appear mimicking rings.

\subsection{Disc with an inward offset in scattered light}
\label{sec_groupC}

\noindent \textbf{HD\,61005 (the moth)}

HD~61005 is a disc with an inner edge at $\sim45$~au, a main ring up to $\sim80$ au and a more diffuse `swept-back' component, best seen with {\em HST} - STIS up to $\sim240$~au \citep{Schneider2014}. No gas has been detected in this system \citep{gas_arks}. With SPHERE, we detect the main ring and the halo up to $\sim90$~au (Fig. \ref{fig_radial_profile_comparison_cont}). It is the only ARKS disc with a significant negative relative offset between scattered light and thermal emission (see Table \ref{tab_offsets}). Both the SPHERE total and polarised intensity images agree with a peak surface density occurring at a radius $-14.1\% \pm4.9\%$ smaller than the ALMA continuum peak surface density. However, one must note that HD~61005 has a relatively broad belt seen in scattered light and the double-power law profile parametrised in  Eq. \ref{eq_dust_density} is not well adapted for broad belts without a clear peak. Secondly, the diffuse dust halo is not reproduced by the simple parametrisation of the dust surface density used in this work, because this halo is not aligned with the main ring. There are several possible explanations for that feature being seen misaligned with the main ring. The most likely is that the dust is pushed back from the disc plane as the system travels through the interstellar medium \citep{Debes2009,Maness2009}. The interstellar medium (ISM) can indeed blow the small grains from a system and  influence the morphology of debris discs \citep{Heras2025}. An alternative scenario invokes a giant collision \citep{Jones2023} although ISM sculpting is preferred for HD\,61005. 

\section{Sensitivity to planets}
\label{sec_sensitivity_to_planets}

Planetesimal belts can be influenced by the presence of giant planets in a system: they can sculpt the inner edge of a belt \cite[as for the Kuiper belt, e.g.][]{Malhotra2019}, carve a gap within a belt or trap planetesimals in resonance \cite[as in TWA~7, see][]{Lagrange2025}, or interact with the dust creating asymmetries or warps \cite[as in \bPic, e.g.][]{Mouillet1997}. 
We used the scattered light images obtained with SPHERE to derive uniform upper limits on the presence of planets among the ARKS sample, complemented by Gaia astrometric data. We present these results in section \ref{sec:constraints_on_planets_obs}, following a brief summary of the known planets in our sample in section \ref{sec:known_planets_obs}. 

\subsection{Known planets in the ARKS sample}
\label{sec:known_planets_obs}

Planets have been confirmed in six ARKS systems, all interior to the cold debris belts. Half of those debris belts are detected in scattered light: \bPic, AU~Mic and HD~10647 ($q^1$ Eri), none of them which have a significant offset between the peak surface brightness seen with ALMA and in scattered light. There are 2 transiting planets orbiting AU~Mic, both Neptune-sized bodies with periods of 8.5 and 18.9 days \citep{Plavchan2020,Gilbert2022}. A radial-velocity planet was detected at $2$ au around HD~10647, with a minimum mass of $\sim1 M_{\text{Jup}}$ \citep{Butler2006,Marmier2013}. The other planets were detected with direct techniques (direct imaging or interferometry, sometimes confirmed with radial-velocity or astrometry). Two giant planets orbit \bPic{} at $\sim10$ and $\sim3$ au, with masses of $10-11 M_{\text{Jup}}$ and $7.8\pm0.4 M_{\text{Jup}}$, respectively \citep{Lagrange2009,Lagrange2020}. 

The other three systems known to host planets do not have their disc detected in scattered light.
HD~95086 host a $4-5 M_{\text{Jup}}$ giant planet orbiting in $\sim51-73$ au range \citep{Rameau2013_discovery,Desgrange2022}. HD~206893 contains a giant planet of $\sim13 M_{\text{Jup}}$ at $\sim3.5$~au and a brown dwarf of $\sim28 M_{\text{Jup}}$ at $\sim10$~au \citep{Hinkley2023}. Four giant planets of $6$ to $12 M_{\text{Jup}}$ orbit the HR~8799 system between $15$ and $65$ au \citep{Marois2008,Thompson2023}. 

\subsection{Constraints on the presence of additional planets}
\label{sec:constraints_on_planets_obs} 

To assess the sensitivity to planets in our sample of stars, we generate detection probability maps (DPMs) based on the achieved contrast limits from archival SPHERE observations. All 24 targets in our sample have archival SPHERE data, and for each system, we select the observation that provides the deepest contrast using the dual-band H23 or broad-band H filters. We refer the reader to Appendix \ref{app_SPHERE_program_ids} for the list of observations and filters. To ensure consistency across the sample, all contrast curves have been uniformly reduced using a Principal Component Analysis algorithm \citep{Soummer2012,Amara2012}, as implemented in the SPHERE High-Contrast Data Center \citep[HC-DC,][]{Delorme2017_DC}. 

The DPMs were computed on a 2D grid of planet mass and semi-major axis using the \emph{MADYS} \citep{Squicciarini2022} and \emph{ExoDMC} \citep{Bonavita2020} tools. Because contrast curves are expressed as a function of projected separation, and DPMs operate in deprojected orbital space, we account for each system's inclination \citep[taken from][and also reported in Table~\ref{tab_targets}]{overview_arks} and sample a range of orbital configurations at every grid point. We also sample planet eccentricity by using a half-Gaussian distribution\footnote{Negative values are discarded} centred at 0 with a standard deviation of 0.1 for all the systems. For each orbit, the planet's projected separation is calculated, and its mass is converted into magnitude using the system's age and the \emph{bex-atmo2023-ceq} planet evolutionary model in \emph{MADYS}, which combines the ATMO \citep{phillips2020} and BEX \citep{linder2019} models as in \citet{carter2021}. A planet is considered detectable if its brightness exceeds the SPHERE contrast limit at the corresponding projected separation. This process is repeated across different ages, accounting for the system's age uncertainty. The detection probability at each grid point is then defined as the fraction of simulated planets that would be detectable.

 \begin{figure*}
    \centering
  \includegraphics[width=0.32\hsize]{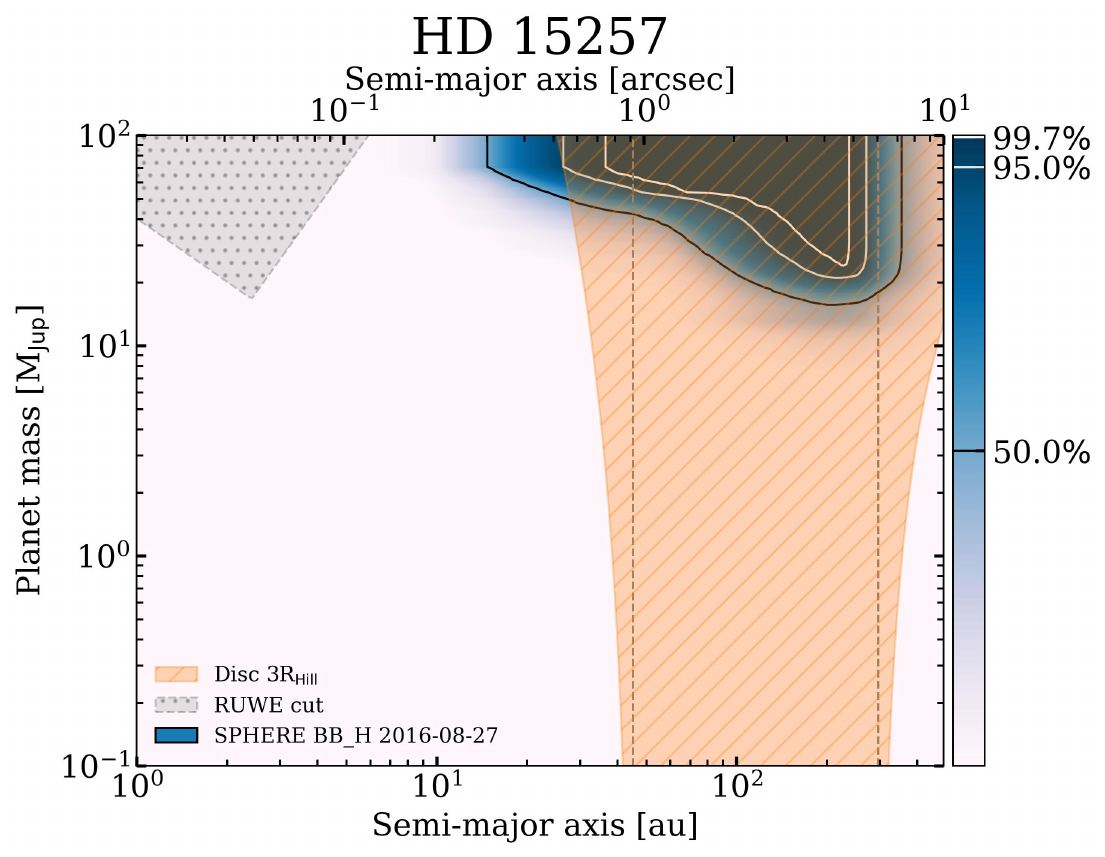}
  \includegraphics[width=0.32\hsize]{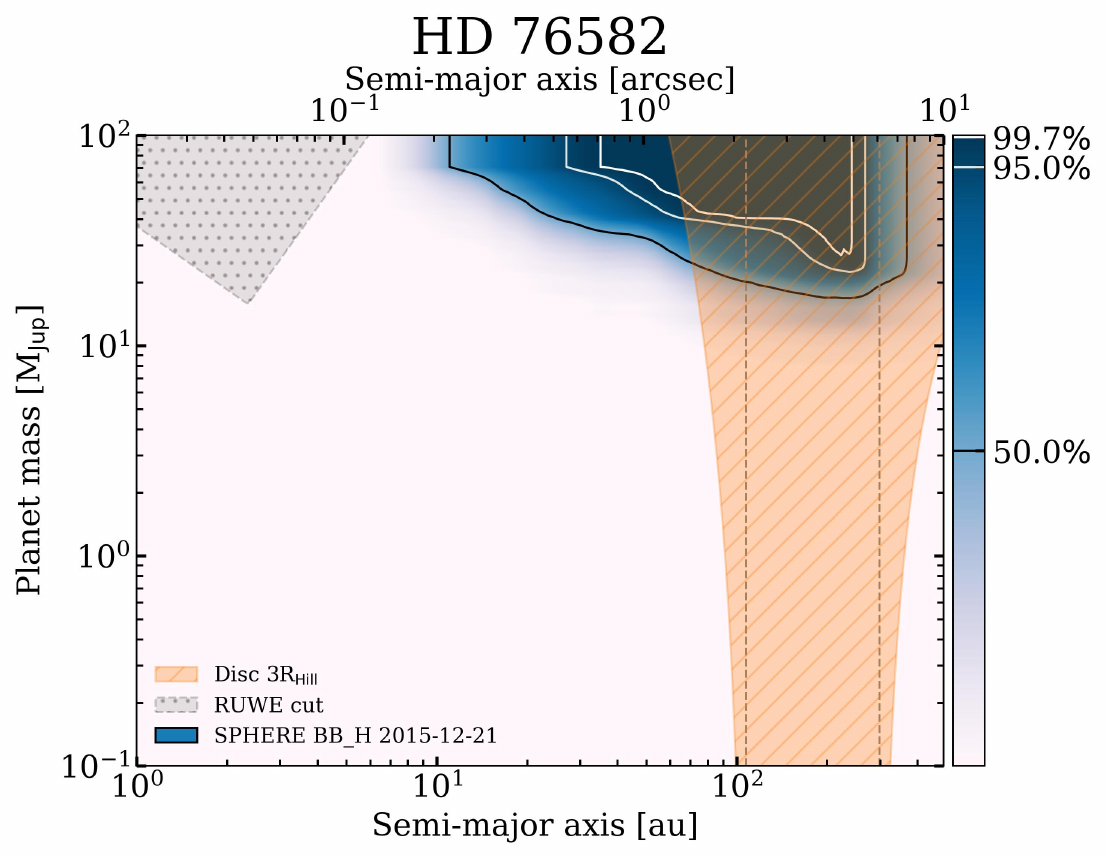}
  \includegraphics[width=0.32\hsize]{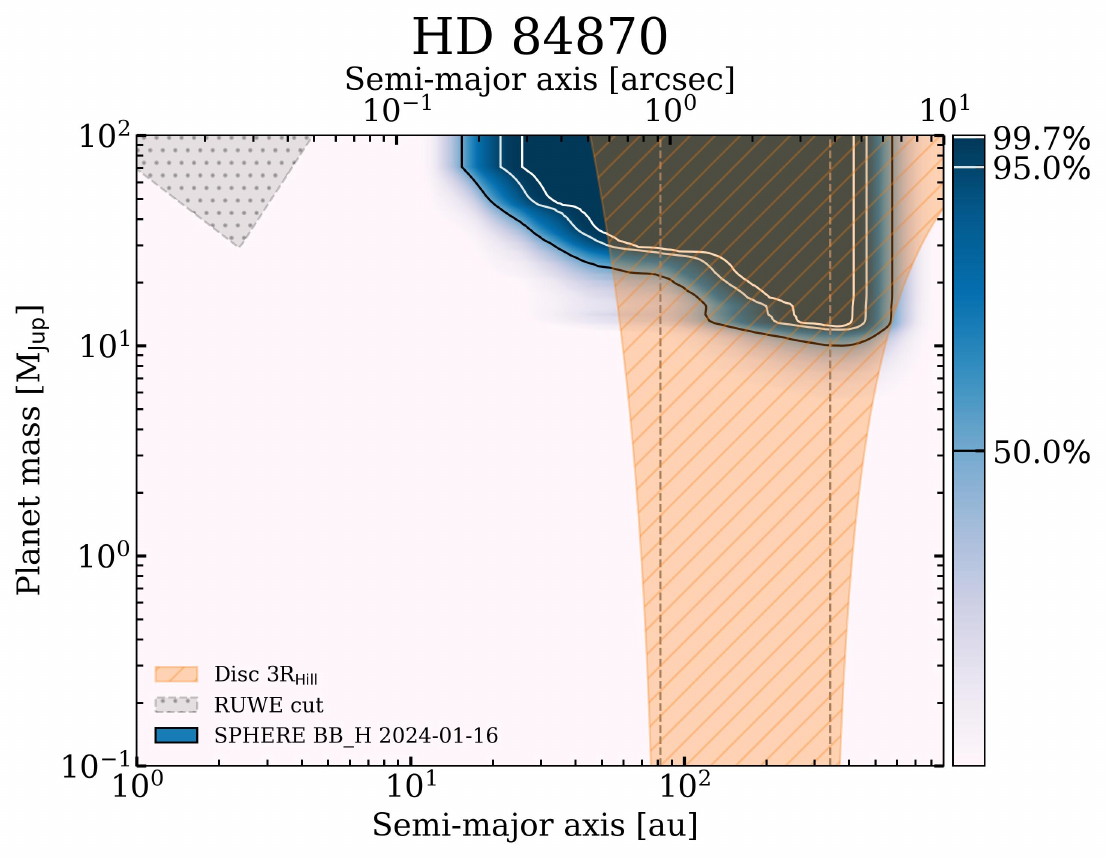}
  \includegraphics[width=0.32\hsize]{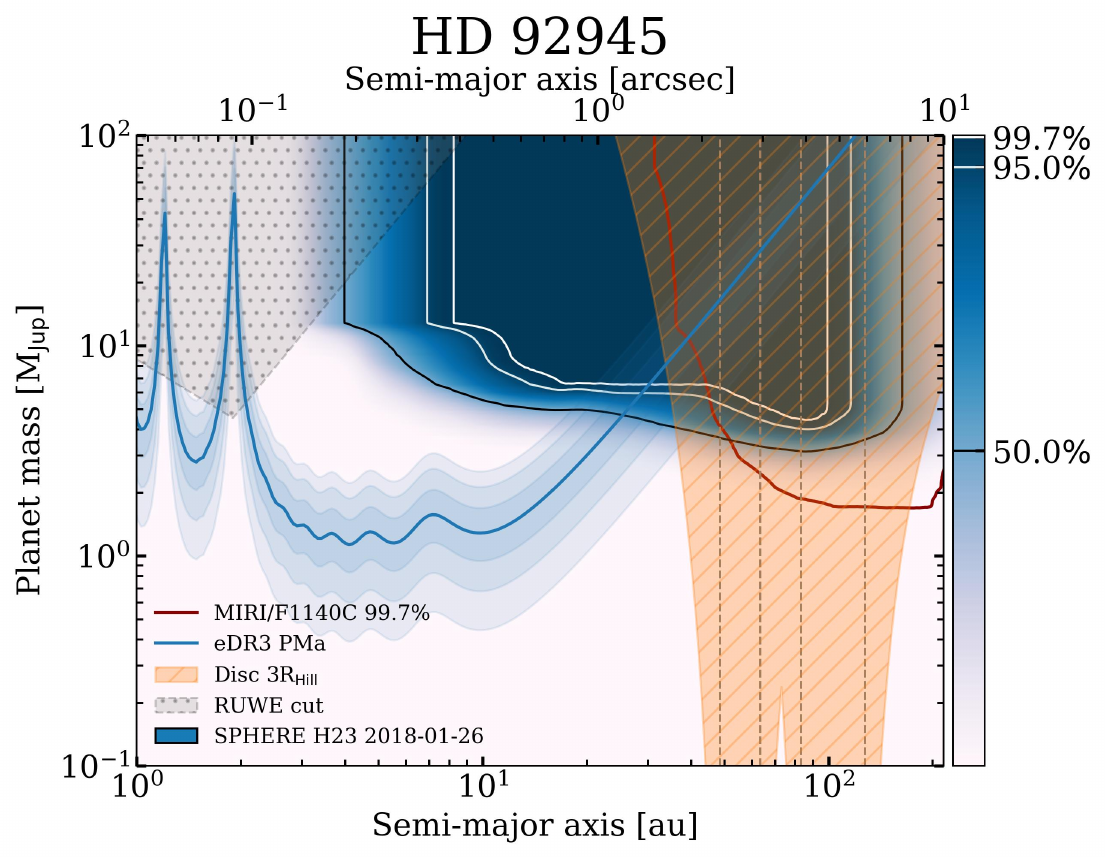}
  \includegraphics[width=0.32\hsize]{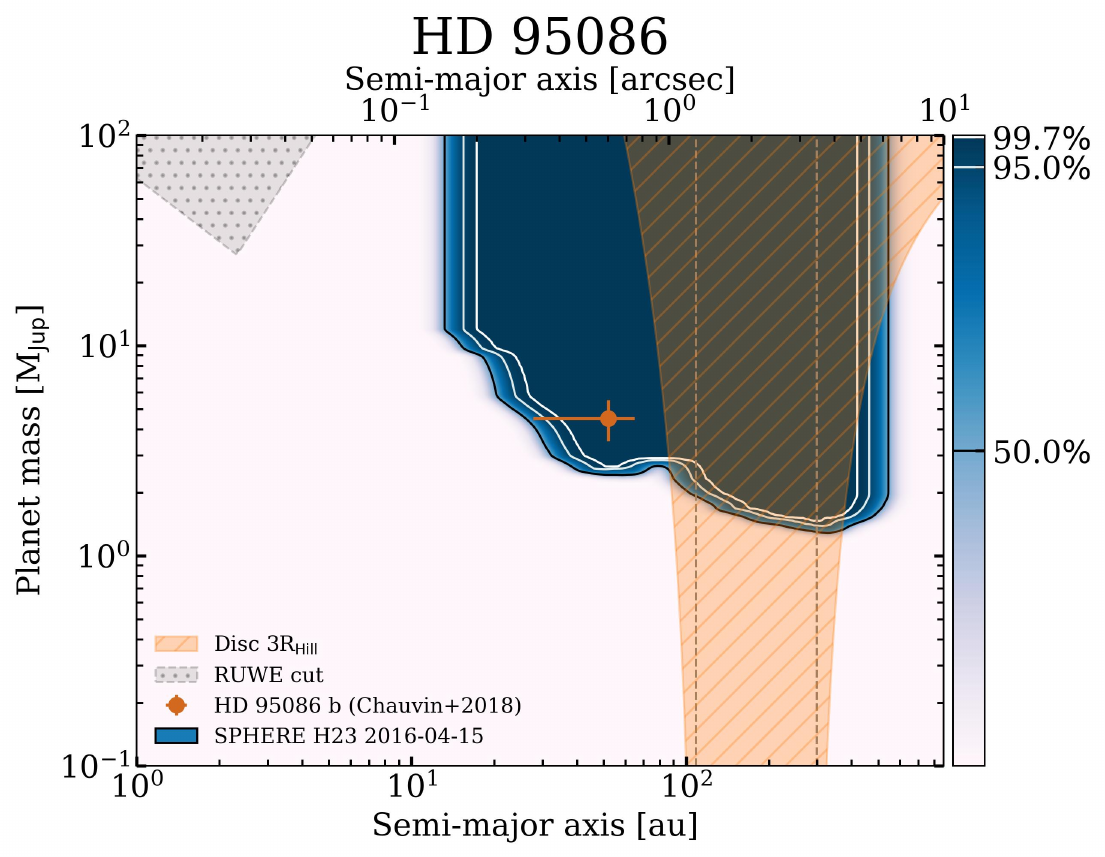}
  \includegraphics[width=0.32\hsize]{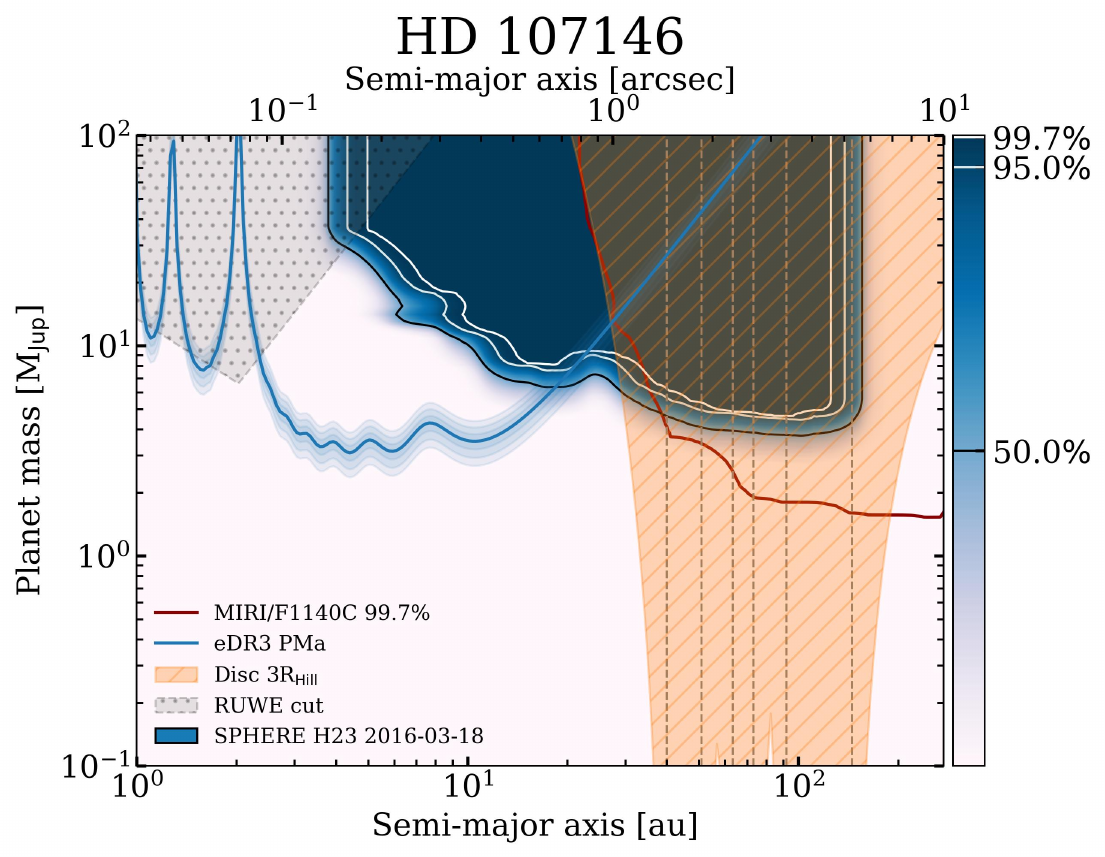}
  \includegraphics[width=0.32\hsize]{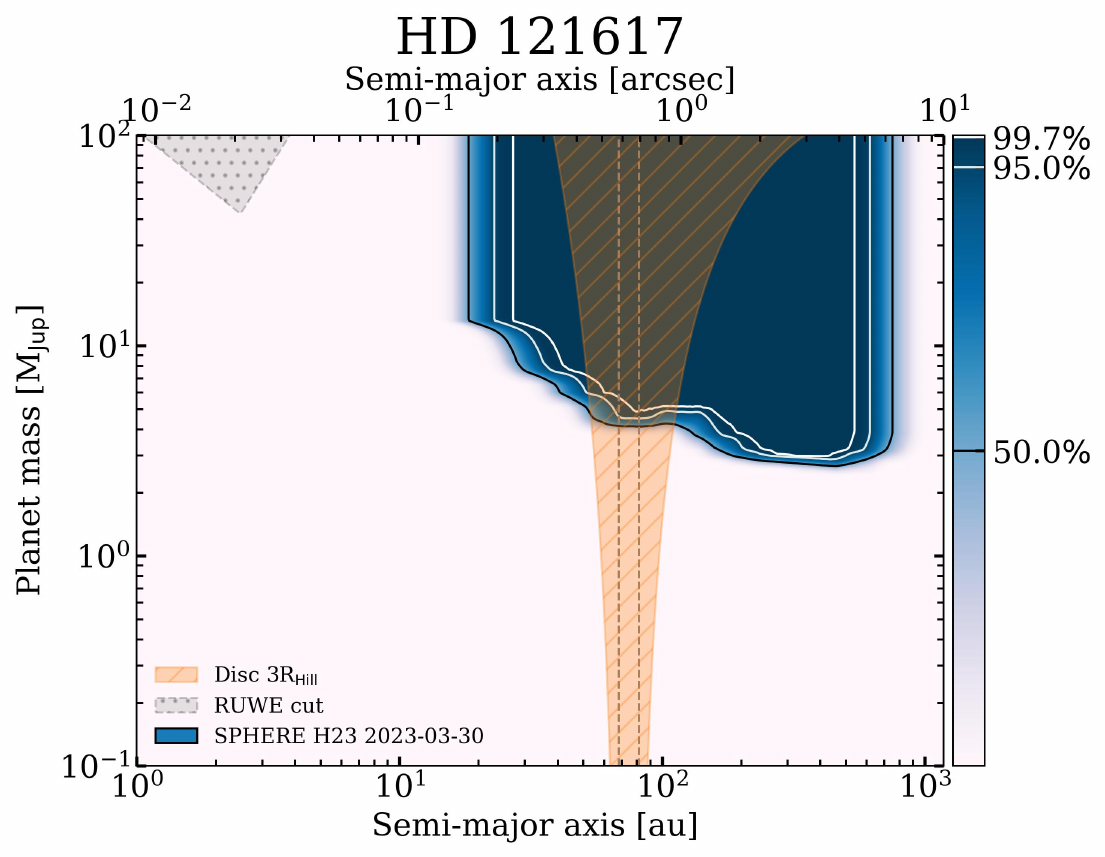}
  \includegraphics[width=0.32\hsize]{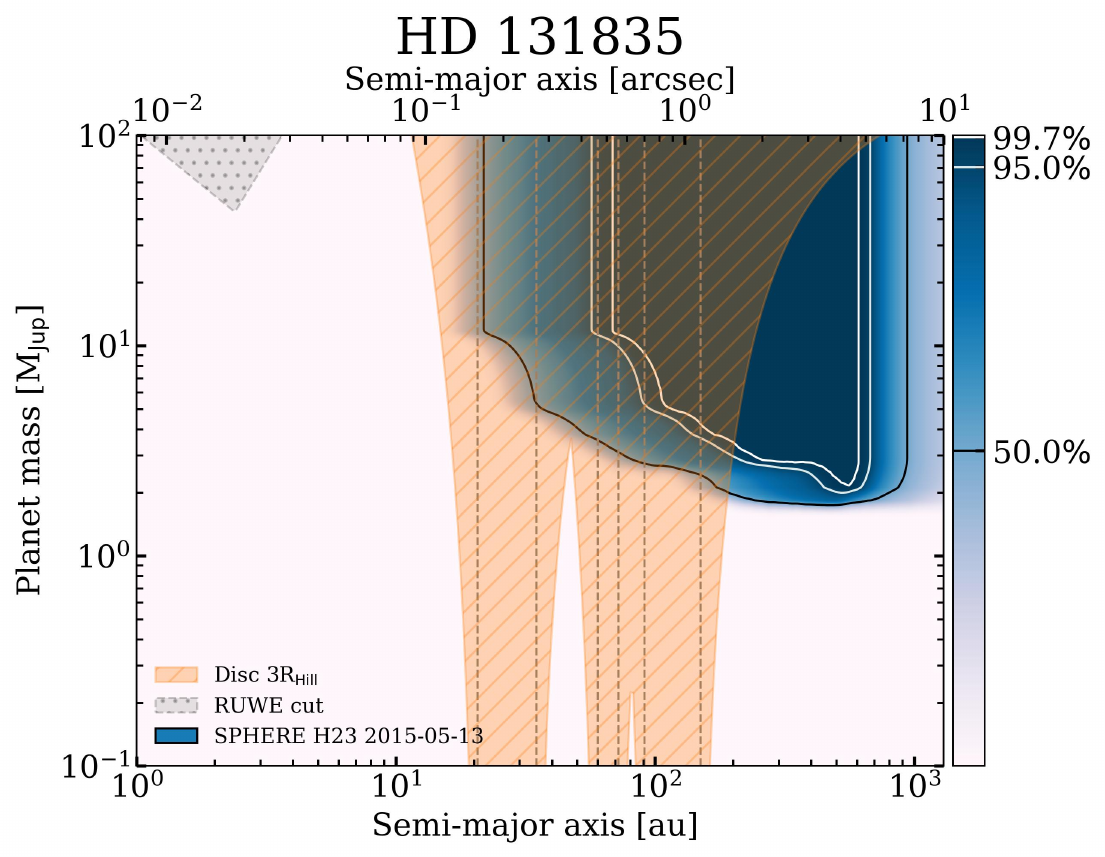}
  \includegraphics[width=0.32\hsize]{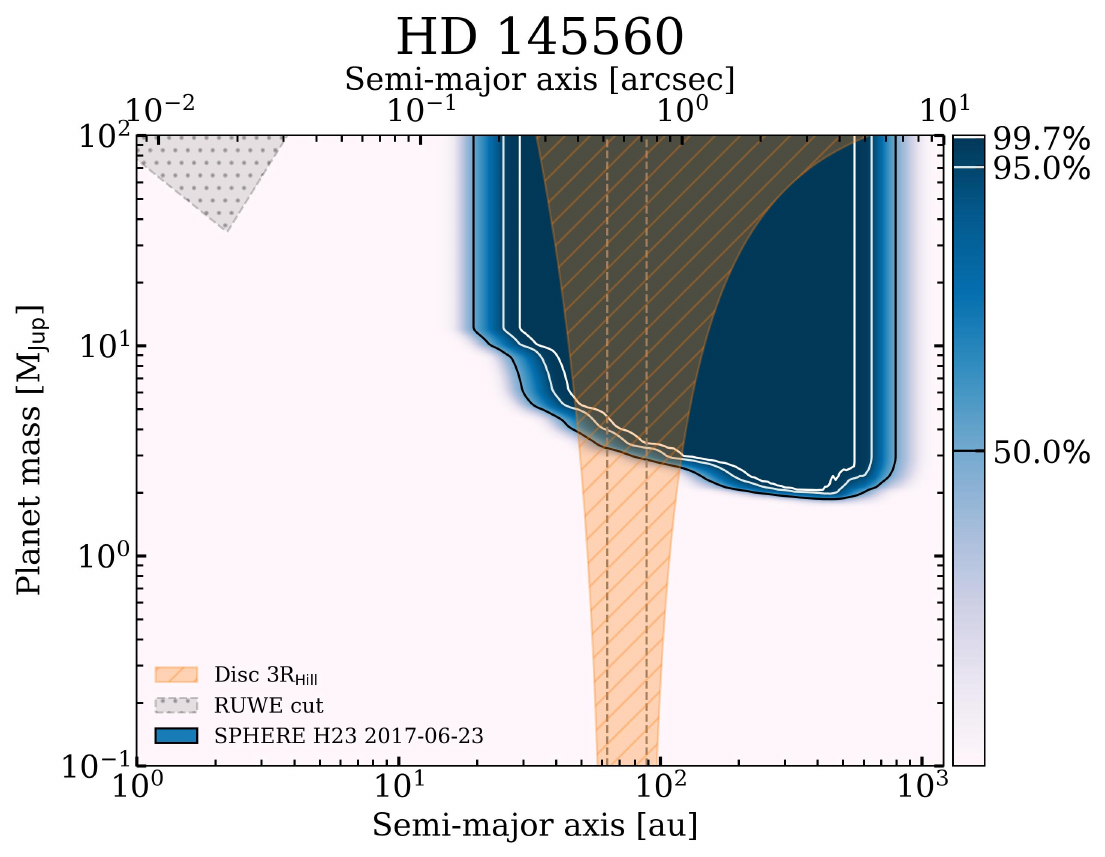}
  \includegraphics[width=0.32\hsize]{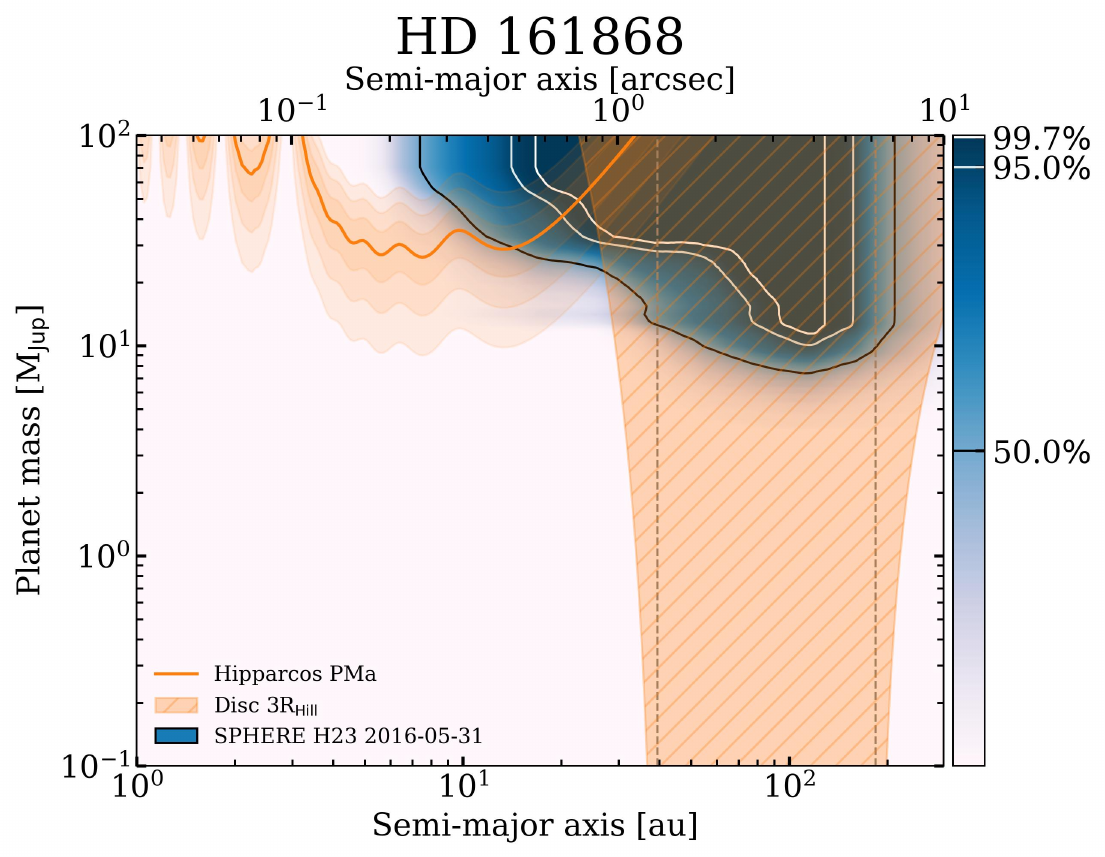}
  \includegraphics[width=0.32\hsize]{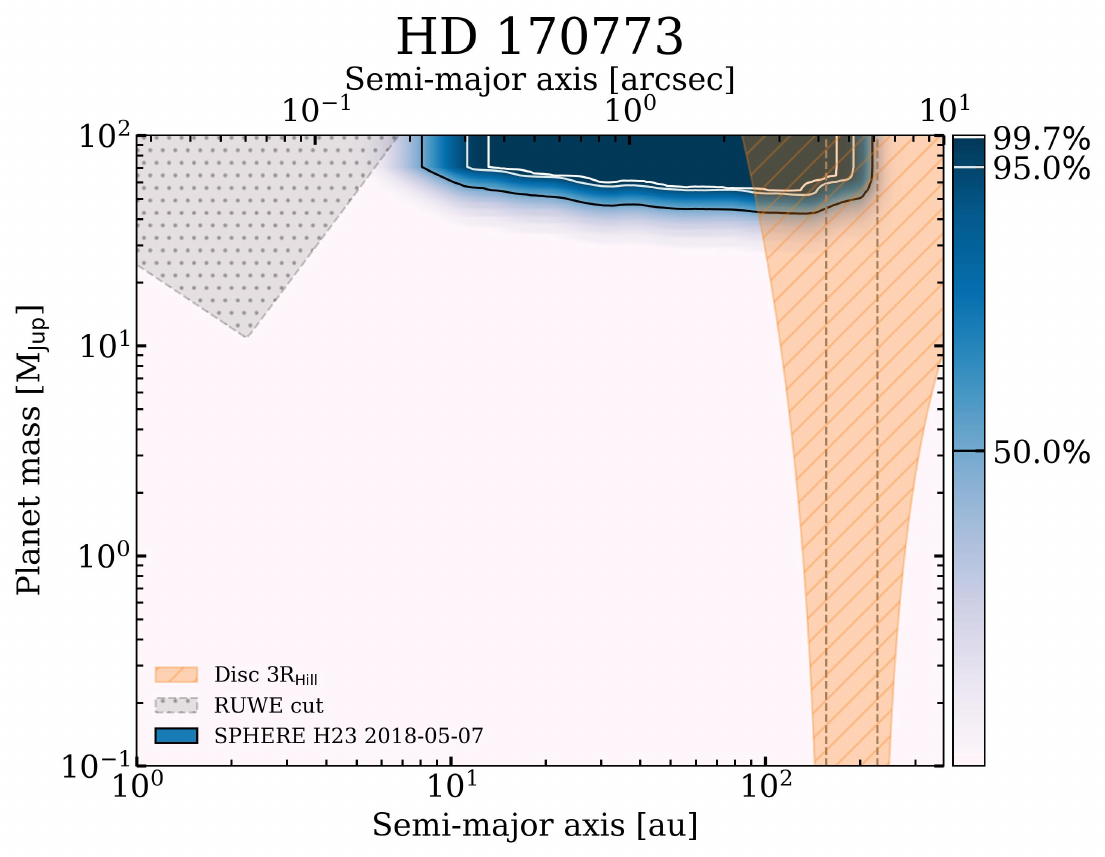}
  \includegraphics[width=0.32\hsize]{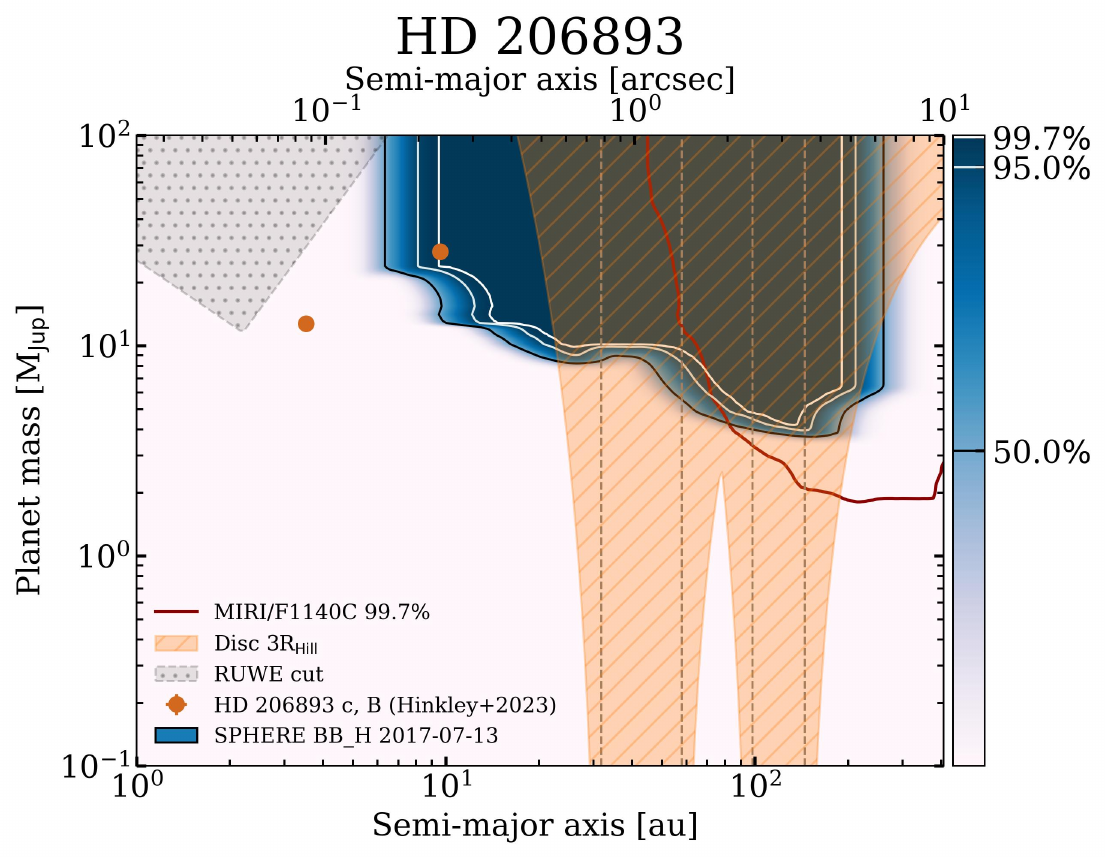}
  \includegraphics[width=0.32\hsize]{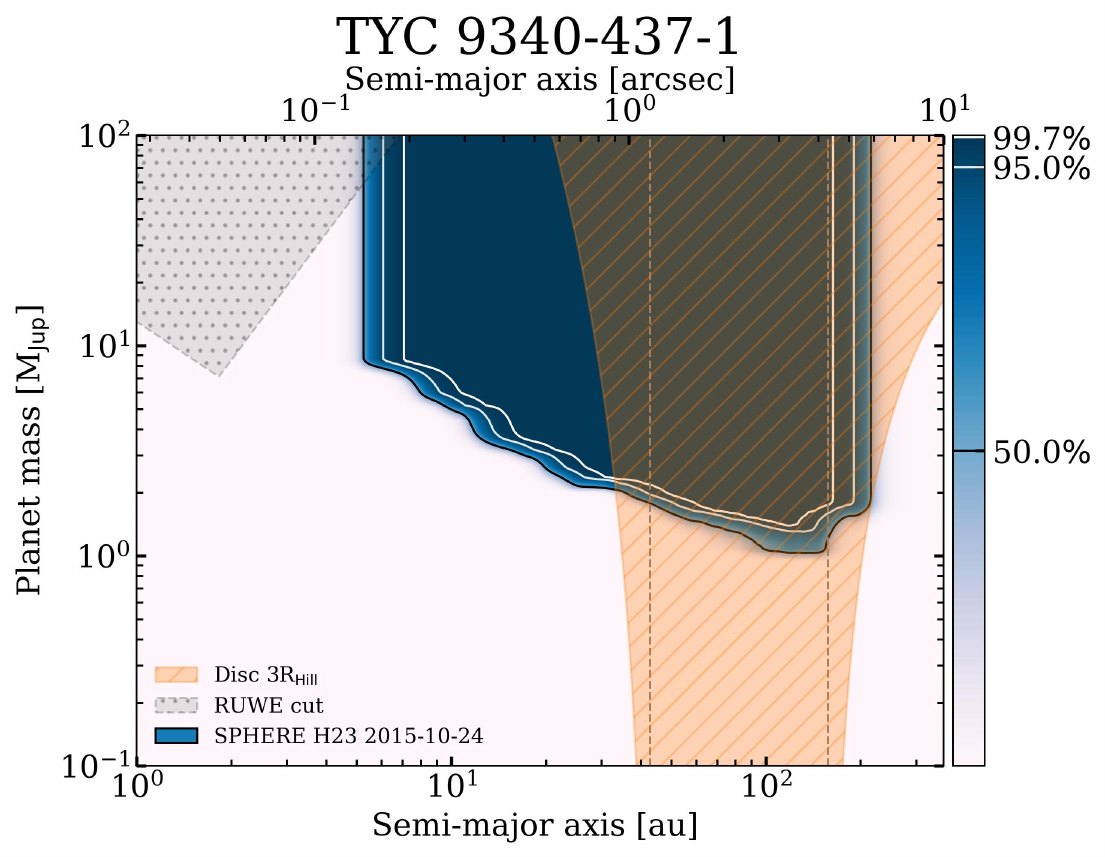}
  \includegraphics[width=0.32\hsize]{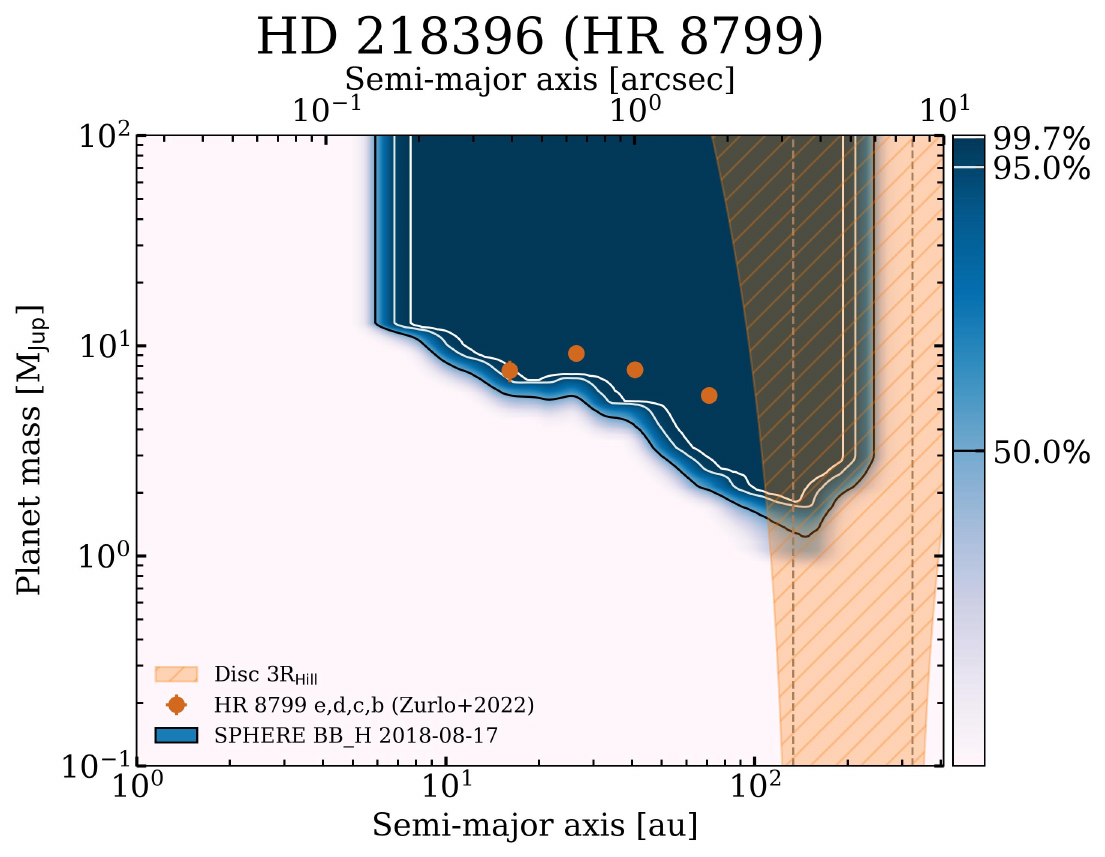}
  \caption{SPHERE DPMs for the moderately inclined systems. The blue regions with contours at 99.7\%, 95\%, and 50\% indicate the probability of detecting a planet at a 5$\sigma$ level with SPHERE. For three systems, HD 92945, HD 107146, and HD 206893, the 99.7\% JWST/MIRI F1140C contour is shown in red for comparison. The orange hatched areas denote the 3R$_{\mathrm{Hill}}$ regions from the disc edges, where 1R$_{\mathrm{Hill}}$ is defined as Eq.~\ref{eq_hill_radius}. Constraints from Gaia RUWE are indicated by grey dotted regions. The light blue curve marks the mass and location of a planet required to explain the significant proper motion anomaly. In systems with known planets, their positions are shown as orange dots.} 
    \label{fig_DPM_1}
    \end{figure*}

 \begin{figure*}
    \centering
  \includegraphics[width=0.32\hsize]{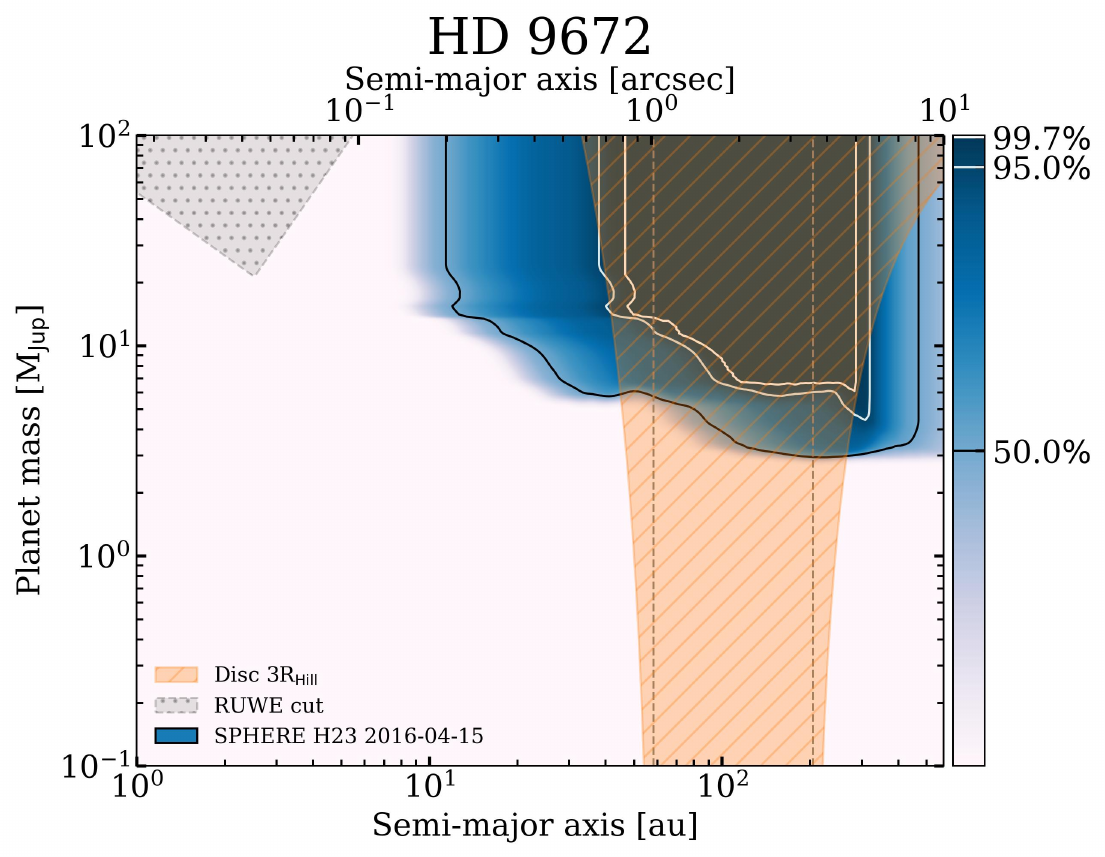}
  \includegraphics[width=0.32\hsize]{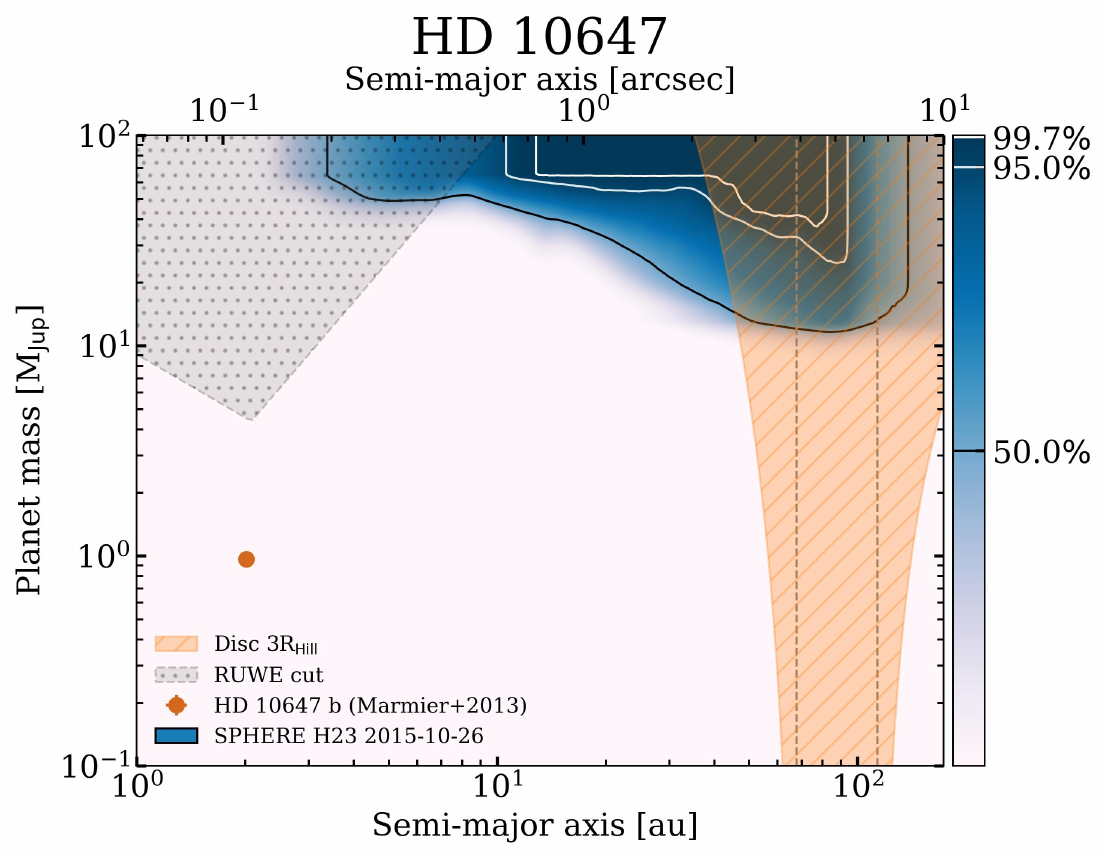}
  \includegraphics[width=0.32\hsize]{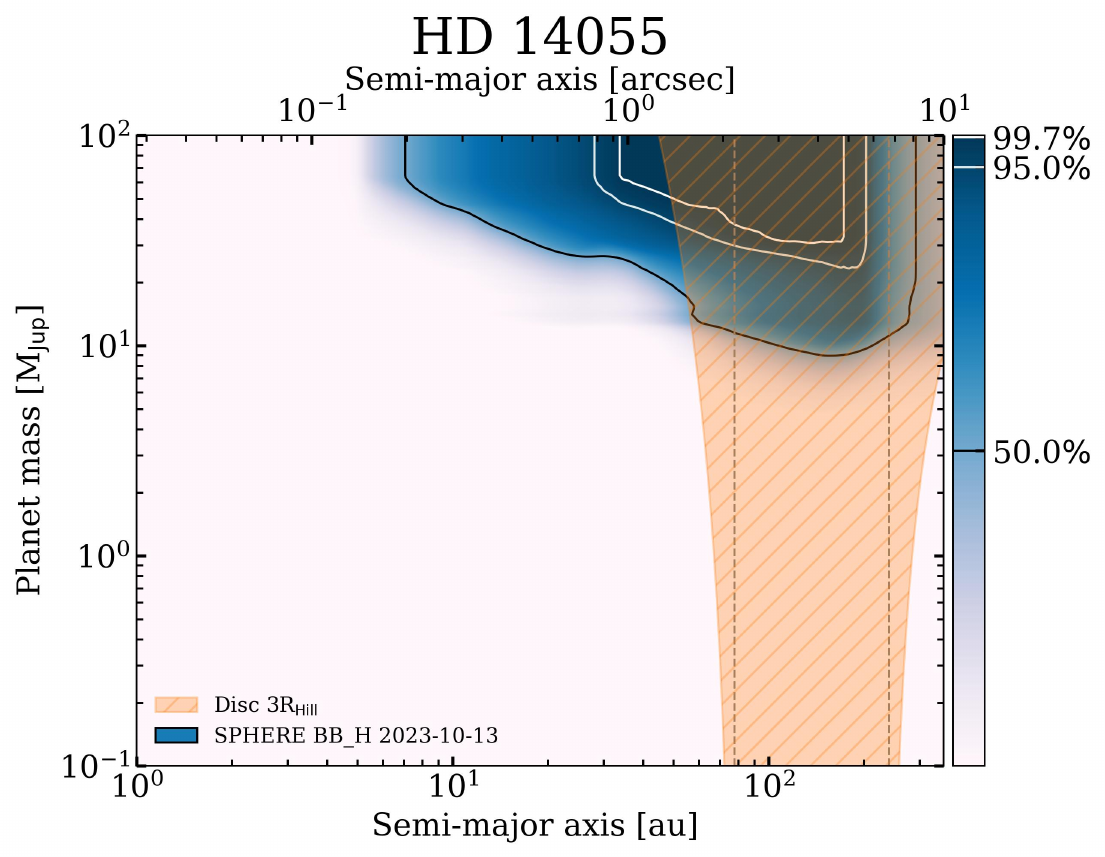}
  \includegraphics[width=0.32\hsize]{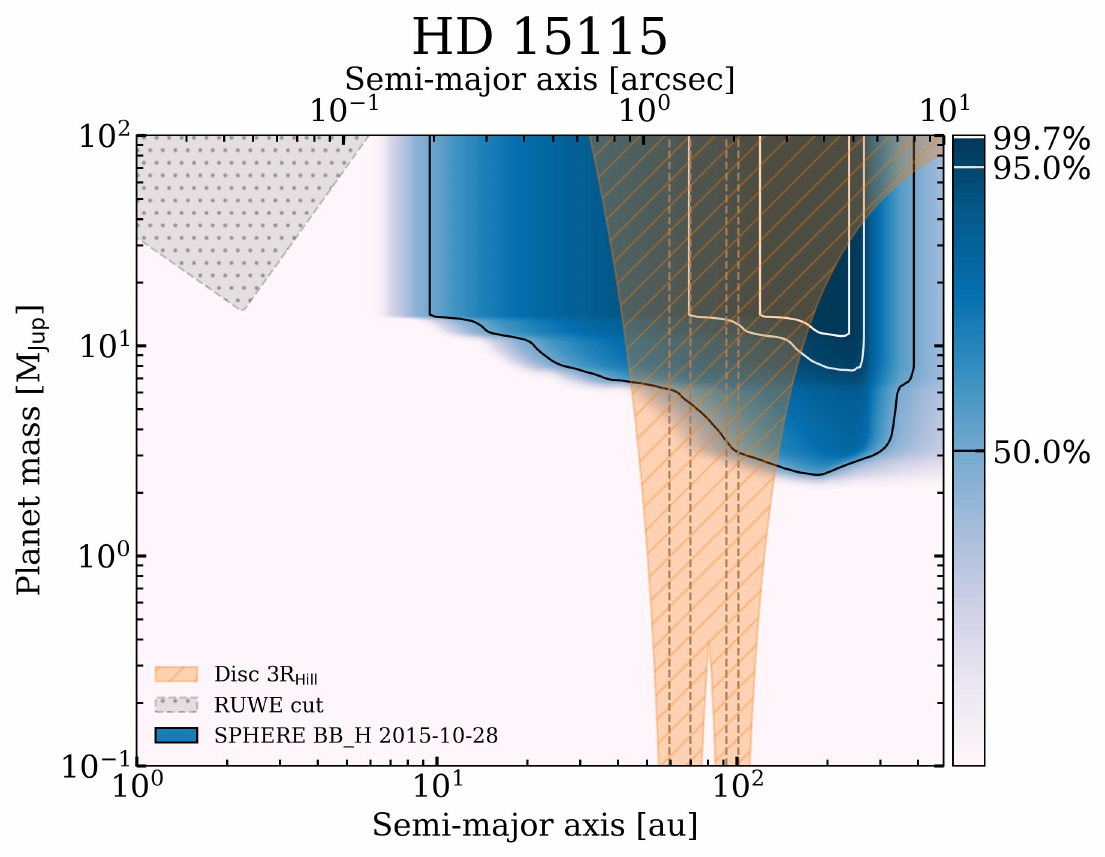}
  \includegraphics[width=0.32\hsize]{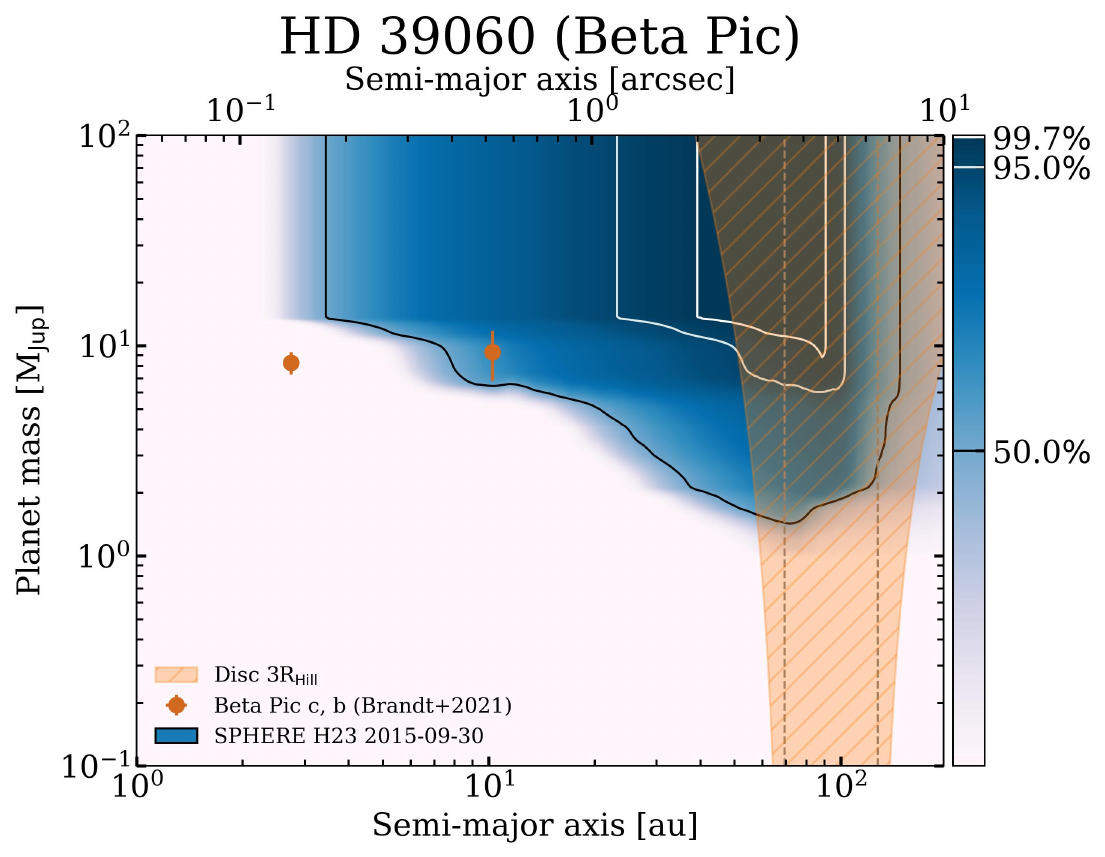}
  \includegraphics[width=0.32\hsize]{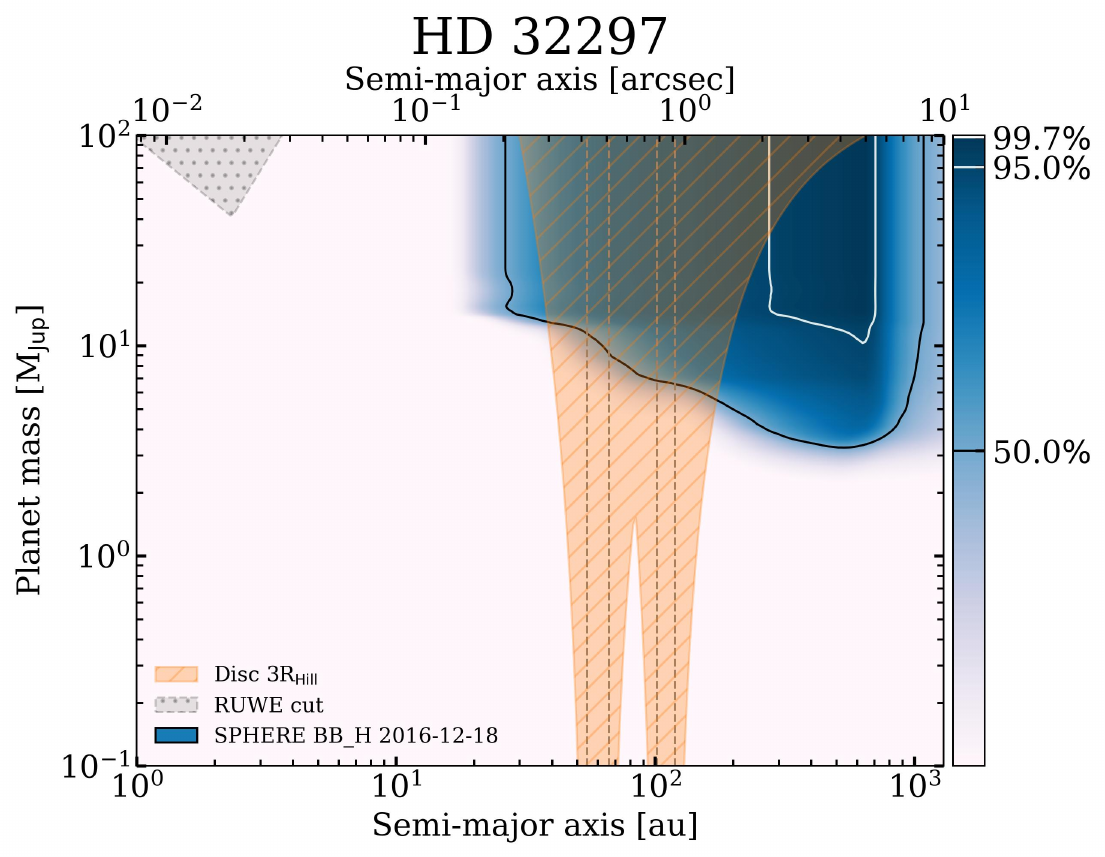}
  \includegraphics[width=0.32\hsize]{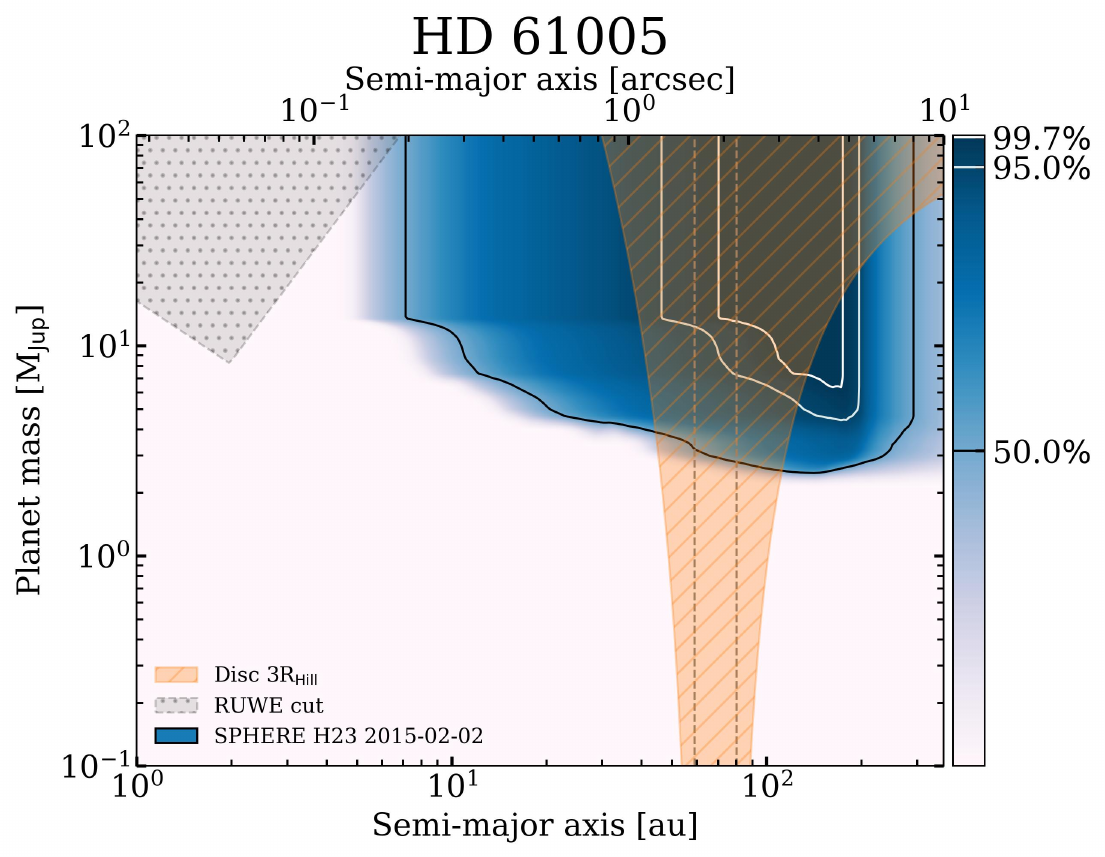}
  \includegraphics[width=0.32\hsize]{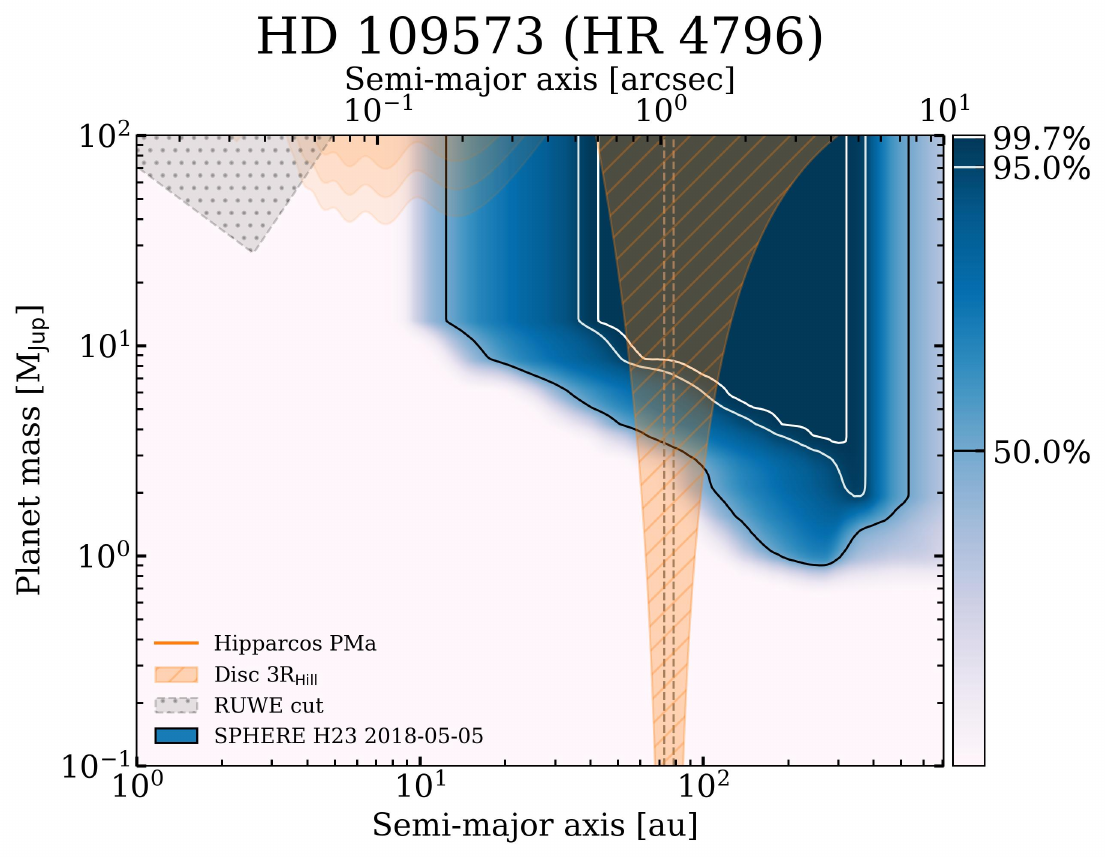}
  \includegraphics[width=0.32\hsize]{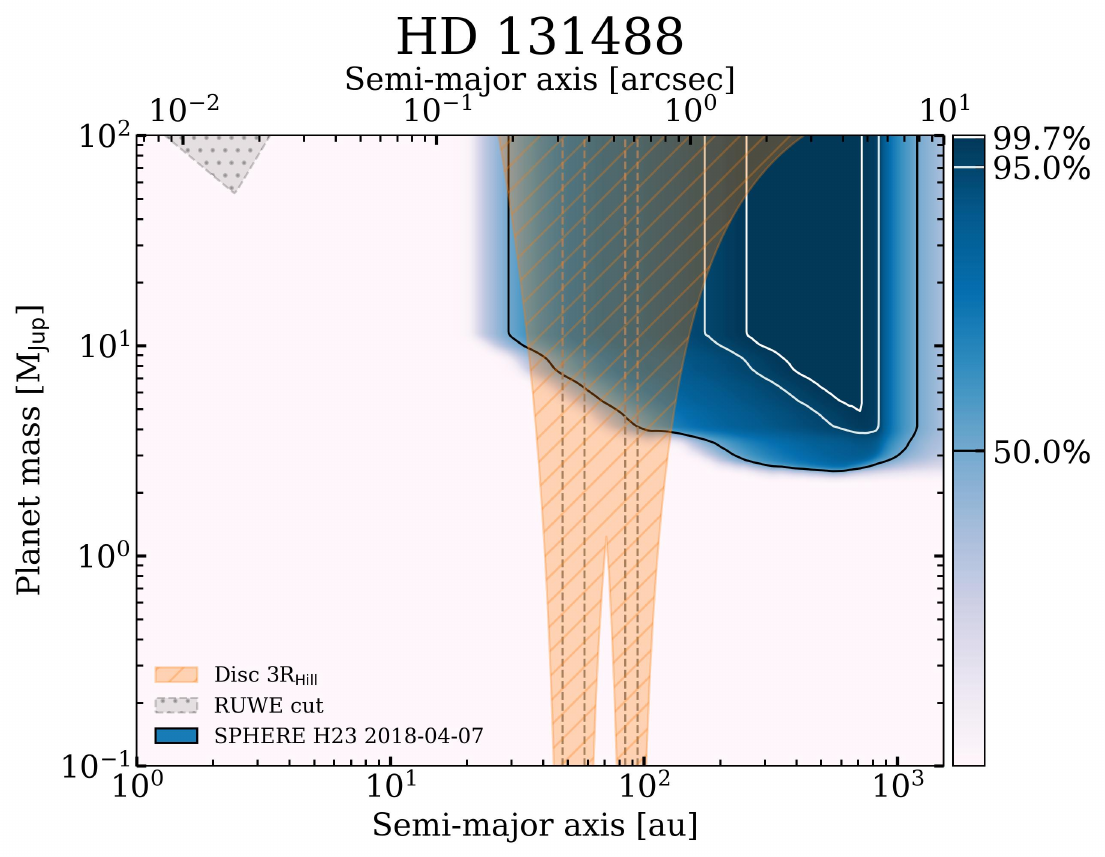}
  \includegraphics[width=0.32\hsize]{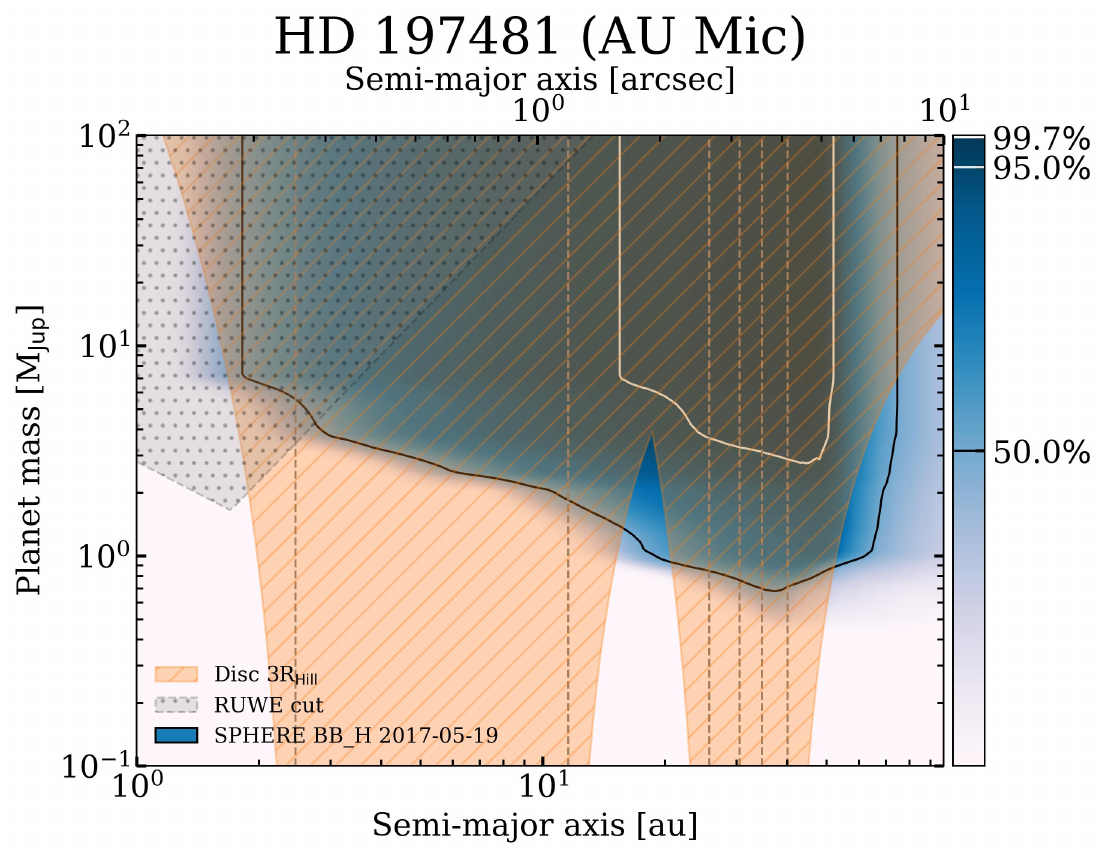}
  \caption{Same as for Fig. \ref{fig_DPM_1} but for the highly inclined systems.} 
    \label{fig_DPM_2}
    \end{figure*}

\begin{table*}
    \centering
    \caption{Planet mass upper limits and corresponding semi-major axes from the DPMs.}
    \renewcommand{\arraystretch}{1.2} 
    \begin{tabular}{lcc cc cc}
        \hline
        \hline
        \multirow{2}{*}{Star} & \multicolumn{2}{c}{Planet carving inner edge} & \multicolumn{2}{c }{Planet in gap(s)} & \multicolumn{2}{c}{PMa companion} \\
        \cmidrule(lr){2-3} \cmidrule(lr){4-5} \cmidrule(lr){6-7}
        & a$_\mathrm{p}$ [au] & m$_\mathrm{p}$ [\Mjup] & a$_\mathrm{p}$ [au] & m$_\mathrm{p}$ [\Mjup] & a$_\mathrm{p}$ [au] & m$_\mathrm{p}$ [\Mjup] \\
        \midrule
        HD\,9672        & -- (45) & -- (6) & x & x & x & x \\
        HD\,10647 (\qEri)& 38 (46) & 60 (13) & x & x & x & x \\
        HD\,14055       & 50 (55) & 50 (17) & x & x & x & x \\
        HD\,15115       & -- (44) & -- (7) & 79 & -- (4.5) & x & x \\
        HD\,15257       & -- (28) & -- (49) & x & x & x & x \\
        HD\,32297       & -- (39) & -- (13) & 80 & -- (8) & x & x \\
        HD\,39060 ($\beta$ Pictoris) & 49 (58) & 13 (1.5) & x & x & $2.74 \pm 0.03$ $^{(2)}$ & $8.3 \pm 1.0$ $^{(2)}$ \\
        HD\,61005       & -- (44) & -- (4) & x & x & x & x \\
        HD\,76582       & 65 (71) & 50 (24) & x & x & x & x \\
        HD\,84870       & 52 (55) & 34 (23) & x & x & x & x \\
        HD\,92945       & 34 (36) & 6.5 (4) & 72 & 4.5 (3) $^{\mathrm{a}}$ & 2--30 & 0.4--5 \\
        HD\,95086       & 87 (85) & 2.9 (2.4) & x & x & x & x \\
        HD\,107146      & 28 (29) & 9 (6) & 57, 79 & 5 (4), 4.5 (3.8) $^{\mathrm{b}}$ & 2.5--20 & 2.5--8 \\
        HD\,109573 (HR\,4796) & 54 (58) & 10 (4) & x & x & 4--38 & 30--400 \\
        HD\,121617      & 52 (54) & 7 (4.5) & x & x & x & x \\
        HD\,131488      & -- (35) & -- (10) & 69 & -- (5.5) & x & x \\
        HD\,131835      & -- (--) & -- (--) & 42, 85 & -- (4.5), 9 (3) & x & x \\
        HD\,145560      & 47 (49) & 5.5 (4) & x & x & x & x \\
        HD\,161868      & 26 (27) & 37 (22) & x & x & 3--25 & 9--100 \\
        HD\,170773      & 92 (95) & 53.5 (42) & x & x & x & x \\
        HD\,197481 (AU Mic) & -- (--) & -- (--) & 17, 32 & -- (1.1), -- (0.7) & x & x \\
        HD\,206893      & 22.5 (23) & 10 (8.5) & 73 & 6.5 (4.5) $^{\mathrm{c}}$ & $3.53\pm0.07$ $^{(1)}$ & $12.7 \pm 1.1$ $^{(1)}$ \\
        TYC\,9340-437-1 & 33 (33) & 2.3 (2) & x & x & x & x \\
        HD\,218396 (HR\,8799) & 109 (110) & 2 (1.5) & x & x & $16.10\pm0.03$ $^{(3)}$ & $7.6\pm0.9$ $^{(3)}$ \\
        \bottomrule
    \end{tabular}
\tablefoot{
All values are quoted at the 99.7\% confidence level, with those in brackets indicating the 50\% level. Dashes indicate regions where the observation lack sensitivity. A cross (x) marks systems with no detected gaps and no significant proper motion anomaly (PMa). Upper limits are provided for three scenarios: a planet carving the disc's inner edge, a planet located at the gap(s), and a planet responsible for the PMa.\\
\tablefoottext{a}{Deeper mass upper limit observed in gap location with JWST - MIRI F1140C: 2~\Mjup~(3$\sigma$); Bendahan-West et al. (under review).}
\tablefoottext{b}{Deeper mass upper limit observed in gap location with JWST - MIRI F1140C: 3~\Mjup, 1.8~\Mjup~(3$\sigma$); Bendahan-West et al. (under review).}
\tablefoottext{c}{Deeper mass upper limit observed in gap location with JWST - MIRI F1140C: 5~\Mjup~(3$\sigma$); Bendahan-West et al. (under review).}
Planet references:
\tablefoottext{1}{\citet{Hinkley2023},}
\tablefoottext{2}{\citet{brandt2021},}
\tablefoottext{3}{ \citet{zurlo2022}.}
}
    \label{tab:planet-constraints}
\end{table*}

The DPMs can be seen in Fig. \ref{fig_DPM_1} for the moderately inclined systems and Fig. \ref{fig_DPM_2} for the very inclined systems. The blue shaded region in these plots indicates the probability of making a 5$\sigma$ detection of a planet, if such a planet existed, with the contours representing the 99.7\%, 95\% and 50\% levels. For three systems: HD 92945, HD 107146, and HD 206893, we also show the 99.7\% contour from JWST MIRI coronagraphy at 11.4\micron{} for comparison (Bendahan-West et al., under review).

Additional planet constraints are overlaid on the SPHERE DPMs. Vertical dashed lines indicate the inner and outer edges of the disc, as well as the boundaries of any observed gaps, where the disc information comes from 
the ALMA modelling detailed in \citet{rad_arks}. Dynamically unstable regions, defined as the zones within three Hill radii (R$_{\mathrm{Hill}}$) of the disc or gap edges, are highlighted in orange hatching, with 1R$_{\mathrm{Hill}}$ defined as
\begin{equation}
    \mathrm{R}_{\mathrm{Hill}} \equiv a_p \left(\frac{m_p}{3m_*}\right)^{1/3},
\label{eq_hill_radius}
\end{equation}
where $a_p$ is the planet semi-major axis, and $m_p$ and $m_*$ are the planet and star masses, respectively. A planet on a circular orbit within 3R$_{\mathrm{Hill}}$ of a disc edge, i.e. within the orange hatched region, would likely disrupt the observed disc morphology \citep[e.g.][]{gladman1993, pearce2014, pearce2022, pearce2024}. We use Gaia astrometry and the renormalised unit weight error (RUWE) to construct the grey dotted region and exclude more planet parameters, only when RUWE $<1.4$. We follow the approach of \citet{limbach2024} for constraining companions with orbital periods shorter than the Gaia DR3's 1038-day baseline, and that of \citet{kiefer2024} for longer-period systems. In systems with significant proper motion anomaly (PMa), i.e. S/N $>3$ in the \citet{kervella2022} catalogue, we include a blue curve showing the planet parameters capable of reproducing the observed PMa. This curve is calculated following the approach described in \citet{kervella2019} and assuming a single planet in a circular orbit as done in \citet{Marino2020_hd206893}. Finally, known planets in the systems are shown as orange dots. Full details on the construction of these DPMs are presented in Bendahan-West et al. (under review).

Interpreting the DPMs enables us to place observational upper limits on undetected planets across our sample. These limits on planet mass and semi-major axis, reported in Table \ref{tab:planet-constraints}, are given at the 99.7\% and 50\% confidence level for three specific scenarios. First, we consider a single planet sculpting the inner edge of the disc via scattering, with upper limits identified by the intersection of the DPM contours and the 3R$_{\mathrm{Hill}}$ dynamically unstable orange region. The corresponding minimum mass for such planet would be defined by the diffusion timescale \citep[e.g.][]{pearce2022, pearce2024}, which in all cases is below the mass upper limits that we report. This means that we cannot rule out the presence of a planet that carved the disc inner edge via scattering. Second, we report observational upper limits on planet mass at the location(s) of the observed gap(s). While the presence of a planet within a gap is a plausible explanation for its origin, many other mechanisms can also produce similar features \citep[e.g.][]{Pearce2015, Zheng2017, Yelverton2018, Sefilian2021, Sefilian2023}; exploring the full range of gap-carving scenarios lies beyond the scope of this work. Third and finally, the full DPMs allow us to place tighter constraints on companions responsible for the significant PMa observed in some systems. For systems with known PMa companions, the corresponding parameters are listed in Table \ref{tab:planet-constraints}.

Except for the two Gigayear-old systems HD~15257 and HD~76582 for which our sensitivity is poor because of their age, direct imaging achieves typical sensitivity between 1 and 10 $M_\text{Jup}$ beyond 10 to 20 au. This region can be further restrained by considering the dynamically unstable regions within 3 Hill radii of the disc edges. All the discovered planets with direct imaging are well within this region. We note that for inclined systems (see Fig. \ref{fig_DPM_2}), the 50\% probability contour and the 95\% are well apart, indicating that single-epoch imaging does not entirely rule out the presence of giant planets even at relatively large separations depending on their projected separation. We obtained the deepest dataset on the young nearby low-mass star AU Mic, achieving a 50\% sensitivity of $0.8M_\text{Jup}$ at $\sim30$ au. 

\section{Discussion}
\label{sec_discussion}

\begin{figure}
    \centering
    \includegraphics[width=\hsize]{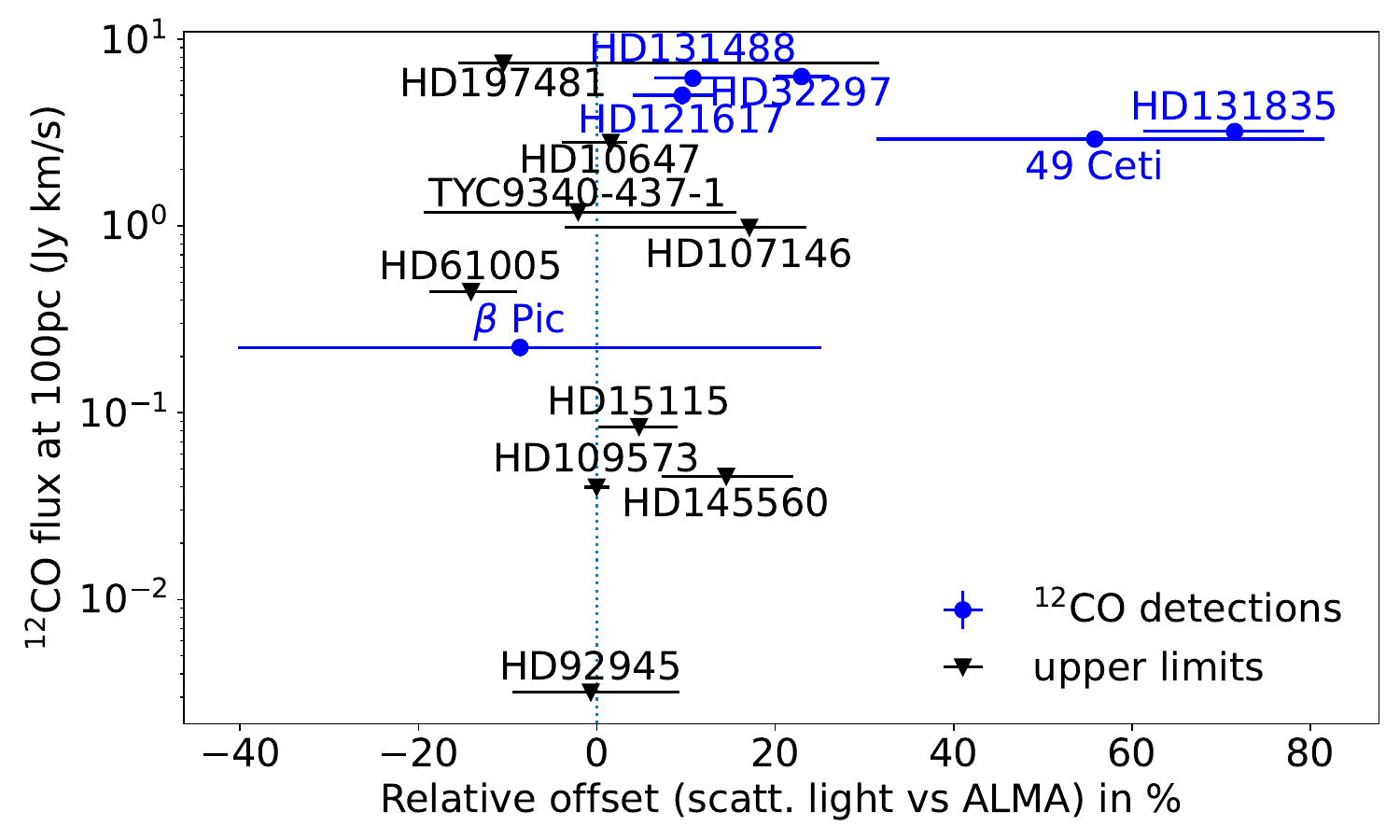}
    \caption{Integrated intensity of $^{12}$CO (J=3-2) scaled at 100~pc \cite[from][]{gas_arks} as a function of the relative offset of the millimetre- and micrometre-sized dust surface densities (positive offset means the micrometre-sized dust peaks further than the millimetre-sized dust). The six discs with a CO detection are shown in blue, while CO non-detections appear in black. The upper limit for AU Mic refers to the $^{12}$CO (J=2-1) as no (J=3-2) is available \citep{gas_arks}.}
    \label{fig_offset_gas_rich}
\end{figure}

For the first time, we can compare near-infrared and sub-millimetre images with comparable angular resolution for a sample of debris discs. This opens a new discovery space, allowing to determine offsets in the peak surface density as probed by scattered light and the ALMA continuum. Determining these offsets is fundamental to reveal the dynamical processes affecting the dust and the possible signatures of unseen planets \citep{Thebault2014}. It requires a careful methodology to take into account the different resolution and sensitivity of both regimes. Depending on the system considered, we are typically sensitive to offsets between 1 and 10 au. Our comparison between the disc surface density as probed by ALMA and scattered light images, summarised in Table \ref{tab_offsets}, shows that there are no significant offsets ($>1\sigma$) between millimetre-sized and micron-sized grains for about half of the ARKS discs detected in scattered light. Interestingly, for five of the six discs with CO gas detection, an offset $>1\sigma$ is measured. There are two exceptions, \bPic{} and HD~145560. \bPic{} is a CO-bearing disc with no significant offset. However, there is a large uncertainty on the offset. This is because the edge-on geometry of this disc makes the scattered light surface density extraction less reliable than less inclined systems, and the presence of an asymmetric CO clump \citep{Matra2017} makes the interpretation of this system more difficult. It is also worth noting that while \bPic{} has gas detections, it is relatively gas poor compared to the other ARKS gas-bearing discs, with a line luminosity a factor 4 to 22 times fainter. We should note that this factor does not account for optical depth which is likely significantly above 1 for the gas-rich ARKS discs \citep{line_arks,gas_arks}, and would push the gas-to-dust mass ratios higher. In that sense, all gas-rich discs in the sample have scattered light profiles that peak further out than ALMA. Conversely, all discs with a further out scattered light peak than ALMA but HD~145560 are gas-rich. HD~145560 is not known to host CO gas and has a significant offset detected. If CO gas is present in this system, the derived upper limit excludes a CO integrated line flux higher than 5\% than that of HD~131835 \citep{gas_arks}, a disc of similar age and distance with a somewhat earlier spectral type. 

Figure \ref{fig_offset_gas_rich} provides a summary of the offsets measured between the peak surface densities of millimetre and micron-sized dust. The vertical axis shows the amount of CO detected in those systems (blue circles) or the CO upper limits for systems without detected CO (black triangles). Despite a large scatter, the offsets seem to be higher for systems with more gas. Part of the scatter can be minimised if the outer ring of HD~131835 seen in scattered light corresponds to the marginally detected outer ring seen with ALMA, resulting in a smaller offset of $14.5^{+5.9}_{-5.7}\%$. 

Considering the individual CO-bearing systems, the peak of the micron-sized dust surface density probed in scattered light usually occurs when the gas intensity has significantly dropped (see Fig. \ref{fig_radial_profile_comparison}). We should keep in mind that this statement would only be stronger if we consider that the $^{12}$CO and $^{13}$CO are optically thick, because the CO surface density will likely have dropped more steeply than the (saturated) intensity profile suggests. This observation is in a good accordance with the theoretical prediction of \citet{Takeuchi2001}. If the gas pressure decreases with radius, as seen in those systems at radii where dust is detected, the gas orbital motion is sub-Keplerian. It orbits at a velocity $v_K\sqrt{1-\eta}$ where $v_K$ is the Keplerian velocity and $\eta>0$ is the ratio between the pressure gradient and the gravitational force. Without any drag force, the dust orbits at a velocity $v_K\sqrt{1-\beta}$, meaning that large grains with small $\beta<\eta$ rotate faster than the gas and small grains with larger $\beta>\eta$ rotate slower than the gas. In this case, the back-wind felt by the small grains causes an outward migration and the head-wind felt by the large grains causes an inward migration.
The migration halts when the grains feel no net torque. For small grains close to the blow-out size ($\beta=0.5$), simulations show that they tend to concentrate in the regions where the gas density is rapidly declining with radius \citep{Takeuchi2001}. 
The case of HD~121617 seems to defy this quantitative explanation looking at the integrated CO line flux in Fig. \ref{fig_radial_profile_comparison}, but because of optical depth effect, the intensity profile is much broader than the actual CO surface density profile, and the $^{12}$CO will indeed have dropped significantly at the location of the scattered light peak.

\section{Conclusions}
\label{sec_conclusions}

Out of the 24 discs included in the ARKS sample, 15 are detected in scattered light, comprising the targets that are most highly inclined or have the greatest fractional luminosities. The reported detections were made with at least one of these three facilities: \emph{VLT} (SPHERE), \emph{Gemini} (GPI), or \emph{HST} (NICMOS, STIS or ACS). Moreover, the entire ARKS sample has been observed by the VLT - SPHERE instrument. \tyc{} is detected for the first time in scattered light with {\em HST} - NICMOS, but follow-up observations with a higher sensitivity are necessary to compare the micron-sized dust surface density with that from millimetre-sized dust seen with ALMA in this system. With a parametric modelling approach, we show that six of the 15 discs show a significant offset in the peak surface density of the dust detected in scattered light with respect to that in thermal emission. The six discs are HD\,32297, HD\,131488, HD\,131835, 49\,Ceti, HD\,121617, and HD\,145560. Such dust segregation by size may happen for dust particles exposed to the gas drag on top of the radiation pressure of the central star, and CO gas has been confirmed in five of the six discs \citep{gas_arks}. This effect was predicted in theoretical works \citep{Takeuchi2001,Krivov2010}. Detailed hydrodynamical modelling is now required to confirm this observation (Olofsson et al. in prep). 

For many systems, interpretation is limited by the low S/N of the scattered light disc image, with bright stellar residuals that are difficult to disentangle from the disc brightness. Higher sensitivity images would allow us to perform a non-parametric surface density extraction, for instance, with the \emph{rave} algorithm, which was recently adapted to also handle polarised scattered light in addition to thermal emission images \citep{Han2022,Han2025}. The comparison to mid-infrared images also represents a promising and complementary future prospect to confirm the offsets measured in this work and probe thermal emission of warmer dust closer to the star compared to ALMA, as shown for Fomalhaut, Vega, or $\epsilon$ Eri with JWST \citep{Gaspar2023,Su2024,Wolff2025}. 

\begin{acknowledgements}

This paper makes use of the following ALMA data: ADS/JAO.ALMA\# 2022.1.00338.L, 2012.1.00142.S, 2012.1.00198.S, 2015.1.01260.S, 2016.1.00104.S, 2016.1.00195.S, 2016.1.00907.S, 2017.1.00167.S, 2017.1.00825.S, 2018.1.01222.S and 2019.1.00189.S. ALMA is a partnership of ESO (representing its member states), NSF (USA) and NINS (Japan), together with NRC (Canada), MOST and ASIAA (Taiwan), and KASI (Republic of Korea), in cooperation with the Republic of Chile. The Joint ALMA Observatory is operated by ESO, AUI/NRAO and NAOJ. The National Radio Astronomy Observatory is a facility of the National Science Foundation operated under cooperative agreement by Associated Universities, Inc. The project leading to this publication has received support from ORP, that is funded by the European Union’s Horizon 2020 research and innovation programme under grant agreement No 101004719 [ORP]. We are grateful for the help of the UK node of the European ARC in answering our questions and producing calibrated measurement sets. This research used the Canadian Advanced Network For Astronomy Research (CANFAR) operated in partnership by the Canadian Astronomy Data Centre and The Digital Research Alliance of Canada with support from the National Research Council of Canada the Canadian Space Agency, CANARIE and the Canadian Foundation for Innovation.
We respectfully acknowledge the University of Arizona is on the land and territories of Indigenous peoples. Today, Arizona is home to 22 federally recognised tribes, with Tucson being home to the O’odham and the Yaqui. Committed to diversity and inclusion, the University strives to build sustainable relationships with sovereign Native Nations and Indigenous communities through education offerings, partnerships, and community service. JM thanks the members of the EPOPEE project (Etude des POussières Planétaires Et Exoplanétaires) for the useful discussions. EPOPEE is a research group funded by the `Programme National de Planétologie' (PNP) of CNRS-INSU in France, from which JM and MB acknowledge financial support. JM thanks Steve Ertel, Stanimir Metchev and Andras Gasp\`ar for providing the HD~107146 {\em HST} - STIS data, which was published by the sorely missed Glenn Schneider \citep{Schneider2014}.
SPHERE is an instrument designed and built by a consortium consisting of IPAG (Grenoble, France), MPIA (Heidelberg, Germany), LAM (Marseille, France), LESIA (Paris, France), Laboratoire Lagrange (Nice, France), INAF Osservatorio di Padova (Italy), Observatoire de Gen\`eve (Switzerland), ETH Zurich (Switzerland), NOVA (Netherlands), ONERA (France) and ASTRON (Netherlands) in collaboration with ESO. SPHERE was funded by ESO, with additional contributions from CNRS (France), MPIA (Germany), INAF (Italy), FINES (Switzerland) and NOVA (Netherlands).  SPHERE also received funding from the European Commission Sixth and Seventh Framework Programmes as part of the Optical Infrared Coordination Network for Astronomy (OPTICON) under grant number RII3-Ct-2004-001566 for FP6 (2004-2008), grant number 226604 for FP7 (2009-2012) and grant number 312430 for FP7 (2013-2016). This work has made use of the High Contrast Data Centre, jointly operated by OSUG/IPAG (Grenoble), PYTHEAS/LAM/CeSAM (Marseille), OCA/Lagrange (Nice), Observatoire de Paris/LESIA (Paris), and Observatoire de Lyon/CRAL, and supported by a grant from Labex OSUG@2020 (Investissements d’avenir – ANR10 LABX56). MB acknowledges funding from the Agence Nationale de la Recherche through the DDISK project (grant No. ANR-21-CE31-0015). RBW was supported by a Royal Society Grant (RF-ERE-221025). JPM acknowledges research support by the National Science and Technology Council of Taiwan under grant NSTC 112-2112-M-001-032-MY3. E.C. acknowledges funding from the European Union (ERC, ESCAPE, project No 101044152).  SMM acknowledges funding by the European Union through the E-BEANS ERC project (grant number 100117693), and by the Irish research Council (IRC) under grant number IRCLA- 2022-3788. Views and opinions expressed are however those of the author(s) only and do not necessarily reflect those of the European Union or the European Research Council Executive Agency. Neither the European Union nor the granting authority can be held responsible for them. Support for BZ was provided by The Brinson Foundation. A.A.S. is supported by the Heising-Simons Foundation through a 51 Pegasi b Fellowship. TDP is supported by a UKRI Stephen Hawking Fellowship and a Warwick Prize Fellowship, the latter made possible by a generous philanthropic donation. SM acknowledges funding by the Royal Society through a Royal Society University Research Fellowship (URF-R1-221669) and the European Union through the FEED ERC project (grant number 101162711). S.E. is supported by the National Aeronautics and Space Administration through the Exoplanet Research Program (Grant No. 80NSSC23K0288, PI: Faramaz). MRJ acknowledges support from the European Union's Horizon Europe Programme under the Marie Sklodowska-Curie grant agreement no. 101064124 and funding provided by the Institute of Physics Belgrade, through the grant by the Ministry of Science, Technological Development, and Innovations of the Republic of Serbia. PW acknowledges support from FONDECYT grant 3220399 and ANID -- Millennium Science Initiative Program -- Center Code NCN2024\_001. LM acknowledges funding by the European Union through the E-BEANS ERC project (grant number 100117693), and by the Irish research Council (IRC) under grant number IRCLA- 2022-3788. AB acknowledges research support by the Irish Research Council under grant GOIPG/2022/1895. CdB acknowledges support from the Spanish Ministerio de Ciencia, Innovaci\'on y Universidades (MICIU) and the European Regional Development Fund (ERDF) under reference PID2023-153342NB-I00/10.13039/501100011033, from the Beatriz Galindo Senior Fellowship BG22/00166 funded by the MICIU, and the support from the Universidad de La Laguna (ULL) and the Consejer\'ia de Econom\'ia, Conocimiento y Empleo of the Gobierno de Canarias. AMH acknowledges support from the National Science Foundation under Grant No. AST-2307920. JBL acknowledges the Smithsonian Institute for funding via a Submillimeter Array (SMA) Fellowship, and the North American ALMA Science Center (NAASC) for funding via an ALMA Ambassadorship. SP acknowledges support from FONDECYT Regular 1231663 and ANID -- Millennium Science Initiative Program -- Center Code NCN2024\_001.

\end{acknowledgements}

\bibliography{biblio}     


\begin{appendix} 

\section{The ARKS sample and the scattered light dataset}

\subsection{Overall ARKS sample}
\label{app_ARKS_sample}

Table~\ref{tab_targets} lists the 24 discs comprising the ARKS sample, along with important stellar properties.

\begin{sidewaystable*}
    \caption{List of the 24 ARKS targets, with the main system parameters from \citet[][see there for uncertainties]{overview_arks} and their detection or non-detection in scattered light.}
    \centering

  \begin{threeparttable}
    \begin{tabular}{ m{3.2cm} r  c c c c c | c c }
    \hline
    \hline 
       Name  & Distance & Spectral type & $L_{\star}$   & Age   & $L_{\rm cold}/L_{\star}$& Incl.$^{(a)}$ & Detected in  scattered light ?\\
                   &  [pc] &      & [$L_{\odot}$] & [Myr] &   [$\times10^{-4}$]   & [$^\circ$]  &       if yes: Instrument / Filter / Mode$^{(b)}$ / Ref$^{(c)}$     \\
    \hline
    \multicolumn{8}{l}{Moderately inclined belts ($i<75\degr$)}\\
HD~15257 & 49.0 & F1V & 14 & $1000^{+800}_{-800}$ & 0.61 & $59.4\pm4.9$ & No \\
HD~161868 & 29.7 & A1V & 24 & $300^{+200}_{-200}$ & 0.65 & $66.1\pm1.5$ & No \\
HD~76582 & 48.9 & A7V & 9.9  & $1200^{+900}_{-900}$ & 2.0 & $72.6 \pm 0.7$ & No \\
HD~206893 & 40.8 & F5V & 2.8 & $160^{+20}_{-20}$ & 2.7 & $45.2\pm2.5$ & No \\
HD~218396 (HR 8799) & 40.9 & A9V & 5.5 & $33^{+7}_{-10}$ & 2.5 & $28.8 \pm 3.3$ & No$^{(d)}$ \\
HD~84870 & 88.8 & A9V & 7.9 & $300^{+200}_{-200}$ & 4.3 & $47.1\pm2.6$ & No \\
HD~170773 & 36.9 & F4V & 3.6  & $200^{+100}_{-100}$ & 4.9 & $32.9\pm1.9$ &  No \\
\rowcolor{gray!20} HD~92945 & 21.5 & K0V & 0.37 & $200^{+100}_{-100}$ & 6.6 & $65.4\pm0.6$ & ACS / F606W and F814W / I / \citet{Golimowski2011,Ren2017} \\
\rowcolor{gray!20} HD~107146 & 27.5 & G0V & 1.0 & $150^{+100}_{-50}$ & 9.8 & $19.3\pm0.6$ & ACS / F606W and F814W / I / \citet{Ardila2004,Ertel2011} \\
\rowcolor{gray!20} \tyc & 36.7 & K7V & 0.19 & $23^{+3}_{-3}$ & 10 & $18.0\pm8.0$ & NICMOS / F160W / I / this work \\
HD~95086 & 86.5 & A9V & 6.4 & $13^{+1}_{-0.6}$ & 16  & $31.6\pm3.5$ & No$^{(e)}$\\
\rowcolor{gray!20} HD~145560 & 121 & F5V & 3.3 & $16^{+2}_{-2}$ & 20 & $47.6\pm0.6$ & GPI / H / pI / \citet{Hom2020}\\
\rowcolor{gray!20} \textbf{HD~131835} & 130 & A8V & 8.9 & $16^{+2}_{-2}$ & 23 & $74.2\pm0.2$ & SPHERE / H / I / \citet{Feldt2017,hd131835_arks} \\
\rowcolor{gray!20} \textbf{HD~121617} & 118 & A1V & 14 & $16^{+2}_{-2}$ & 51 & $44.1\pm0.6$ & SPHERE / J / pI / \citet{Perrot2023} \\
\hline
\multicolumn{8}{l}{Highly inclined belts ($i>75\degr$)} \\
HD~14055 ($\gamma$ Tri) & 35.7 & A1V & 29 & $300^{+200}_{-200}$ & 0.89 & $80.7\pm0.2$ & No$^{(f)}$ \\
\rowcolor{gray!20} HD~10647 (q$^{1}$ Eri) & 17.3 & F8V & 1.6 & $1700^{+600}_{-600}$ & 2.6 & $77.8\pm0.1$ & ACS / F606W / I / \citet{Lovell2021}\\
\rowcolor{gray!20} HD~197481 (AU Mic) & 9.71 & M1V & 0.098  & $23^{+3}_{-3}$ & 3.9 & $88.3\pm0.1$ & SPHERE / H / I and pI / \citet{Boccaletti2015,Boccaletti2018}\\
\rowcolor{gray!20} HD~15115 & 48.8 & F3V & 3.7 & $45^{+5}_{-5}$ & 4.7 & $86.7\pm0.0$ & SPHERE / J / I \citep{Engler2019} \\
\rowcolor{gray!20} \textbf{HD~9672 (49 Ceti)} & 57.2 & A2V & 15 & $45^{+5}_{-5}$ & 7.2 & $78.7\pm0.2$ & SPHERE / H / I / \citet{Choquet2017} \\
 \rowcolor{gray!20}                   &        &         &        &        &                          &       & NICMOS / F110W / I / \citep{Choquet2017} \\
\rowcolor{gray!20} \textbf{HD~39060 ($\beta$ Pic)} & 19.6 & A4V & 8.7  & $23^{+3}_{-3}$ & 25 & $86.4\pm0.2$ & SPHERE  / H / pI / \citet{vanHolstein2021} \\
\rowcolor{gray!20} HD~61005 & 36.5 & G6V & 0.62  & $45^{+5}_{-5}$ & 25 & $85.9\pm0.1$ & SPHERE / H / I and pI  / \citet{Olofsson2016}\\
\rowcolor{gray!20} \textbf{HD~131488} & 152 & A3V & 12 & $16^{+2}_{-2}$ & 26 & $85.0\pm0.1$ & SPHERE, H, I \citet{Pawellek2024} \\
\rowcolor{gray!20} HD~109573 (HR 4796) & 70.8 & B9.5V & 25 & $10^{+3}_{-3}$ & 42 & $76.6\pm0.1$ & SPHERE / H / I  / \citet{Milli2017}\\
\rowcolor{gray!20}                    &        &         &        &        &                          &       & SPHERE / J / pI  / Bonduelle et al. (in prep)\\
\rowcolor{gray!20} \textbf{HD~32297} & 130 & A8V & 7.0  & $30^{+10}_{-10}$ & 61 & $88.3\pm0.0$ & SPHERE / H / I and pI / \citet{Bhowmik2019}\\
\hline
    \end{tabular}
\tablefoot{
The list is grouped into moderately and highly inclined discs, and within each group, the discs are listed in increasing fractional luminosity $L_{\rm cold}/L_{\star}$. The discs detected in scattered light and analysed in this paper are highlighted in grey. The discs in bold face have been detected in CO with ALMA. \\
\tablefoottext{a}{Inclination of the disc, as estimated from the ARKS ALMA image \citep{overview_arks}.}
\tablefoottext{b}{ I stands for total intensity while pI stands for polarised intensity. }
\tablefoottext{c}{Reference corresponding to the scattered light dataset used in this work. Additional scattered light dataset may exist.}
\tablefoottext{d}{\citet{Engler2025} detected the inner warm disc with SPHERE in polarisation, but the outer cold disc remains elusive.}
\tablefoottext{e}{A marginal detection in polarised light with SPHERE in the J band is claimed in \citet[][see their Fig. 15]{Chauvin2018}. However, our reanalysis of the data does not recover the signal.}
\tablefoottext{f}{HD~14055 was observed with IRDIS in the H band in polarised light on 2023-10-14 (PI: Moor) but not detected (private com.).}
}
\end{threeparttable}
\label{tab_targets}
\end{sidewaystable*}

\subsection{SPHERE data sets}
\label{app_SPHERE_program_ids}
 
 Table \ref{tab_SPHERE_prog_id} details the VLT - SPHERE datasets used for the modelling of the ARKS discs detected in scattered light and for the derivation of the planet sensitivity limits.
 
 \begin{table*}
    \caption{List of the 24 targets comprising the ARKS sample. Included are the ESO program ID of the SPHERE observations  (if the disc was detected with SPHERE) and the corresponding IRDIS filter.}
    \label{tab_SPHERE_prog_id}
    \centering
    \begin{tabular}{ l c c c c}
    \hline
    \hline 
       Name  & \multicolumn{2}{c}{Planet search dataset} & \multicolumn{2}{c}{Disc dataset}     \\
             & ESO program ID  & Filter$^a$                  &  ESO program ID  & Filter           \\
    \hline
    \multicolumn{5}{l}{Moderately inclined belts ($i<75\degr$)}\\
HD~15257 &  096.C-0388(A) & BB\_H & - & - \\  
HD~161868 & 097.C-0865(B) & DB\_H23 & - & - \\  
HD~76582 & 096.C-0388(A) & BB\_H & - & - \\ 
HD~206893 & 099.C-0708(A) & BB\_H & - & - \\  
HD~218396 (HR 8799) & 0101.C-0315(A) & BB\_H & - & - \\  
HD~84870 & 112.25YJ.001 & DP\_0\_BB\_H & - & - \\ 
HD~170773 & 1100.C-0481(G) & DB\_H23 & - & - \\  
\rowcolor{gray!20} HD~92945 & 1100.C-0481(D) & DB\_H23 & - & - \\  
\rowcolor{gray!20} HD~107146 & 095.C-0374(A) & DB\_H23 & - & - \\  
\rowcolor{gray!20} \tyc & 096.C-0241(A) & DB\_H23 & - & - \\ 
HD~95086 & 097.C-0865(A) & DB\_H23 & - & - \\  
\rowcolor{gray!20} HD~145560 & 099.C-0177(A) & DB\_H23 & - & - \\  
\rowcolor{gray!20} HD~131835 & 095.C-0298(A) & DB\_H23 & 095.C-0298(A) & DB\_H23 \\
\rowcolor{gray!20} HD~121617 & 1101.C-0092(J) & DB\_H23 & 0101.C-0420(A) & DP\_0\_BB\_J \\
\hline
\multicolumn{5}{l}{Highly inclined belts ($i>75\degr$)} \\
HD~14055 ($\gamma$ Tri) & 112.25YJ.001 & DP\_0\_BB\_H & - & - \\ 
\rowcolor{gray!20} HD~10647 (q$^{1}$ Eri) & 097.C-0750(A) & DB\_H23 & - & - \\  
\rowcolor{gray!20} HD~197481 (AU Mic) & 598.C-0359(B) & BB\_H & 598.C-0359(F) & DP\_0\_BB\_H \\
\rowcolor{gray!20} HD~15115 & 096.C-0640(A) & BB\_H & 098.C-0686(B) & DP\_0\_BB\_J \\
\rowcolor{gray!20} HD~9672 (49 Ceti) & 097.C-0865(A) & DB\_H23 & 096.C-0388(A) & BB\_H \\
\rowcolor{gray!20} HD~39060 ($\beta$ Pic) & 097.C-0865(A) & DB\_H23 & 0102.C-0916(B) & DP\_0\_BB\_H \\
\rowcolor{gray!20} HD~61005 & 095.C-0298(H) & DB\_H23 & 095.C-0273(A) & DP\_0\_BB\_H \\
\rowcolor{gray!20} HD~131488 & 0101.C-0753(B) & DB\_H23 & 0.101.C-0753(B) & DB\_H23 \\
\rowcolor{gray!20} HD~109573 (HR 4796) & 1100.C-0481(G) & DB\_H23 & 0104.C-0436(B) & DP\_0\_BB\_J \\
\rowcolor{gray!20} & & & 098.C-0686(A)  & BB\_H \\
\rowcolor{gray!20} \multirow{-2}{*}{HD~32297} & \multirow{-2}{*}{098.C-0686(A)} & \multirow{-2}{*}{BB\_H} & 098.C-0686(B) & DP\_0\_BB\_J \\
\hline
    \end{tabular}
    
\tablefoot{
The discs detected in scattered light and analysed in this paper are highlighted in grey.\\
\tablefoottext{a}{BB, DB and DP\_0 stand respectively for broadband, dual-band and dual-polarimetric. These acronyms correspond to the `\emph{ESO INS COMB IFLT}` keyword in the SPHERE fits headers.}
}
\end{table*}
  
\section{TYC 9340-437-1}
\label{app_TYC}

\tyc{} is an M dwarf debris disc member of the \bPic{} moving group at a distance of 36.7~pc. It was spatially resolved for the first time by \emph{Herschel}-PACS at a wavelength of $100\micron$ \citep{Tanner2020} suggesting a face-on inclination. ALMA revealed a higher resolution view of the disc at a wavelength of 1.33~mm, suggesting the disc architecture is a broad ring at 130~au (100~au width) and moderate inclination ($i=40^{\circ}$ $^{+10^{\circ}}_{-20^{\circ}}$). 

{\em HST} - NICMOS observations were carried out on UT date 2004-09-15, with the F160W filter ($\lambda=1.601$ \micron, $\Delta\lambda=0.390$ \micron{}). Although \citet{Tanner2020} reported a non-detection from {\em HST} - NICMOS observations, we revisited the NICMOS data that were obtained as part of the coronagraphic program \# 10176 (PI: I. Song). The total on-source integration time was $1855$ seconds. Eight frames were taken at two different spacecraft orientations $30^\circ$ apart. Instead of using this angular diversity to subtract the stellar Point-Spread Function (PSF), we used the multi-reference star differential imaging PSF-subtraction method developed for the ALICE program  \citep{Choquet2014_ALICE_processing,Hagan2018}. We assembled and registered large and homogeneous libraries of 33 coronagraphic images from the NICMOS archive, gathering images from multiple reference stars observed as part of several {\em HST} programs. These libraries were used to subtract the star Point-Spread Function (PSF) from each exposure of the science target with the Principal Component Analysis (PCA) KLIP algorithm \citep{Soummer2012}. This technique has proven successful in detecting many faint discs in scattered light \citep{Soummer2014,Choquet2016,Choquet2017,Choquet2018,Marshall2018,Marshall2023}.

\begin{figure}
    \centering
    \includegraphics[width=0.48\textwidth,trim = {0 2.5cm 0 0}]{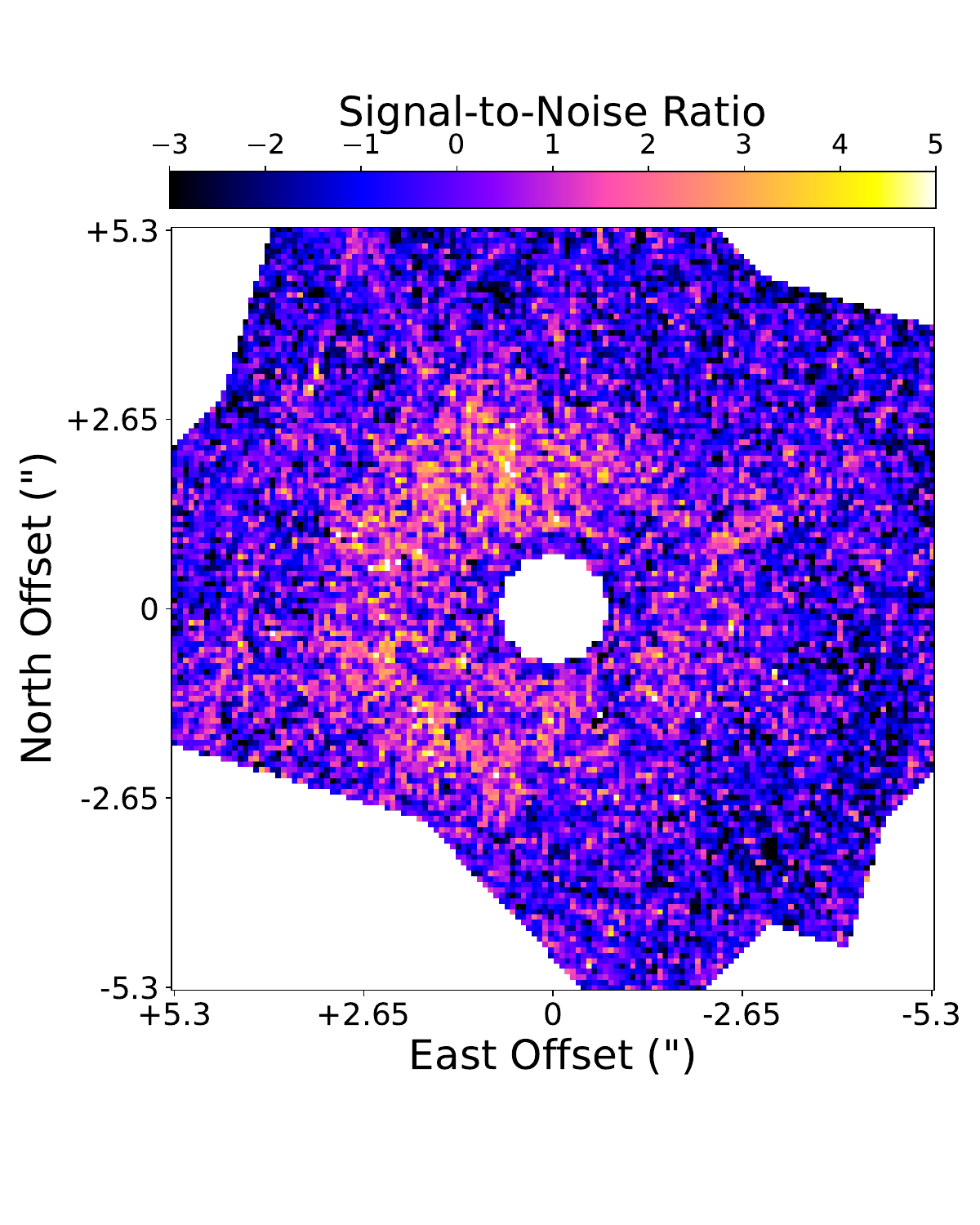}
    \caption{Signal-to-noise map of the {\em HST} - NICMOS detection of the debris disc of TYC 9340-437-1. Orientation: north is up, east is left.}
    \label{fig:tyc9340_snr}
\end{figure}

We present the signal-to-noise (S/N) map of the {\em HST} - NICMOS observation of \tyc{} in Figure \ref{fig:tyc9340_snr}. The noise maps used to build the S/N were calculated with the same method as in \citet{Choquet2018}. 136 reference star images from the PSF libraries were processed with the same method and reduction parameters as the science images. The PSF-subtracted libraries were then partitioned into 17 sets with the same number of frames as the science targets, rotated with the target image orientations, and combined. The noise maps were computed from the pixel-wise standard deviation across these sets of processed reference star images. The disc is detected at low significance with a peak S/N $\simeq 3$ (per pixel). The scattered light from the disc appears asymmetric in the image with the eastern side being substantially brighter. Assuming the same orientation (position angle, inclination) in both images, the disc extent and architecture are consistent between the scattered light and millimetre wavelengths (see Table \ref{tab_offsets}).

 \section{Parametrisation}
\label{app_parametrisation}

\subsection{Scattered light parametric modelling}
\label{app_parametrisation_scatt_light}

The dust parametrisation follows the model introduced in \citet{Augereau1999} \cite[and originally presented in][]{Artymowicz1989}. In a cylindrical coordinate system $(r,\theta,z)$, the dust volume density $\rho(r,\theta,z)$ is expressed as
\begin{equation}
\label{eq_dust_density}
\rho(r,\theta,z) = \rho_0 \left( \left( \frac{r}{r_0} \right)^{-2\alpha_\text{in}} + \left( \frac{r}{r_0} \right)^{-2\alpha_\text{out}} \right)^{-1/2} \times e^{- z^2 / 2(r \tan \psi)^2  }
,\end{equation}
where $\alpha_\text{in}$ and $\alpha_\text{out}$ are the inner/outer slopes of the volume density distribution and $h=r \tan \psi$ is the vertical scale height of the disc (linearly increasing with $r$), parametrised by the opening angle $\psi$.

In case of an elliptical disc (case of HR~4796), the reference radius $r_0$ depends on the azimuthal angle $\theta$. Therefore, the  parameter $r_0$ is superseded by the semi-major axis of the disc $a_0$, the eccentricity $e$ and the argument of pericenter $\omega$, such that
\begin{equation}
\label{eq_eccentric_ref_radius}
r_0(\theta) = \frac{a_0(1-e^2)}{1+e \cos(\omega + \theta)}
.\end{equation}

The surface density $\Sigma(r)$ can be written as 
\begin{equation}
\label{eq_surf_density}
\Sigma(r) = \int_{-\infty}^{+\infty} \rho(r,z) \,dz = \Sigma_0 \left( \left( \frac{r}{r_0} \right)^{-2\alpha_\text{in}} + \left( \frac{r}{r_0} \right)^{-2\alpha_\text{out}} \right)^{-1/2} \times \frac{r}{r_0}
,\end{equation}
where $\Sigma_0$ is a constant. 

For such a parametrisation, the radius of the maximum of the surface density $\Sigma(r)$ is not $r_0$ but also depends on $\alpha_\text{in}$, $\alpha_\text{out}$. \citet{Augereau1999} showed that it is

\begin{equation}
\label{eq_max_surface_density}
r_{\text{max},\Sigma} = r_0 \left( -\frac{\alpha_\text{in}+1}{\alpha_\text{out}+1} \right)^{1/2(\alpha_\text{in}-\alpha_\text{out})}
.\end{equation}

\subsection{ALMA continuum parametric modelling}
\label{app_parametrisation_alma}

As explained in Sect. \ref{subsec_ALMA} and in more details in \citet{rad_arks}, discs brightness profiles are converted to surface density profiles assuming a blackbody equilibrium temperature profile, then a parametric modelling using a double power-law was fitted to the ALMA continuum surface density profile:
\begin{equation}
\label{eq_surf_density_alma}
\Sigma_\text{ALMA}(r,\theta,z) = \Sigma_0 \left( \left( \frac{r}{r_0} \right)^{-2\alpha_\text{in}} + \left( \frac{r}{r_0} \right)^{-2\alpha_\text{out}} \right)^{-1/2}
.\end{equation}
We note here that this functional form parameterises the \emph{volume density} in scattered light (Eq. \ref{eq_dust_density}) but the \emph{surface density} for the ALMA continuum data (Eq. \ref{eq_surf_density_alma}), resulting in a different expression for the location of the maximum radius of the ALMA surface density:
\begin{equation}
\label{eq_max_surface_density_alma}
r_{\text{max},\Sigma_\text{ALMA}} = r_0 \left( -\frac{\alpha_\text{in}}{\alpha_\text{out}} \right)^{1/2(\alpha_\text{in}-\alpha_\text{out})}
.\end{equation}
For the same reason, the parameters $\alpha_\text{in}$, $\alpha_\text{out}$ and $r_0$ cannot be directly compared between the parametric modelling done in scattered light and with ALMA. However, given that we have used a vertical scale height linearly increasing with the radius $r$, the two parameterisations are rigorously equivalent.

\section{Modelling results}
\label{app_modelling_results}

The best parameters from the disc forward modelling are presented in Table \ref{tab_disc_parameters}. The two discs HD~107146 and HD~92945 do not appear in this table because the surface density was extracted in previous papers with a  different parametrisation \citep{Ertel2011,Golimowski2011}. 

We have made the scattered light data and best models publicly available on our dedicated website \href{https://arkslp.org}{arkslp.org} and \href{https://dataverse.harvard.edu/dataverse/arkslp/}{ARKS dataverse}.

The comparison of the ALMA and scattered light surface density profile is presented in Sect. \ref{sec_comparison}. The comparison of the vertical profiles of the discs is done in the companion paper \citet{ver_arks}. A comparison to the position angle and inclination extracted from the ARKS ALMA images is done in the last two columns of Table \ref{tab_modelling_results} and a discussion is provided below in Sect. \ref{sec_comp_PA_incl}. The best scattering phase functions derived for the discs' models are shown below in Sect. \ref{sec_SPF}.

\subsection{Comparison of the position angle and inclination between scattered light and ALMA}
\label{sec_comp_PA_incl} 

An accurate estimation of the inclination for edge-on or highly inclined systems is difficult in scattered light, as this parameter can be slightly degenerate with the vertical scale height and because the semi-minor axis which is critical to constrain the inclination is often hidden behind the coronagraphic mask or located in a noisy area of the image. The inclination differs by more than $3\sigma$ for five systems. 

For AU Mic and HD~32297, the inclination derived from total intensity is incompatible with ALMA, but that derived from polarised intensity is compatible. Polarimetry gives more weight to the $90^\circ$ scattering angles (e.g. larger projected distances to the star), while total intensity gives more weight to the small scattering angles (e.g. small projected separations). Therefore given the distortions in the AU Mic disc \citep{Boccaletti2015,Boccaletti2018}, such a difference likely reflects the complex structure of the disc. 

For HD~32297, the value derived from polarised intensity is compatible with that derived from GPI in the same mode \citep{Duchene2020}, which tends to show that there is no real discrepency between ALMA and scattered light. For \bPic, in polarised intensity, both the inclination and position angle are incompatible with ALMA. This difference also likely reflects the complex structure of the disc at short separations, with an inner warp inclined by several degrees from the main disc \citep{Apai2015}. 

For HD~131835, there is a $2^\circ$ difference in inclination between the scattered light and ALMA models, corresponding to a $3.2\sigma$ difference. As the semi-minor axis of the ring is poorly seen in scattered light, this may explain this difference. 

For HD~10647, the disc inclination in scattered light is $7^\circ$ less than with ALMA. As the vertical opening angle $\psi=0.1\pm0.07$~rad appears poorly constrained, we also tried to fix this value to a small ($\psi=0.04$~rad) or a large ($\psi=0.14$~rad) value, but that did not change the best disc inclination. Because the {\em HST} / ACS inner working angle is large, we could only use disc signals beyond 3.5\arcsec{} (61~au) which may explain the difference in inclination found with ALMA. However we cannot rule out a true misalignment between micrometre- and millimetre-size grains, as suggested for instance on the disc around $\eta$~Crv \citep{Lovell2022}.

The position angles are typically very consistent between scattered light and ALMA, except for the two edge-on systems \bPic{} already discussed above and HD~15115.  
For the latter, the position angle is $0.7^\circ$ higher in scattered light, corresponding to a $3.5\sigma$ difference with ALMA. This position angle differs by $0.5^\circ$ with that derived in \citet{Engler2019} from the same dataset but with a different technique (ellipse fitting). The structural features seen inside the belt and described in \citep{Engler2019} may therefore have an impact on the determination of the position angle. 

\begin{sidewaystable*}
\begin{threeparttable}
\caption{Best disc parameters.}
\label{tab_modelling_results}
    \centering
    \begin{tabular}{c c c c c c c c c c c c c c}
\hline
\hline
Target & Mode$^{(a)}$  & $r_0$$^{(b)}$     & $\alpha_\text{in}$ & $\alpha_\text{out}$ & $\psi$ & $i$           & PA       & e & $\omega$  & $i$ (ALMA) & PA (ALMA)  \\
               &                      &  [mas] &                          &                    &[rad$\times 10^{-2}$]$^{(c)}$ & [$^\circ$] & [$^\circ$] &  $\times 10^{-2}$ & [$^\circ$] & [$^\circ$] & [$^\circ$] \\
\hline
HD9672 & I & $2186\pm169$ & $2.6^{2.1}_{-1.3}$ & $-2.1\pm0.5$ & $3.5$ & $73\pm3$ & $109\pm2$ & - & - & $78.7\pm0.2$ & $107.9\pm0.2$  \\
HD~10647 & I & $3895\pm10$ & $9.9\pm0.2$ & $-1.1\pm0.1$ & $10\pm7$ & \cellcolor{gray!20}$71.8\pm0.1$ & $56.9\pm0.1$ & - & - & $78.8\pm0.1$ & $57.3\pm0.1$  \\
HD~15115 & I & $2042\pm100$ & $4.9\pm2.5$ & $-5.4\pm1.3$ & $2.8\pm0.6$ & $86.4\pm0.4$ & \cellcolor{gray!20} $-80.8\pm0.2$ & - & - & $86.7 \pm0.0$ & $-81.5\pm0.0$ \\
\multirow{2}{*}{HD~32297} & I & $1057\pm23$ & $1.7\pm0.3$ & $-5.0\pm0.2$ & $0.8\pm0.3$ & \cellcolor{gray!20} $87.1\pm0.1$ & $-132.7\pm0.1$ & - & - & \multirow{2}{*}{ $88.3\pm0.0$ } & \multirow{2}{*}{ $-132.5\pm0.0$ } \\
 & pI & $1000\pm36$ & $2.6\pm0.5$ & $-3.6\pm0.2$ & $1.3\pm0.3$ & $88.2\pm0.2$ & $-132.3\pm0.1$ & - & - & & \\
HD~39060 & pI  & $5007\pm88$ & $0.4\pm0.1$ & $-17.0\pm2.4$ & $10.0\pm0.5$ & \cellcolor{gray!20} $84.6\pm0.03$ & \cellcolor{gray!20} $-148.6\pm0.2$ & - & - & $86.4\pm0.2$ & $-150.0\pm0.1$ \\
\multirow{2}{*}{HD~61005} & I & $1445\pm59$ & $15.8\pm3.0$ & $-1.3\pm0.1$ & $3.9\pm0.4$ & $86.1\pm0.3$ & $70.5\pm0.2$ & - & - &\multirow{2}{*}{$85.9\pm0.1$} & \multirow{2}{*}{$70.4\pm0.1$}  \\
 & pI & $1399\pm69$ & $14.7\pm3.2$ & $-1.2\pm0.1$ & $4.9\pm1.2$ & $83.9\pm0.8$ & $71.9\pm0.8$ & - & - & &  \\
\multirow{2}{*}{HD~109573} & I & $1062\pm1$ & $34.7\pm0.3$ & $-18.6\pm0.5$ & $0.2\pm0.1$ & $76.6\pm0.1$ & $-153.5\pm0.1$ & $3\pm1$ & $-137\pm2$ & \multirow{2}{*}{$76.6\pm0.1$} & \multirow{2}{*}{$-153.5\pm0.0$} \\
 & pI & $1066\pm2$ & $34.3\pm0.7$ & $-17.9\pm0.5$ & $1.7\pm0.2$ & $77.1\pm0.2$ & $-153.3\pm0.1$ & $7\pm1$ & $-109\pm3$ & &  \\
HD~121617 & pI & $668\pm11$ & $28.7\pm5.0$ & $-6.9\pm1.1$ & $4.1\pm2.9$ & $44.8\pm1.4$ & $-119.5\pm2.1$ & $3\pm1$ & $-10\pm36$ & $44.1\pm0.6$ & $-121.3\pm0.7$  \\
HD~131488 & I & $619\pm24$  & $15.1\pm3.7$ & $-3.8\pm0.6$ & $1.8\pm0.8$ & $84.9\pm0.3$ & $-82.9\pm0.2$  & - & - & $85.0\pm0.1$ & $-82.8\pm0.0$  \\
HD~131835 & I & $785\pm24$ & $6.0\pm1.5$  & $-2.4\pm0.5$ & $3.5\pm1.4$ & \cellcolor{gray!20} $76.2\pm0.6$ & $-120.6\pm0.3$ & - & - & $74.2\pm0.2$ & $-120.8\pm0.2$  \\
HD~145560 & pI & $686\pm30$ & $4.3\pm0.9$ & $-3.8\pm0.7$ & $5.8\pm3.3$ & $46.8\pm2.0$ & $36.8\pm2.0$ & - & - & $47.6\pm0.6$ & $39.4\pm0.9$ \\
\multirow{2}{*}{HD~197481} & I & $3336\pm234$ & $1.8\pm0.5$ & $-3.5\pm0.3$ & $2.7\pm0.1$ & \cellcolor{gray!20} $89.0\pm0.1$ & $128.1\pm0.1$ & - & - & \multirow{2}{*}{$88.3\pm0.1$} & \multirow{2}{*}{$128.6\pm0.1$} \\
 & pI & $3406\pm234$ & $2.7\pm0.6$ & $-4.5\pm1.5$ & $1.0\pm0.7$ &  $86.5\pm0.8$ & $128.2\pm0.5$  & - & -   & & \\
\tyc{} & I & $2460\pm119$ & $5.4\pm1.7$ & $-7.9\pm2.1$ & $18.0\pm4.2$ & $18^{(d)}$ & $149.2^{(d)}$ & - & -  & $18.0\pm8.0$ & $149.2\pm20.2$ \\
\end{tabular}

\label{tab_disc_parameters}    
\tablefoot{
The uncertainty is given at a $1\sigma$ confidence level and comes from the MCMC alone, except for the position angle, where a systematic uncertainty of $0.04^\circ$ from the true north calibration of VLT - SPHERE \citep{Maire2016} was added quadratically (when SPHERE was used). The last two columns show, as a comparison, the best-fit inclination and position angle as extracted from the ALMA images in \citet{overview_arks} and the corresponding scattered light values are highlighted in grey if they are incompatible at a $>3\sigma$ confidence level. Values without uncertainty are fixed parameters in the fit.\\
\tablefoottext{a}{The observing mode is either I for total intensity or pI for polarised intensity.}
\tablefoottext{b}{If the disc is eccentric, the reference radius $r_0$ is replaced with the reference semi-major axis $a_0$ described in Eq. \ref{eq_eccentric_ref_radius}, and the eccentricity $e$ and the argument of pericentre $\omega$ are specified.}
\tablefoottext{c}{When expressed in radians, the opening angle, $\psi$, can also be understood as the ratio between the vertical scale height, $h$, and the reference radius, $r_0$. Therefore, the numerical value expressed in this column in rad$\times 10^{-2}$ can be understood as a percentage.}
\tablefoottext{d}{The inclination and position angle for \tyc{} were fixed parameters set to the best values derived from the modelling of the ALMA images \citep{overview_arks} because the S/N was too low in scattered light to leave them as free parameters.}

}

\end{threeparttable}
\end{sidewaystable*}

\subsection{Scattering phase functions}
\label{sec_SPF}

\begin{figure}[htb]
    \centering
    \includegraphics[width=0.5\textwidth]{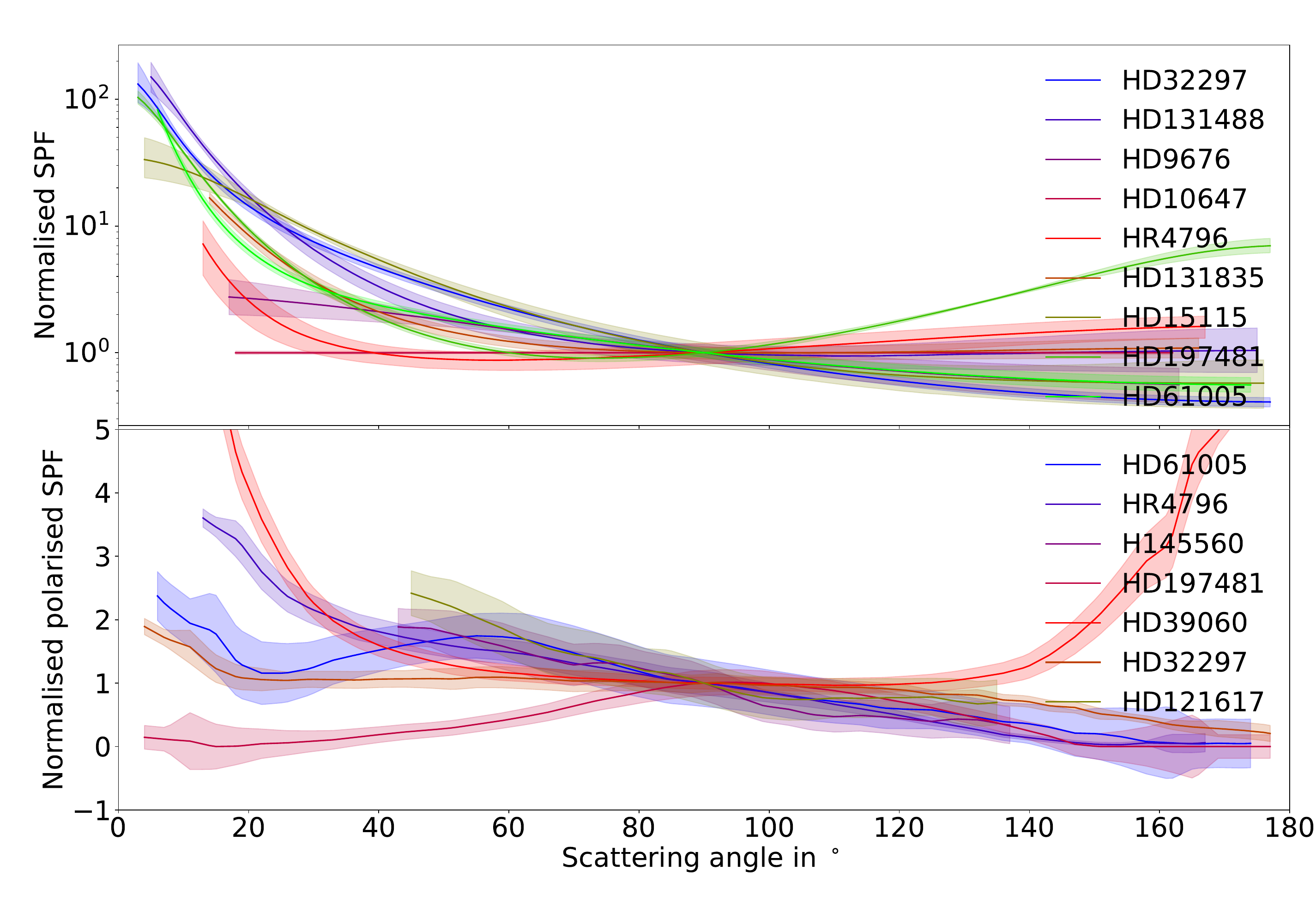}
    \caption{Scattering phase functions (top) and polarised scattering phase functions (bottom) of the ARKS discs modelled in this work.}
    \label{fig_SPF}
\end{figure}

As a by-product of the modelling, we also show in Fig. \ref{fig_SPF}, the scattering phase functions (SPF) and polarised SPF (pSPF). In total intensity, the SPF is parametrised with a two-component Henyey-Greenstein phase function \cite[][see the top panel in Fig. \ref{fig_SPF}]{Henyey1941}. In polarised intensity, the pSPF is non-parametric and estimated empirically following the method described in \citet{Olofsson2020} (bottom panel in Fig. \ref{fig_SPF}).

\section{Modelling of the inner and outer rings in the HD~131835 disc}
\label{app_HD131835}

Even though the outer ring of HD~131835 is brighter than the inner ring in the scattered light image post-processed with ADI, self-subtraction inherent to this post-processing technique may bias this conclusion. Self-subtraction is typically stronger at short separations \citep{Milli2012}. We therefore tried to constrain the intrinsic brightness ratio between the inner and the outer ring seen in scattered light. To limit the number of degrees of freedom, we assumed a similar inclination and position angle for the inner ring as for the outer ring, and varied the semi-major axis, brightness ratio and the scattering anisotropy parameter. This exercise showed that we can confidently rule out models with a surface brightness of the inner ring as bright as that of the outer ring in scattered light (as measured along the semi-major axis of the disc). This constraint is used in the dedicated companion paper \citep{hd131835_arks}.

\end{appendix}
 
\end{document}